\documentclass[journal]{IEEEtran}

\usepackage{cite} 

\usepackage[caption=false,font=footnotesize]{subfig}	
\usepackage{graphicx}

\usepackage{algorithm}
\usepackage{algorithmicx}
\usepackage{algpseudocode}

\usepackage{amsmath,amsfonts,amssymb,amsthm}
\usepackage{mathtools}
\usepackage{mathrsfs}	
\usepackage{subdepth}	

\usepackage{array}	
\usepackage{tabularx}   

\usepackage{hyperref}
\usepackage[sort,compress,capitalise]{cleveref}	
\usepackage{url}

\usepackage[dvipsnames]{xcolor} 
\usepackage{xspace} 

\usepackage{tikz}
\usetikzlibrary{babel}  
\usepackage[straightvoltages,nooldvoltagedirection]{circuitikz}

{
\theoremstyle{plain}
\newtheorem{Hypothesis}{Hypothesis}

\newtheorem{Definition}{Definition}
}

{
\theoremstyle{definition}

}

\crefname{Hypothesis}{Hyp.}{Hyps.}
\Crefname{Hypothesis}{Hyp.}{Hyps.}

\crefname{Lemma}{Lemma}{Lemmata}
\Crefname{Lemma}{Lemma}{Lemmata}

\crefname{Definition}{Def.}{Defs.}
\Crefname{Definition}{Def.}{Defs.}

\newcommand{\CIGRE}{CIGR{\'E}\xspace}

\newcommand{\DFT}[1][]{DFT\xspace}  

\newcommand{\RMS}{RMS\xspace}   

\newcommand{\FF}[1][]{\texttt{FF#1}}
\newcommand{\FB}[1][]{\texttt{FB#1}}
\newcommand{\FFB}[1][]{\texttt{FFB#1}}
\newcommand{\FT}[1][]{\texttt{FT#1}}

\newcommand{\ctrlPI}{\textup{PI}\xspace}

\newcommand{\ctrlFB}{\textup{FB}\xspace}   
\newcommand{\ctrlFF}{\textup{FF}\xspace}   
\newcommand{\ctrlFT}{\textup{FT}\xspace}   

\newcommand{\LTI}{LTI\xspace}	
\newcommand{\LTP}{LTP\xspace}	

\newcommand{\TE}[1][]{\textup{TE#1}\xspace}   

\newcommand{\AC}{AC\xspace} 
\newcommand{\DC}{DC\xspace}	

\newcommand{\DER}[1][]{DER#1\xspace}    
\newcommand{\CIDER}[1][]{CI\DER[#1]}    

\newcommand{\TDS}{TDS\xspace}	
\newcommand{\HPF}{HPF\xspace}	

\newcommand{\PLL}[1][]{PLL#1\xspace}		

\newcommand{\KPI}[1][]{KPI#1\xspace}    
\newcommand{\THD}[1][]{THD#1\xspace}    
\newcommand{\DFS}[1][]{DFS#1\xspace}	

\newcommand{\Abs}[1]{\left|#1\right|}
\newcommand{\Arg}[1]{\arg\left(#1\right)}

\newcommand{\Exp}[1]{\exp\left(#1\right)}

\newcommand{\Set}[1]{\mathcal{#1}}

\newcommand{\diag}{\operatorname{diag}}

\newcommand{\col}{\operatorname{col}}

\newcommand{\grd}{\gamma}
\newcommand{\pwr}{\pi}
\newcommand{\ctrl}{\kappa}
\newcommand{\act}{\alpha}
\newcommand{\fltr}{\varphi}

\newcommand{\spt}{\sigma}

\newcommand{\V}{\mathbf{V}}	
\newcommand{\I}{\mathbf{I}}	
\newcommand{\Y}{\mathbf{Y}}     
\newcommand{\HG}{\mathbf{H}}    

\newcommand{\phases}{\Set{P}}
\newcommand{\nodes}{\Set{N}}

\newcommand{\phsABC}{\texttt{ABC}}

\newcommand{\cmpD}{\texttt{D}}
\newcommand{\cmpQ}{\texttt{Q}}
\newcommand{\cmpDQ}{\texttt{DQ}}

\newcommand{\harmonics}{\Set{H}}
\newcommand{\formers}{\Set{S}}
\newcommand{\followers}{\Set{R}}

\newcommand{\VT}{\mathbf{v}}
\newcommand{\IT}{\mathbf{i}}
\newcommand{\XT}{\mathbf{x}}
\newcommand{\UT}{\mathbf{u}}
\newcommand{\WT}{\mathbf{w}}

\newcommand{\YT}{\mathbf{y}}

\newcommand{\AP}{\mathbf{A}}
\newcommand{\BP}{\mathbf{B}}
\newcommand{\CP}{\mathbf{C}}
\newcommand{\DP}{\mathbf{D}}
\newcommand{\EP}{\mathbf{E}}
\newcommand{\FP}{\mathbf{F}}

\newcommand{\RP}{\mathbf{R}}
\newcommand{\LP}{\mathbf{L}}
\newcommand{\GP}{\mathbf{G}}
\newcommand{\KP}{\mathbf{K}}

\newcommand{\PWM}{PWM\xspace}	

\begin{document}

\title{\huge{%
    Modelling of AC/DC Interactions of Converter-Interfaced Resources for Harmonic Power-Flow Studies in Microgrids
}}

\author{%
	Johanna~Kristin~Maria~Becker,~\IEEEmembership{Student Member,~IEEE},
	Andreas~Martin~Kettner,~\IEEEmembership{Member,~IEEE},\\
	Yihui~Zuo,~\IEEEmembership{Member,~IEEE},
	Federico~Cecati,~\IEEEmembership{Student Member,~IEEE},
	Sante~Pugliese,~\IEEEmembership{Member,~IEEE},\\
	Marco~Liserre,~\IEEEmembership{Fellow,~IEEE},
	and~Mario~Paolone,~\IEEEmembership{Fellow,~IEEE}%
	\thanks{J. Becker, Y. Zuo and M. Paolone are with the Distributed Electrical Systems Laboratory at the {\'E}cole Polytechnique F{\'e}d{\'e}rale de Lausanne (EPFL) in CH-1015 Lausanne, Switzerland (E-mail: \{johanna.becker, yihui.zuo, mario.paolone\}@epfl.ch).}%
	\thanks{A. Kettner is with PSI NEPLAN AG, 8700 Küsnacht, Switzerland (E-mail: andreas.kettner@neplan.ch).}%
	\thanks{F. Cecati, S. Pugliese and M. Liserre are with the Chair of Power Electronics at the Christian-Albrechts-Universit{\"a}t zu Kiel (CAU) in DE-24143 Kiel, Germany (E-mail: \{fc, sapu, ml\}@tf.uni-kiel.de).}%
	\thanks{This work was funded by the Schweizerischer Nationalfonds (SNF, Swiss National Science Foundation) via the National Research Programme NRP~70 ``Energy Turnaround'' (projects nr. 173661 and 197060) and by the Deutsche Forschungsgemeinschaft (DFG, German Research Foundation) via the Priority Programme DFG~SPP~1984 ``Hybrid and Multimodal Energy Systems'' (project nr. 359982322).}%
}

\maketitle







\begin{abstract}
    Modern power distribution systems experience a large-scale integration of \emph{Converter-Interfaced Distributed Energy Resources} (\CIDER[s]).
    As acknowledged by recent literature, the interaction of individual \CIDER components and different \CIDER[s] through the grid can lead to undesirable amplification of harmonic frequencies and, ultimately, compromise the distribution system stability.
    In this context, the interaction of the \DC and \AC sides of \CIDER[s] has been shown to have a significant impact.
    In order to analyze and support the mitigation of such phenomena, the authors of this paper recently proposed a \emph{Harmonic Power-Flow} (\HPF) framework for polyphase grids with a high share of \CIDER[s].
    The framework considers the coupling between harmonics, but ignores the \DC-side response of the \CIDER[s].
    Modelling the \DC side and the \AC/\DC converter
    introduces a nonlinearity into the \CIDER model that needs to be approximated for the numerical solution of the \HPF.
    This paper extends the \CIDER model and \HPF framework to address this aspect, whose inclusion is non-trivial.
    The extended \HPF method is applied to a modified version of the \CIGRE low-voltage benchmark microgrid.
    The results are compared to (i)~time-domain simulations with Simulink, (ii)~the predecessor of the extended \HPF which neglects the \DC side, and (iii)~a classical decoupled \HPF.
\end{abstract}





\section*{Nomenclature}

\begin{center}
    
\begin{tabularx}{\columnwidth}{p{3cm}p{5cm}}
    \hline
    \multicolumn{2}{c}{Generic Model of the $n$-th \CIDER}\\
    \hline
\end{tabularx}
\begin{IEEEdescription}[\IEEEusemathlabelsep\IEEEsetlabelwidth{$\formers\cup\followers$}]
    \item[$\mathbf{x}_{n}(t)$]
        The state vector of a state-space model
    \item[$\mathbf{u}_{n}(t)$]
        The input vector of a state-space model
    \item[$\mathbf{y}_{n}(t)$]
        The output vector of a state-space model
    \item[$\mathbf{w}_{n}(t)$]
        The disturbance vector of a state-space model
    \item[$\mathbf{A}_{n}(t)$]
        The system matrix of a \emph{Linear Time-Periodic} (\LTP) system
    \item[$\mathbf{B}_{n}(t)$]
        The input matrix of an \LTP system
    \item[$\mathbf{C}_{n}(t)$]
        The output matrix of an \LTP system
    \item[$\mathbf{D}_{n}(t)$]
        The feed-through matrix of an \LTP system
    \item[$\mathbf{E}_{n}(t)$]
        The input disturbance matrix of an \LTP system
    \item[$\mathbf{F}_{n}(t)$]
        The output disturbance matrix of an \LTP system
    \item[$\Sigma_{n}$]
        The \LTP system of a generic \CIDER
    \item[$f$] 
        An arbitrary frequency
    \item[$f_1\coloneqq\frac{1}{T}$]
        The fundamental frequency w.r.t. period $T$
    \item[$h\in\harmonics$] 
        The harmonic order ($\harmonics\coloneqq\{-h_{\max},\ldots,h_{\max}\}$)
    \item[$f_{h}$]
        The harmonic frequency of order $h$ ($f_{h}\coloneqq h\cdot f_{1}$)
    \item[$\mathbf{X}_{n,h}$] 
        The Fourier coefficients of $\mathbf{x}_{n}(t)$ ($h\in\harmonics$)
    \item[$\hat{\mathbf{X}}_{n}$] 
        The column vector composed of the $\mathbf{X}_{n,h}$
    \item[$\mathbf{A}_{n,h}$] 
        The Fourier coefficients of $\mathbf{A}_{n}(t)$ ($h\in\harmonics$)
    \item[$\hat{\mathbf{A}}_{n}$] 
        The Toeplitz matrix composed of the $\mathbf{A}_{n,h}$
    \item[$\grd$]
        The power grid
    \item[$\pwr$]
        The power hardware of a \CIDER
    \item[$\ctrl$]
        The control software of a \CIDER
    \item[$\spt$] 
        The generic setpoint of a \CIDER
    \item[$\epsilon$] 
        The \DC equivalent of a \CIDER
    \item[$\delta$] 
        The \DC filter stage of a \CIDER
    \item[$\act$]
        The actuator of a \CIDER
    \item[$\mathbf{T}_{\ctrl|\pwr,n}$] 
        The transformation from $\pwr$ to $\ctrl$
    \item[$\Hat{\mathbf{G}}_{n}$] 
        The harmonic-domain closed-loop gain
    \item[$\Hat{\mathbf{Y}}_{n}$] 
        The harmonic-domain internal response of the \CIDER
    \item[$\Hat{\mathbf{Y}}_{\grd,n}$] 
        The harmonic-domain grid response of the \CIDER
    \item[$\mathbf{y}_{o,n}(t)$]
        The vector of the operating point vector of a linearized \CIDER model
    \item[$S_\spt$]
        The power setpoint of a grid-following \CIDER
    \item[$v_{\grd,\cmpD}(t)$]
        The direct component of $v_{\grd}$
    \item[$v_{\grd,\cmpQ}(t)$]
        The quadrature component of $v_{\grd}$
    \item[$V_{\grd,\cmpD,h}$]
        The Fourier coefficient of $v_{\grd,\cmpD}(t)$ ($h\in\harmonics$)
    \item[$\mathbf{e}_i$]
        The unit vector in $\mathbb{R}^2$ with all entries zero except the $i$th element.
\end{IEEEdescription}
\hrule

\end{center}

\begin{center}
    
\begin{tabularx}{\columnwidth}{p{3cm}p{5cm}}
    \hline
    \multicolumn{2}{c}{Grid Model and \HPF Algorithm}\\
    \hline
\end{tabularx}
\begin{IEEEdescription}[\IEEEusemathlabelsep\IEEEsetlabelwidth{$\formers\cup\followers$}]
    \item[$n\in\nodes$]
        A three-phase node ($\nodes\coloneqq\{1,...,N\}$)
    \item[$\formers\cup\followers$]
        A partition of $\nodes$ ($\nodes=\formers\cup\followers$, $\formers\cap\followers=\emptyset$)
    \item[$\formers$]
        The nodes with grid-forming \CIDER[s]
    \item[$\followers$] 
        The nodes with grid-following \CIDER[s]
    \item[$\I_{\formers}$]
        The phasors of the injected currents at all $s\in\formers$
    \item[$\V_{\followers}$]
        The phasors of the nodal voltages at all $r\in\followers$
    \item[$\HG$] 
        The compound nodal hybrid matrix (w.r.t. $\formers,\followers$)
    \item[$\HG_{\formers\times\followers}$] 
        The block of $\HG$ linking $\V_{\formers}$ and $\V_{\followers}$
    \item[$\hat{\V}_{\formers}$] 
        The column vector composed of the Fourier coefficients of $\V_{\formers}$
    \item[$\hat{\HG}_{\formers\times\followers}$] 
        The Toeplitz matrix of the Fourier coefficients of $\HG_{\formers\times\followers}$ (i.e., $\HG_{\formers\times\followers}(f)$ evaluated at $f=f_{h}$)
    \item[$\Delta\Hat{\mathbf{V}}_{\formers}$]
        The mismatch equations w.r.t. $\Hat{\mathbf{V}}_{\formers}$
    \item[$\Delta\Hat{\mathbf{I}}_{\followers}$] 
        The mismatch equations w.r.t. $\Hat{\mathbf{I}}_{\followers}$
    \item[$\partial_{\formers}$]
        The partial derivative w.r.t. $\Hat{\mathbf{I}}_{\formers}$
    \item[$\partial_{\followers}$] 
        The partial derivative w.r.t. $\Hat{\mathbf{V}}_{\followers}$
\end{IEEEdescription}
\hrule

\end{center}

\begin{center}
    
\begin{tabularx}{\columnwidth}{p{3cm}p{5cm}}
    \hline
    \multicolumn{2}{c}{Validation}\\
    \hline
\end{tabularx}
\begin{IEEEdescription}[\IEEEusemathlabelsep\IEEEsetlabelwidth{$(\cdot)_{+/-/0}$}]
    \item[$\mathbf{X}_{h}$]
        The Fourier coefficient of a polyphase electrical quantity ($h\in\harmonics$)
    \item[$e_{\textup{abs}}(\mathbf{X}_{h})$]
        The maximum absolute error over all phases w.r.t. $\Abs{\mathbf{X}_{h}}$ between \HPF and \TDS
    \item[$e_{\textup{arg}}(\mathbf{X}_{h})$]
        The maximum absolute error over all phases w.r.t. $\angle{\mathbf{X}_{h}}$ between \HPF and \TDS
\end{IEEEdescription}
\hrule
\end{center}

\section{Introduction}
\label{sec:intro}

%
%
%




\IEEEPARstart{M}{odern} power distribution systems host a large number of \emph{Converter-Interfaced Distributed Energy Resources} (\CIDER[s]), such as renewable generation, energy storage systems, and electric-vehicle charging stations.
In such systems excessive harmonic distortion may occur~\cite{Jrn:PSE:PEC:2004:Enslin} due to interactions between \AC/\DC converters and their components (i.e., \DC-link capacitors, \AC-side filters, and the associated controls).
Unsatisfactory performance and instability of the \CIDER[s] due to inadequate tuning~\cite{Jrn:Lu:2018,Rep:PSE:SA:2018:Canizares} or inaccurate modelling of the \DC-side~\cite{Jrn:Yazdani:2005} have been reported.
Moreover, it has been shown that the interactions between \AC and \DC side of the \CIDER[s] have an significant impact on the propagation of harmonics~\cite{Jrn:Lian:2006}.
While the component sizes of the \DC-link capacitors and \AC-side filters are tightly restricted by volume, weight, and cost \cite{Jrn:M:2013:Friedli}, the controller design offers considerable freedom for mitigating such problems.
For the controller design process, it is crucial to understand the creation, propagation and coupling of harmonics due to the converter and \AC/\DC interactions.

\subsection{Literature Review}
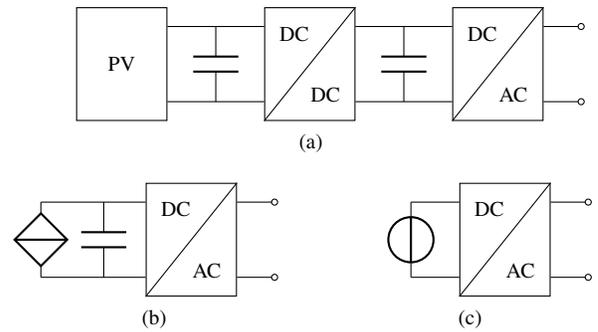
\begin{figure}[t]
    \centering
    \subfloat[]
    {%
        \centering
    	{

\ctikzset{bipoles/length=1.0cm}
\ctikzset{bipoles/diode/height=.2}
\ctikzset{bipoles/diode/width=.15}
\tikzstyle{block}=[rectangle, draw=black,fill=white, minimum size=8mm, inner sep=0pt]
\tikzstyle{dot}=[circle, draw=black, fill=black, minimum size=2pt, inner sep=0pt]
\tikzstyle{measurement}=[rectangle,draw=black,minimum size=1mm,inner sep=0pt]
\tikzstyle{signal}=[-latex]

\footnotesize

\begin{circuitikz}
    
    \def\x{1.0}
    \def\y{1.0}
    
    \node[rectangle,draw=black,fill=white,minimum height=15mm,minimum width=12mm,inner sep=0pt] (C) at (0,0) {};
    \draw (C.south west) to (C.north east);
    \node at ($(C.north west)+(0.4,-0.35)$) {DC}; 
    \node at ($(C.south east)+(-0.4,0.35)$) {AC};
    
    \coordinate (CEN) at ($(C.east)-(0*\x,0.5*\y)$);
    \coordinate (CEP) at ($(C.east)+(0*\x,0.5*\y)$);
    
    \coordinate (CWN) at ($(C.west)-(0*\x,0.5*\y)$);
    \coordinate (CWP) at ($(C.west)+(0*\x,0.5*\y)$);
    
    \draw (CEP) to[short,-o] ($(CEP)+(0.5*\x,0)$);
    \draw (CEN) to[short,-o] ($(CEN)+(0.5*\x,0)$);
    
    \node[rectangle,draw=black,fill=white,minimum height=15mm,minimum width=12mm,inner sep=0pt] (D) at (-2.5*\x,0) {};
    \draw (D.south west) to (D.north east);
    \node at ($(D.north west)+(0.4,-0.35)$) {DC}; 
    \node at ($(D.south east)+(-0.4,0.35)$) {DC};
    
    \coordinate (DEN) at ($(D.east)-(0*\x,0.5*\y)$);
    \coordinate (DEP) at ($(D.east)+(0*\x,0.5*\y)$);
    
    \coordinate (DWN) at ($(D.west)-(0*\x,0.5*\y)$);
    \coordinate (DWP) at ($(D.west)+(0*\x,0.5*\y)$);
    
    \draw (DEP) to[short] (CWP);
    \draw (DEN) to[short] (CWN);
    
    \draw ($0.5*(DEP)+0.5*(CWP)$) to[capacitor] ($0.5*(DEN)+0.5*(CWN)$);
   
   \node[rectangle,draw=black,fill=white,minimum height=15mm,minimum width=12mm,inner sep=0pt] (P) at (-5*\x,0) {PV};
    
    \coordinate (PEN) at ($(P.east)-(0*\x,0.5*\y)$);
    \coordinate (PEP) at ($(P.east)+(0*\x,0.5*\y)$);
    
    \coordinate (PWN) at ($(P.west)-(0*\x,0.5*\y)$);
    \coordinate (PWP) at ($(P.west)+(0*\x,0.5*\y)$);
    
    \draw (DWP) to[short] (PEP);
    \draw (DWN) to[short] (PEN);
    
    \draw ($0.5*(DWP)+0.5*(PEP)$) to[capacitor] ($0.5*(DWN)+0.5*(PEN)$);
        
\end{circuitikz}

}
        \label{fig:DCSide:det}
    }
    
    \subfloat[]
    {%
        \centering
    	{

\ctikzset{bipoles/length=1.0cm}

\footnotesize

\begin{circuitikz}
    
    \def\x{1.0}
    \def\y{1.0}
    
    \node[rectangle,draw=black,fill=white,minimum height=15mm,minimum width=12mm,inner sep=0pt] (C) at (0,0) {};
    \draw (C.south west) to (C.north east);
    \node at ($(C.north west)+(0.4,-0.35)$) {DC}; 
    \node at ($(C.south east)+(-0.4,0.35)$) {AC};
    
    \coordinate (CEN) at ($(C.east)-(0*\x,0.5*\y)$);
    \coordinate (CEP) at ($(C.east)+(0*\x,0.5*\y)$);
    
    \coordinate (CWN) at ($(C.west)-(0*\x,0.5*\y)$);
    \coordinate (CWP) at ($(C.west)+(0*\x,0.5*\y)$);
    
    \draw (CEP) to[short,-o] ($(CEP)+(0.5*\x,0)$);
    \draw (CEN) to[short,-o] ($(CEN)+(0.5*\x,0)$);
    
    \coordinate (D) at (-2*\x,0);
    
    \coordinate (DEN) at ($(D.east)-(0*\x,0.5*\y)$);
    \coordinate (DEP) at ($(D.east)+(0*\x,0.5*\y)$);
    
    \coordinate (DWN) at ($(D.west)-(0*\x,0.5*\y)$);
    \coordinate (DWP) at ($(D.west)+(0*\x,0.5*\y)$);
    
    \draw (DEP) to[short] (CWP);
    \draw (DEN) to[short] (CWN);
    
    \draw ($0.4*(DEP)+0.6*(CWP)$) to[capacitor] ($0.4*(DEN)+0.6*(CWN)$);
   
    \draw (DEN) to[cI] (DEP);
        
\end{circuitikz}

}
        \label{fig:DCSide:CS}
    }
    \subfloat[]
    {%
        \centering
    	{

\ctikzset{bipoles/length=1.0cm}
\ctikzset{bipoles/diode/height=.2}
\ctikzset{bipoles/diode/width=.15}
\tikzstyle{block}=[rectangle, draw=black,fill=white, minimum size=8mm, inner sep=0pt]
\tikzstyle{dot}=[circle, draw=black, fill=black, minimum size=2pt, inner sep=0pt]
\tikzstyle{measurement}=[rectangle,draw=black,minimum size=1mm,inner sep=0pt]
\tikzstyle{signal}=[-latex]

\footnotesize

\begin{circuitikz}
    
    \def\x{1.0}
    \def\y{1.0}
    
    \node[rectangle,draw=black,fill=white,minimum height=15mm,minimum width=12mm,inner sep=0pt] (C) at (0,0) {};
    \draw (C.south west) to (C.north east);
    \node at ($(C.north west)+(0.4,-0.35)$) {DC}; 
    \node at ($(C.south east)+(-0.4,0.35)$) {AC};
    
    \coordinate (CEN) at ($(C.east)-(0*\x,0.5*\y)$);
    \coordinate (CEP) at ($(C.east)+(0*\x,0.5*\y)$);
    
    \coordinate (CWN) at ($(C.west)-(0*\x,0.5*\y)$);
    \coordinate (CWP) at ($(C.west)+(0*\x,0.5*\y)$);
    
    \draw (CEP) to[short,-o] ($(CEP)+(0.5*\x,0)$);
    \draw (CEN) to[short,-o] ($(CEN)+(0.5*\x,0)$);
    
    \coordinate (D) at (-1.25*\x,0) {};
    
    \coordinate (DEN) at ($(D.east)-(0*\x,0.5*\y)$);
    \coordinate (DEP) at ($(D.east)+(0*\x,0.5*\y)$);
    
    \draw (DEP) to[short] (CWP);
    \draw (DEN) to[short] (CWN);
 
    \draw (DEN) to[V] (DEP);
        
\end{circuitikz}

}
        \label{fig:DCSide:VS}
    }
    \caption
    {%
        Different approximations of the \DC side of a \CIDER.
        Detailed representation \cref{fig:DCSide:det}, the current-source representation \cref{fig:DCSide:CS} and the voltage-source representation \cref{fig:DCSide:VS}.
    }
    \label{fig:DCSide}
\end{figure}

In the recent past, the modelling of \CIDER[s] for frequency ranges beyond the power-system fundamental component has been a prominent research topic. 
To this end, different levels of abstraction can be applied w.r.t. the DC-side model~\cite{Jrn:DeCarne:2022}.
In reality, the \DC side is typically composed of the following components (see \cref{fig:DCSide:det}): a \DC source, a \DC/\DC converter (e.g., a boost converter), and a \DC-link capacitor connected to the \AC/\DC converter (e.g., \cite{Jrn:Wu:2011,Jrn:Kjaer:2005,Jrn:Blaabjerg:2004}).
However, this detailed model is unnecessarily complex for many studies.
For most purposes, the dynamics of the \DC source and the \DC/\DC converter can be neglected.
In this case, as shown in \cref{fig:DCSide:CS} the elements are approximated by a current source, which emulates their aggregate behaviour (e.g., active power-point tracking).
This representation is commonly used for harmonic analysis of \CIDER[s] \cite{Jrn:Lu:2018,Jrn:Gao:2021}.
If the \DC-link capacitor is sufficiently large (i.e., the \DC-link ripples are negligible), the model can be simplified further.
Namely, the entire \DC side can be represented by a \DC voltage source connected directly to the \AC/\DC converter (see \cref{fig:DCSide:VS}).
This simplistic model is often used for impedance modelling and stability analysis of \CIDER[s] \cite{Jrn:Cespedes:2013,Jrn:Rygg:2017} and power grids \cite{Jrn:Pogaku:2007}.

When modelling the \DC side of \CIDER[s], the \AC/\DC converter needs to be included.
In this respect, different levels of abstraction can be applied.
For instance, one can either consider, or neglect, the effects of switching in the converter.
If the switching needs to be considered in the harmonic analysis, a \emph{Double Fourier Series} (\DFS) results from the convolution of the spectra of the switching signals and the electrical quantities \cite{Jrn:TIA:2009:McGrath}.
The calculation of the \DFS involves Bessel functions, whose evaluation is non-trivial.
In the recent literature, it has been proposed to use look-up tables in combination with harmonic state-space modelling for this purpose \cite{Jrn:TPD:2020:Fu}.
However, this approach incurs significant computational burden and compromises the scalability of the model. 
If at least the high-frequency contributions of the switching action can be neglected, the \AC/\DC converters can be represented by average models \cite{Jrn:TPD:2010:Chiniforoosh,Dis:Peralta:2013}.

Recently, \emph{Linear Time-Periodic} (\LTP) systems theory~\cite{Ths:CSE:LTP:1991:Wereley} has been employed to analyse power-system harmonics in grids with high shares of \CIDER[s] \cite{Jrn:Rico:2003,Jrn:Kwon:2016}.
For instance, harmonic-domain models of \CIDER[s] which consider in detail the impact of the power control loop~\cite{Jrn:Gao:2021} or a \PLL-based grid synchronization~\cite{Jrn:Cespedes:2013} have been developed.
Moreover, a small-signal model of a \CIDER including the \DC-side dynamics has been proposed in~\cite{Jrn:Kwon:2016}.
All these studies focus on the analysis of individual \CIDER[s] and, by consequence, ignore any interactions via the grid.
On the other hand, most system-oriented approaches for harmonic analysis do not consider the coupling between harmonics.
Such a decoupled \HPF was proposed for instance in \cite{Ulinuha:AUPEC:2007}, where the resources are represented as independent and superposed harmonic current sources and the system equations are solved independently at each harmonic frequency.
By contrast, \cite{Bathurst:TPD:2000} proposes a framework which is capable to handle nonlinear active devices (e.g., power converters) and coupling effects between harmonics.
This approach is both modular and generally applicable, but only at the system level.
That is, it does not provide any information about how typical devices such as \CIDER[s] can be represented in a general, modular, and accurate way.
Hence, there is a need for an \HPF method which combines the aforementioned advantages (i.e., generality, modularity, and accuracy) of device- and system-oriented approaches.

\subsection{Contributions of this Paper}
The authors of this paper recently proposed a 
framework for the \emph{Harmonic Power-Flow} (\HPF) study of polyphase grids with a high share of \CIDER[s] \cite{jrn:2020:kettner-becker:HPF-1,jrn:2020:kettner-becker:HPF-2}.
The method combines the following features: i) generality and modularity of the underlying modelling framework, ii) high accuracy of the underlying models and iii) consideration of the coupling between harmonics.
The \HPF method is based on polyphase circuit theory and \LTP systems theory.
More precisely, the \HPF problem is formulated based on the closed-loop transfer functions of the \CIDER[s] and the hybrid nodal equations of the grid.
Notably, the \CIDER model in \cite{jrn:2020:kettner-becker:HPF-2} represents the \AC-side components and their associated controllers in detail (incl. coupling between harmonics), but does not consider the \DC side.

This paper extends the aforementioned \HPF framework proposed in \cite{jrn:2020:kettner-becker:HPF-1,jrn:2020:kettner-becker:HPF-2} to include the \DC side of the \CIDER[s].
Therefore, the extended \HPF method is capable of analysing \AC/\DC interactions in addition to converter interactions through the grid.
In this respect, the previously mentioned current-source model is employed.
That is, the \DC side is represented by a current source and a \DC-link capacitor which are connected to the \AC/\DC converter.
The \AC/\DC converter itself is represented by an average model.
This introduces a nonlinearity into the model, which needs to be approximated for the numerical solution of the \HPF problem.
To this end, a suitable linearization is incorporated into the \HPF algorithm.
In summary, the contributions of the paper are as follows.
\begin{itemize}
    \item 
        The extended \CIDER model, which includes the \DC side modelling, is presented and validated through time-domain simulations with Simulink.
    \item 
        The \HPF method is extended to account for the extended \CIDER model and validated based on a modified version of the \CIGRE low-voltage benchmark microgrid.
    \item
        The impact of \AC/\DC interactions on the propagation of harmonics is investigated based on a comparison of the \HPF including and excluding the \DC side of the \CIDER[s].
    \item
        The proposed \HPF method is benchmarked w.r.t. an existing decoupled \HPF that solves the problem independently at each harmonic frequency.
\end{itemize}

The rest of this paper is organized as follows.
\cref{sec:model-rsc} gives a generic description of the \CIDER model including the linearization.
\cref{sec:algo} shows how the proposed description is incorporated into the \HPF framework.
The specific model of a grid-following \CIDER including the \DC-side dynamics is thoroughly illustrated in \cref{sec:model-DC}.
\Cref{sec:val-rsc} and \cref{sec:val-sys} show the validation of the extended framework for an individual resource and for an entire system, respectively.
The conclusions are drawn in \cref{sec:conclusion}.

\section{Generic Model of the \CIDER}
\label{sec:model-rsc}

As mentioned in \cref{sec:intro}, this paper extends the \HPF framework proposed in \cite{jrn:2020:kettner-becker:HPF-1,jrn:2020:kettner-becker:HPF-2}.
Unless otherwise stated, all hypotheses formulated in \cite{jrn:2020:kettner-becker:HPF-1,jrn:2020:kettner-becker:HPF-2} and briefly summarized in Appendix~\ref{app:HPF_hyp} hold unchanged.

\subsection{Primer on Linear Time-Periodic Systems}

Harmonic analysis can be performed by means of \emph{Linear Time-Periodic} (\LTP) systems theory, which is a generalization of \emph{Linear Time-Invariant} (\LTI) systems theory \cite{Ths:CSE:LTP:1991:Wereley}.
This concept has recently been employed to analyze harmonics in power systems dominated by \CIDER[s] (e.g., \cite{Jrn:Wang:2018,jrn:2020:kettner-becker:HPF-1,jrn:2020:kettner-becker:HPF-2}).

Throughout this paper, all quantities are assumed to be time-periodic w.r.t. an underlying period $T$, which is the inverse of the fundamental frequency $f_{1}$ (i.e., $T=\frac{1}{f_{1}}$).
\footnote{%
Notably, the fundamental frequency does not need to be $50$~Hz.
In power systems, it is set by controllers that act on time-scales substantially longer than the controllers of the \CIDER[s].
Thus, the fundamental frequency used for the \HPF analysis can be fixed a priori in an independent analysis (i.e., using the power balance equations of the system).
}
As known from Fourier analysis, any time-periodic signal (i.e., real- or complex-valued) can be represented by a Fourier series as
\begin{align}
	    \mathbf{x}(t)
    &=	\sum\limits_{h\in\harmonics}\mathbf{X}_{h}\Exp{j h 2\pi f_{1} t}
    \label{eq:TP:matrix}
\end{align}
where $\mathbf{X}_{h} \in \mathbb{C}$ is the complex Fourier coefficient at the $h$-th harmonic of the fundamental frequency $f_{1}$, with $h\in\harmonics\subset\mathbb{Z}$.
In case the signal $\mathbf{x}(t)$ is real-valued, the positive and negative spectrum are complex conjugates of each other.
By consequence:
\begin{equation}
    \mathbf{X}_{h} = \mathbf{X}_{-h}^{*}
\end{equation}

As known, the multiplication of two signals in time domain results in a convolution of their spectra in frequency domain.
\begin{equation}
                    \mathbf{A}(t)\mathbf{x}(t)
    \leftrightarrow \mathbf{A}(f)*\mathbf{X}(f)
    =               \hat{\mathbf{A}}\hat{\mathbf{X}}
    \label{eq:TP:time2freq}
\end{equation}
where $\hat{\mathbf{A}}$ is the Toeplitz matrix of the Fourier coefficients $\mathbf{A}_{h}$, and $\hat{\mathbf{X}}$ the column vector of the Fourier coefficients $\mathbf{X}_{h}$ \cite{Ths:CSE:LTP:1991:Wereley}
\begin{align}
	    \hat{\mathbf{A}}	&:~	\hat{\mathbf{A}}_{\mathit{mk}}=\mathbf{A}_{h},~m,k\in\mathbb{N},~h=m-k\in\harmonics
	    \label{eq:TP:constr}\\
	\hat{\mathbf{X}}	&=	\col_{h\in\harmonics}(\mathbf{X}_{h})
\end{align}

\subsection{Model of a \CIDER Excluding the \DC Side}

\begin{figure*}
	\centering
    	{

\tikzstyle{system}=[rectangle, draw=black, minimum width=1cm, minimum height=0.75cm, inner sep=0pt]
\tikzstyle{block}=[rectangle, draw=black, minimum size=0.5cm, inner sep=0pt]
\tikzstyle{dot}=[circle, draw=black, fill=black, minimum size=0.1cm, inner sep=0pt]

\tikzstyle{signal}=[-latex]

\definecolor{myRed}{rgb}{1 0 0}
\definecolor{myGreen}{rgb}{0 0 1}

\footnotesize

\begin{tikzpicture}

	\def\x{1.0}
	\def\y{1.0}
	
	
	
	\draw[dashed,draw=myRed] (3*\x,3.75*\y) to (3*\x,-2.5*\y);
	\draw[signal,draw=myRed] (2.75*\x,3.5*\y) to (2.0*\x,3.5*\y);
	\node[text=myRed] at (0.6*\x,3.5*\y)
	{%
	    \begin{tabular}{c}
	        Internal Response\\
	        $\Hat{\mathbf{Y}}_{n} = \Hat{\mathbf{G}}_{n}\,\Hat{\mathbf{W}}_{n}$
	    \end{tabular}
	};
	
	\node[system] (P) at (0,0.75*\y)
	{%
	$\Sigma_{\pwr,n}$
	};
	
	\node[system] (C) at (0,-0.75*\y)
	{%
	$\Sigma_{\ctrl,n}$
	};
	
	\node[rectangle,thick,draw=black,minimum width=4.3cm,minimum height=3cm,inner sep=0pt](N) at (-0.2*\x,0) {};
	\node (SigmaN) at ($(N.north)-(1.8*\x,0.25*\y)$) {$\Sigma_{n}$};

	\coordinate (yn) at ($0.5*(P.east)+0.5*(C.east)+(1*\x,0)$);
	\draw[thick] ($(yn)+(0*\x,-0.4*\y)$) to ($(yn)+(0*\x,0.4*\y)$);
	
	\draw[signal] ($(P.east)+(0*\x,0*\y)$)
	    to node[midway,above,sloped]{$\mathbf{y}_{\pwr,n}$} ($(yn)+(0*\x,0.2*\y)$);
	\draw[signal] ($(C.east)+(0*\x,0*\y)$)
	    to node[midway,above,sloped]{$\mathbf{y}_{\ctrl,n}$} ($(yn)+(0*\x,-0.2*\y)$);
	
	\coordinate (wn) at ($(P.west)-(1*\x,0)$);
	\draw[thick] ($(wn)+(0*\x,-0.4*\y)$) to ($(wn)+(0*\x,0.4*\y)$);
	
	\draw[signal] ($(wn)+(0*\x,0.2*\y)$)
	    to node[midway,above]{$\mathbf{w}_{\pwr,n}$} ($(P.west)+(0*\x,0.2*\y)$);
	\draw[signal] ($(wn)+(0*\x,-0.1*\y)$)
	    to node[pos=0.25,above,sloped]{$\mathbf{w}_{\ctrl,n}$} ($(C.west)+(0*\x,0.2*\y)$);
	
	\coordinate (un) at ($(C.west)-(1*\x,0)$);
	\draw[thick] ($(un)+(0*\x,-0.4*\y)$) to ($(un)+(0*\x,0.4*\y)$);
	
	\draw[signal] ($(un)+(0*\x,-0.2*\y)$)
	    to node[midway,below]{$\mathbf{u}_{\ctrl,n}$} ($(C.west)+(0*\x,-0.2*\y)$);
	\draw[signal] ($(un)+(0*\x,0.1*\y)$)
	    to node[pos=0.25,below,sloped]{$\mathbf{u}_{\pwr,n}$} ($(P.west)+(0*\x,-0.2*\y)$);
	
    \coordinate (wnn) at ($(wn)-(1.5*\x,0*\y)$);
	\draw[signal] (wnn) to node[near start,above]{$\mathbf{w}_{n}$} (wn);
	
 	\node[block] (Tn) at ($(C)+(0*\x,-1.25*\y)$) {$\mathbf{T}_{n}$};
   
    \node[dot] (ynn) at ($(yn)+(1*\x,0*\y)$) {};

	\draw[-] (yn) 
	    to (ynn);
	\draw[signal] (ynn)
	    to ($(ynn)+(0,-2*\y)$)
	    to ($(Tn.east)$);

    \coordinate (unn) at ($(un)-(1.5*\x,0*\y)$);
	
	\draw[signal] ($(Tn.west)$)
	    to ($(un)-(1.5*\x,1.25*\y)$)
	    to (unn) 
	    to node[near start,above]{$\mathbf{u}_{n}$} (un);
	    
	

	\draw[dashed,draw=myGreen] (8.25*\x,3.75*\y) to (8.25*\x,-2.5*\y);
	\draw[signal,draw=myGreen] (8*\x,3.5*\y) to (7.25*\x,3.5*\y);
	\node[text=myGreen] at (5.6*\x,3.5*\y)
	{%
	    \begin{tabular}{c}
	        Grid Response\\
	        $ (\Hat{\mathbf{W}}_{\grd,n}\,\Hat{\mathbf{W}}_{\spt,n}) \longrightarrow \Hat{\mathbf{Y}}_{\grd,n}$
	    \end{tabular}
	};
	
	\coordinate (wng) at ($(wnn)+(7*\x,1.5*\y)$);
	\draw[thick] ($(wng)+(0*\x,-0.4*\y)$) to ($(wng)+(0*\x,0.4*\y)$);
	
	\draw[-] (wng) 
	    to node[near start,above,text=myRed]{$\mathbf{w}_{n}$} ($(wng)+(-1.4*\x,0)$)
	    to ($(wnn)+(0,1.5*\y)$)
	    to (wnn);
    
    \node[rectangle, draw=black, minimum size=8mm] (Tpg) at ($(wng)+(3.5*\x,0.2*\y)$) {$\mathbf{T}_{\pwr|\grd,n}$};
    \node[dot] (wpn) at ($0.5*(wng)+0.5*(Tpg)+(0.25*\x,0.1*\y)$) {};
    \node[rectangle, draw=black, minimum size=5mm] (R) at ($(wng)+(2.0*\x,-1.5*\y)$) {$\mathbf{r}(\cdot,\cdot)$};
    \node[rectangle, draw=black, minimum size=8mm] (Tkp) at ($0.5*(R.north)+0.5*(wpn)$) {$\mathbf{T}_{\ctrl|\pwr,n}$};
    
    \draw[signal] (Tpg.west) to node[midway,above]{$\mathbf{w}_{\pwr,n}$} ($(wng)+(0*\x,0.2*\y)$);
    
    \draw[signal] (wpn) to (Tkp.north);
    \draw[signal] (Tkp.south) to (R.north);
    \draw[signal] (R.west) to node[midway,above,sloped]{$\mathbf{w}_{\ctrl,n}$} ($(wng)-(0*\x,0.2*\y)$);
	
	\coordinate (ws) at ($(R.east)+(2.5*\x,0)$);
	\draw[signal] (ws) to node[at start,above,text=myGreen]{$\mathbf{w}_{\spt,n}$} (R.east);
	\coordinate (wg) at ($(Tpg.east)+(1*\x,0)$);
	\draw[signal] (wg) to node[at start,above,text=myGreen]{$\mathbf{w}_{\grd,n}$} (Tpg.east);
	
    \node[rectangle, draw=black, minimum size=8mm] (Tgn) at ($(ynn)+(5*\x,0*\y)$) {$\mathbf{T}_{\grd|\pwr,n}$};
    
    \coordinate (yng) at ($(ynn)+(1.4*\x,0*\y)$);
	\draw[thick] ($(yng)+(0*\x,-0.4*\y)$) to ($(yng)+(0*\x,0.4*\y)$);
	
	\draw[signal] (ynn) to node[near end,above,text=myRed]{$\mathbf{y}_{n}$} (yng);
	\draw[signal] (yng) to node[midway,above]{$\mathbf{y}_{\pwr,n}$} (Tgn.west);

	\coordinate (yg) at ($(Tgn.east)+(1*\x,0)$);
	\draw[signal] (Tgn.east) to node[at end,above,text=myGreen]{$\mathbf{y}_{\grd,n}$} (yg);

\end{tikzpicture}

}
	\caption
	{%
	    General structure of a \CIDER.
	    The power hardware $\Sigma_{\pwr,n}$ and control software $\Sigma_{\ctrl,n}$ form the open-loop system $\Sigma_{n}$.
	    The internal response is derived by closing the loop through the feedback matrix $\mathbf{T}_{n}$ (see \eqref{eq:intresp:harm}).
	    The grid response additionally includes the reference calculation $r(\cdot,\cdot)$ and the required transformation matrices  (see \eqref{eq:CIDER:transfer:outer}).
	    The transformation matrix $\mathbf{T}_{\ctrl|\pwr,n}$ represents a change of coordinates between power hardware and control software.
	   $\mathbf{T}_{\pwr|\grd,n}$ and $\mathbf{T}_{\grd|\pwr,n}$  represent a change of coordinates or circuit configuration between grid and power hardware.
	}
	\label{fig:CIDER:Structure}
\end{figure*}
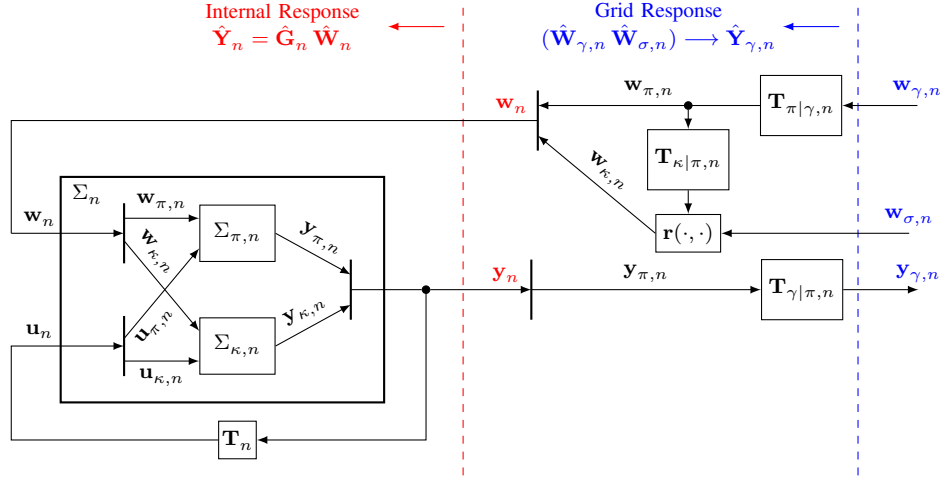

In \cite{jrn:2020:kettner-becker:HPF-1}, the \LTP models of the \CIDER[s] are developed in the time domain.
Let $n$ be a generic \CIDER.
Its closed-loop model can be derived from the generic structure depicted in \cref{fig:CIDER:Structure}.
It consists of the \emph{power hardware} $\pwr$ and the \emph{control software} $\ctrl$, which are represented by \LTP systems $\Sigma_{\pwr,n}$ and $\Sigma_{\ctrl,n}$, respectively, and the \emph{reference calculation}, which is represented by a function $\mathbf{r}(\cdot,\cdot)$.
$\Sigma_{\pwr,n}$ and $\Sigma_{\ctrl,n}$ jointly form the open-loop system $\Sigma_{n}$, which is described by
\begin{align}
\Sigma_{n}:\left\{
	\begin{aligned}
            \dot{\mathbf{x}}_{n}(t)
        &=      \mathbf{A}_{n}(t)\mathbf{x}_{n}(t)
        	+   \mathbf{B}_{n}(t)\mathbf{u}_{n}(t)
        	+   \mathbf{E}_{n}(t)\mathbf{w}_{n}(t)\\
            \mathbf{y}_{n}(t)
        &=      \mathbf{C}_{n}(t)\mathbf{x}_{n}(t)
            +   \mathbf{D}_{n}(t)\mathbf{u}_{n}(t)
        	+   \mathbf{F}_{n}(t)\mathbf{w}_{n}(t)
    \end{aligned}
    \right.
    \label{eq:CIDER:time}
\end{align}
where $\mathbf{x}_{n}(t)$, $\mathbf{u}_{n}(t)$, $\mathbf{y}_{n}(t)$, and $\mathbf{w}_{n}(t)$ are the \emph{state}, \emph{input}, \emph{output}, and \emph{disturbance vector}, respectively.
Accordingly, $\mathbf{A}_{n}(t)$, $\mathbf{B}_{n}(t)$, $\mathbf{C}_{n}(t)$, $\mathbf{D}_{n}(t)$, $\mathbf{E}_{n}(t)$, and $\mathbf{F}_{n}(t)$ are the \emph{system}, \emph{input}, \emph{output}, \emph{feed-through}, \emph{input disturbance}, and \emph{output disturbance matrix}, respectively.

As shown in \cref{fig:CIDER:Structure}, the power hardware $\Sigma_{\pwr,n}$ and the control software $\Sigma_{\ctrl,n}$ form a closed-loop system via the feedback matrix $\mathbf{T}_{n}$:
\begin{equation}
    \mathbf{u}_{n}(t) = \mathbf{T}_{n}\mathbf{y}_{n}(t)
\end{equation}
The internal response of the \CIDER describes the behaviour of this closed-loop system:
\begin{Definition}\label{def:CIDER:intresp}
    The \emph{internal response} of the \CIDER describes the relation from $\mathbf{w}_{n}$ to $\mathbf{y}_{n}$.
    It is derived as the combination of the open-loop system $\Sigma_{n}$ with the feedback matrix $\mathbf{T}_{n}$.
\end{Definition}
\noindent
The feedback matrix describes the coordinate transformations which are applied to the measurements and control signals as they pass from power hardware to control software and vice versa.
As explained in \cite{jrn:2020:kettner-becker:HPF-1}, the internal response is described in the harmonic domain by the following linear expression:
\begin{equation}
    \Hat{\mathbf{Y}}_{n}(\Hat{\mathbf{W}}_{n}) = \Hat{\mathbf{G}}_{n}\,\Hat{\mathbf{W}}_{n}
    \label{eq:intresp:harm} 
\end{equation}
where $\Hat{\mathbf{Y}}_{n}$ and $\Hat{\mathbf{W}}_{n}$ are the vectors of the Fourier coefficients of $\mathbf{y}_{n}$ and $\mathbf{w}_{n}$ in \eqref{eq:CIDER:time}, respectively, and $\Hat{\mathbf{G}}_{n}$ is the closed-loop gain of the \CIDER derived in the harmonic domain.

The \emph{grid response} additionally includes the reference calculation $\mathbf{r}(\cdot,\cdot)$ (i.e., grid-following or grid-forming control laws) as well as transformations $\mathbf{T}_{\ctrl|\pwr,n}$ (i.e., a change of coordinates between power hardware and control software) plus $\mathbf{T}_{\grd|\pwr,n}$ and $\mathbf{T}_{\pwr|\grd,n}$ (i.e., change of circuit configuration between grid and power hardware).
As explained in detail in \cite{jrn:2020:kettner-becker:HPF-1,jrn:2020:kettner-becker:HPF-2}, the reference calculation $\mathbf{r}(\cdot,\cdot)$ translates any type of setpoint $\mathbf{w}_{\spt,n}(t)$ from a higher-level controller (e.g., a system-level controller) into a voltage or current setpoint $\mathbf{w}_{\ctrl,n}(t)$ for the lower-level controller (i.e. the device-level controller).
Note that, as shown in \cref{fig:CIDER:Structure}, the grid-side quantities are the grid disturbance $\mathbf{w}_{\grd,n}(t)$, the setpoint $\mathbf{w}_{\spt,n}(t)$, and the grid output $\mathbf{y}_{\grd,n}(t)$%
\footnote
{%
    For a grid-following \CIDER, the grid disturbance is the nodal voltage and the grid output is the injected current.
    For a grid-forming \CIDER, the grid disturbance is the injected current and the grid output is the nodal voltage.
}
Accordingly, the grid response is defined as follows.
\begin{Definition}\label{def:CIDER:grdresp}
	The \emph{grid response} of the \CIDER describes the relation from $\mathbf{w}_{\grd,n}$ and $\mathbf{w}_{\spt,n}$ to $\mathbf{y}_{\grd,n}$.
	It is derived through the combination of the internal response, the reference calculation $r(\cdot,\cdot)$ and the grid-side transformations $\mathbf{T}_{\pwr|\grd,n}$ and $\mathbf{T}_{\grd|\pwr,n}$.
\end{Definition}
\noindent
As shown in \cite{jrn:2020:kettner-becker:HPF-1}, the grid response can be expressed in the harmonic domain as:
\begin{align}
        \hspace{-1mm}\Hat{\mathbf{Y}}_{\grd,n}(\Hat{\mathbf{W}}_{\grd,n},\Hat{\mathbf{W}}_{\spt,n})
    &=  \Hat{\mathbf{T}}_{\grd|\pwr,n}
        \Hat{\mathbf{Y}}_{\pwr,n}
        (
            \underbrace{\Hat{\mathbf{T}}_{\pwr|\grd,n}\Hat{\mathbf{W}}_{\grd,n}}_{\Hat{\mathbf{W}}_{\pwr,n}},
            \Hat{\mathbf{W}}_{\spt,n}
        )
        \label{eq:gridresp:harm}
\end{align}
Regarding the function $\Hat{\mathbf{Y}}_{\pwr,n}(\cdot,\cdot)$, note in \cref{fig:CIDER:Structure} and recall from \eqref{eq:intresp:harm} that the mapping from $\Hat{\mathbf{W}}_{\pwr,n}$ to $\Hat{\mathbf{Y}}_{\pwr,n}$ is linear.
Indeed, this is why the gain $\hat{\mathbf{G}}_{n}$ can be calculated analytically.
By contrast, the mapping from $\Hat{\mathbf{W}}_{\spt,n}$ to $\Hat{\mathbf{Y}}_{\pwr,n}$ includes the reference calculation, which may be nonlinear.
However, since this potential nonlinearity is not part of the closed-loop structure, it does not necessitate any linearization (see \cite{jrn:2020:kettner-becker:HPF-1}).

\subsection{Model of a \CIDER Including the \DC Side}
\label{sec:model-rsc:nonlin}

In the \HPF framework proposed in \cite{jrn:2020:kettner-becker:HPF-1, jrn:2020:kettner-becker:HPF-2} the \CIDER model covers only the \AC-side components.
That is, the \AC/\DC converter and \DC side are represented by a controlled \AC voltage source.
Since the parts of the model related to the internal response are all linear, they can be represented by \LTP models without further approximations.
Notably, one can use \cref{eq:intresp:harm} to find the internal response of a generic \CIDER, which is characterized by the closed-loop gain $\Hat{\mathbf{G}}_{n}$.

If the \DC side of the \CIDER shall be analyzed in detail, the \DC-side components as well as the \AC/\DC converter need to be included in the \HPF model.
The latter introduces a nonlinearity into the model.
As will be shown shortly, this is due to the power balance equation used in the average model of the converter.
Due to the aforementioned nonlinearity, it is in general not possible to find an explicit solution for the internal response because the equations describing the closed-loop system cannot be solved analytically.
However, one can locally approximate the internal response through linearization w.r.t. the operating point and apply the same methodology as before.
The operating point needs to be explicitly considered as a variable in the linearized version of the open-loop model \eqref{eq:CIDER:time} for the numerical solution of the \HPF problem.

Let $\mathbf{y}_{o,n}(t)$ denote the operating point.
Via linearization w.r.t. the operating point, one obtains a model that is analogous to \eqref{eq:CIDER:time}, in which $\mathbf{y}_{o,n}(t)$ appears as an additional variable:
\begin{alignat}{3}
		\AP_{n}:~
	&   \AP_{n}(t,\mathbf{y}_{o,n}(t) ),~
    &   \BP_{n}~\text{etc. analogous}	
\end{alignat}
In order to facilitate the formulation and solution of the \HPF problem, the following decomposition is proposed:
\begin{align}
		\AP_{n}(t,\mathbf{y}_{o,n}(t) )
	=       \tilde{\AP}_{n}(t)+\bar{\AP}_{n}(t,\mathbf{y}_{o,n}(t))
\end{align}
That is, the matrices can be separated into a term $\tilde{\AP}_{n}$ which is invariant w.r.t. the operating point, and another term $\bar{\AP}_{n}$ which is a function of it.
By consequence, in case the operating point changes, only the second term needs to be recalculated.

Based on the linearized open-loop \CIDER model, one can derive the closed-loop model both in time and frequency domain.
In doing so, keep in mind that the associated matrices in time and frequency domain are functions of $\mathbf{y}_{o,n}(t)$ and $\hat{\mathbf{Y}}_{o,n}$, respectively.
The same applies to the internal and grid response.
The closed-loop gain which characterizes the internal response is of the form (cf. \eqref{eq:intresp:harm})
\begin{equation}
        \Hat{\mathbf{Y}}_{n}(\Hat{\mathbf{W}}_{n},\hat{\mathbf{Y}}_{o,n})
    =   \Hat{\mathbf{G}}_{n}(\hat{\mathbf{Y}}_{o,n})\,\Hat{\mathbf{W}}_{n}
\end{equation}
Analogously, the grid response is written as follows (cf. \eqref{eq:gridresp:harm}):
\begin{align}
        \Hat{\mathbf{Y}}_{\grd,n}&(\Hat{\mathbf{W}}_{\grd,n},\Hat{\mathbf{W}}_{\spt,n},\hat{\mathbf{Y}}_{o,n})\notag\\
    &=  \Hat{\mathbf{T}}_{\grd|\pwr,n}
        \Hat{\mathbf{Y}}_{\pwr,n}
        (
            \underbrace{\Hat{\mathbf{T}}_{\pwr|\grd,n}\Hat{\mathbf{W}}_{\grd,n}}_{\Hat{\mathbf{W}}_{\pwr,n}},
            \Hat{\mathbf{W}}_{\spt,n},
            \hat{\mathbf{Y}}_{o,n}
        )
    \label{eq:CIDER:transfer:outer}
\end{align}

Without loss of generality, it is assumed that the operating point is a subset of the output vector of the internal response:
\begin{equation}
    \mathbf{y}_{o,n}(t) \subset \mathbf{y}_{n}(t)
\end{equation}
Note that, when defining the open-loop model of the \CIDER, any block of the vectors $\mathbf{x},\mathbf{u}$, and $\mathbf{w}$ can be included in the output equation if needed.
Accordingly, there exists a matrix $\mathbf{T}_{o,n}$ such that
\begin{align}
    \mathbf{y}_{o,n}(t)  = \mathbf{T}_{o,n}\mathbf{y}_{n}(t)
    \label{eq:operpt:time}
\end{align}
Respectively, in the harmonic domain
\begin{equation}
    \hat{\mathbf{Y}}_{o,n}  = \hat{\mathbf{T}}_{o,n}\hat{\mathbf{Y}}_{n}
    \label{eq:operpt:harm}
\end{equation}
This definition will be used later in the formulation of the Newton-Raphson method for solving the \HPF problem.

\section{Extension of the Solution Algorithm for the Harmonic Power-Flow Problem}
\label{sec:algo}


In the \HPF framework proposed in \cite{jrn:2020:kettner-becker:HPF-1}, the power system is described by two sets of nodal equations, which are formulated from the point of view of the grid and the resources, respectively.
The set of all nodes $\nodes$ is partitioned into two disjoint sets $\formers$ and $\followers$
\begin{equation}
    \nodes=\formers\cup\followers,~\formers\cap\followers=\emptyset
\end{equation}
In this respect, the unknowns are taken to be the nodal injected currents $\Hat{\mathbf{I}}_{\formers}$ at the nodes $\formers$ with grid-forming \CIDER[s] and the nodal phase-to-ground voltages $\Hat{\mathbf{V}}_{\followers}$ at the nodes $\followers$ with grid-following \CIDER[s].

From the point of view of the grid, the nodal equations are formulated using hybrid parameters:
\begin{align}
        \Hat{\mathbf{V}}_{\formers}(\Hat{\mathbf{I}}_{\formers},\Hat{\mathbf{V}}_{\followers})
    &=      \Hat{\mathbf{H}}_{\formers\times\formers}\Hat{\mathbf{I}}_{\formers}
        +   \Hat{\mathbf{H}}_{\formers\times\followers}\Hat{\mathbf{V}}_{\followers}
    \label{eq:grid:form}
    \\
        \Hat{\mathbf{I}}_{\followers}(\Hat{\mathbf{I}}_{\formers},\Hat{\mathbf{V}}_{\followers})
    &=      \Hat{\mathbf{H}}_{\followers\times\formers}\Hat{\mathbf{I}}_{\formers}
        +   \Hat{\mathbf{H}}_{\followers\times\followers}\Hat{\mathbf{V}}_{\followers}
    \label{eq:grid:follow}
\end{align}
where $\Hat{\mathbf{H}}_{\formers\times\formers}$, $\Hat{\mathbf{H}}_{\formers\times\followers}$, $\Hat{\mathbf{H}}_{\followers\times\formers}$ and $\Hat{\mathbf{H}}_{\followers\times\followers}$ are the blocks of the hybrid matrix $\Hat{\mathbf{H}}$ associated with $\formers$ and $\followers$.

From the point of view of the \CIDER[s], the nodal equations are described by the grid responses:
\begin{align}
        s\in\formers:\quad
    &   \Hat{\mathbf{V}}_{s}(\Hat{\mathbf{I}}_{s},V_{\spt,s},f_{\spt,s},\Hat{\mathbf{Y}}_{o,s})\notag\\
    &   =\Hat{\mathbf{Y}}_{\grd,s}(\Hat{\mathbf{I}}_{s},V_{\spt,s},f_{\spt,s},\Hat{\mathbf{Y}}_{o,s})
    \label{eq:resource:form}\\
        r\in\followers:\quad
    &   \Hat{\mathbf{I}}_{r}(\Hat{\mathbf{V}}_{r},S_{\spt,r},\Hat{\mathbf{Y}}_{o,r})
        =\Hat{\mathbf{Y}}_{\grd,r}(\Hat{\mathbf{V}}_{r},S_{\spt,r},\Hat{\mathbf{Y}}_{o,r})
    \label{eq:resource:follow}
\end{align}
Note that grid-forming \CIDER[s] take voltage and frequency as setpoints (i.e., $V_{\spt,s}$ and $f_{\spt,s}$), and grid-following \CIDER[s] active and reactive powers (i.e., $S_{\spt,r}=P_{\spt,r}+jQ_{\spt,r}$).

In equilibrium, the mismatch equations between \eqref{eq:grid:form}--\eqref{eq:grid:follow} and \eqref{eq:resource:form}--\eqref{eq:resource:follow} must be zero (i.e., they yield identical results):
\begin{align}
        \Delta\Hat{\mathbf{V}}_{\formers}
        (\Hat{\mathbf{I}}_{\formers},\Hat{\mathbf{V}}_{\followers},\mathbf{V}_{\spt},\mathbf{f}_{\spt},\Hat{\mathbf{Y}}_{o,\formers})
    &=  \mathbf{0}
    \label{eq:residual:form}\\
        \Delta\Hat{\mathbf{I}}_{\followers}
        (\Hat{\mathbf{I}}_{\formers},\Hat{\mathbf{V}}_{\followers},\mathbf{S}_{\spt},\Hat{\mathbf{Y}}_{o,\followers})
    &=  \mathbf{0}
    \label{eq:residual:follow}
\end{align}
where $\mathbf{V}_{\spt}$, $\mathbf{f}_\spt$, and $\mathbf{S}_{\spt}$ are column vectors built of the setpoints $V_{\spt,s}$, $f_{\spt,s}$ ($s\in\formers$) and $S_{\spt,r}$ ($r\in\followers$), respectively.
This system of equations can be solved by means of the Newton-Raphson method.



As known from numerical analysis, the Newton-Raphson method requires the calculation (and inversion) of a Jacobian matrix $\Hat{\mathbf{J}}$, which is computationally intensive.
In the original \HPF framework proposed in \cite{jrn:2020:kettner-becker:HPF-1}, most blocks of the Jacobian matrix are constant, because the grid equations \eqref{eq:grid:form}--\eqref{eq:grid:follow} are exactly linear, and the \CIDER responses \eqref{eq:gridresp:harm} are linear except for the reference calculation.
In the extended formulation \eqref{eq:residual:form}--\eqref{eq:residual:follow} of the \HPF problem, the operating points $\Hat{\mathbf{Y}}_{o,\formers}$ and $\Hat{\mathbf{Y}}_{o,\followers}$ of the \CIDER[s] appear as additional unknowns.
Therefore, the entire Jacobian matrix has to be recalculated in each iteration of the Newton-Raphson method.
In doing so, the fact that the operating points are a subset of the internal responses -- as postulated in \eqref{eq:operpt:harm} -- needs to be taken into account.
As a result, one obtains the extended Newton-Raphson algorithm which is shown in \cref{alg:newton-raphson}.
The differences w.r.t. to the original method are highlighted in blue.

Like the unknowns related to the grid (i.e., $\hat{\mathbf{V}}_{\formers}$ and $\hat{\mathbf{I}}_{\followers}$), the operating points of the \CIDER[s] (i.e., $\hat{\mathbf{Y}}_{o,\formers}$ and $\hat{\mathbf{Y}}_{o,\followers}$) need to be initialized at the start  (see \cref{algo:init:form,algo:init:follow}, and \cref{algo:opt:init:form,algo:opt:init:follow} of \cref{alg:newton-raphson}, respectively), and then updated during each iteration.
For the initialization, one can make an ``educated guess'' based on the fact that the operating points are \AC and \DC quantities.
In this respect, two different approaches are used.
First, \AC quantities are initialized assuming balanced, sinusoidal conditions (i.e., a positive-sequence component at the fundamental tone) and using nominal values (e.g., magnitude 1~p.u. and angle 0~rad for voltages) or setpoints.
This is known as ``flat start''.
Second, \DC quantities are initialized assuming ideal steady-state conditions (i.e., constant values) and using nominal values or setpoints.
For the update, \eqref{eq:operpt:harm} is used: that is, the operating points are retrieved via the output equations of the \CIDER (see \cref{algo:opt:updt:form,algo:opt:updt:follow} of \cref{alg:newton-raphson}).

\textcolor{black}{
\begin{algorithm}[t]
	\centering

{
\renewcommand{\arraystretch}{1.1}

\begin{algorithmic}[1]
    \Procedure{\texttt{\HPF}}
    {%
        $\Delta\Hat{\mathbf{V}}_{\formers}(\cdot,\cdot,\cdot)$,
        $\Delta\Hat{\mathbf{I}}_{\followers}(\cdot,\cdot,\cdot)$,
        $\mathbf{V}_{\spt}$,
        $\mathbf{f}_{\spt}$,
        $\mathbf{S}_{\spt}$
    }
        \State
        {%
            \texttt{\# Initialization}
        }
        \State
        {%
            $\Hat{\I}_{\formers} \gets \mathbf{0}$
        }   \label{algo:init:form}
        \State
        {%
            $\Hat{\V}_{\followers} \gets \texttt{flat\_start()}$
        }   \label{algo:init:follow}
        \State
        {%
            \textcolor{blue}{$\Hat{\mathbf{Y}}_{o,\formers} \gets \texttt{initialize\_operating\_point()}$}
        }   \label{algo:opt:init:form}
        \State
        {%
            \textcolor{blue}{$\Hat{\mathbf{Y}}_{o,\followers} \gets \texttt{initialize\_operating\_point()}$}
        }   \label{algo:opt:init:follow}
        \While
        {$
            \max(|\Delta\Hat{\mathbf{V}}_{\formers}|,|\Delta\Hat{\mathbf{I}}_{\followers}|)
            \geqslant\epsilon
        $}
            \State{\texttt{\# Residuals}}
            \State
            {$
                        \Delta\Hat{\mathbf{V}}_{\formers}
                \gets   \Delta\Hat{\mathbf{V}}_{\formers}
                        (\Hat{\mathbf{I}}_{\formers},\Hat{\mathbf{V}}_{\followers},\mathbf{V}_{\spt},\mathbf{f}_{\spt},\textcolor{blue}{\Hat{\mathbf{Y}}_{o,\formers}})
            $}
            \State
            {$
                        \Delta\Hat{\mathbf{I}}_{\followers}
                \gets   \Delta\Hat{\mathbf{I}}_{\followers}
                        (\Hat{\mathbf{I}}_{\formers},\Hat{\mathbf{V}}_{\followers},\mathbf{S}_{\spt},\textcolor{blue}{\Hat{\mathbf{Y}}_{o,\followers}})
            $}
            \State{\texttt{\# Jacobian matrix}}
            \State
            {$
                        \Hat{\mathbf{J}}_{\formers\times\formers}
                \gets   \partial_{\formers}\Delta\Hat{\mathbf{V}}_{\formers}
                        (\Hat{\mathbf{I}}_{\formers},\Hat{\mathbf{V}}_{\followers},\mathbf{V}_{\spt},\mathbf{f}_{\spt},\textcolor{blue}{\Hat{\mathbf{Y}}_{o,\formers}})
            $}
            \State
            {$
                        \Hat{\mathbf{J}}_{\formers\times\followers}
                \gets   \partial_{\followers}\Delta\Hat{\mathbf{V}}_{\formers}
                        (\Hat{\mathbf{I}}_{\formers},\Hat{\mathbf{V}}_{\followers},\mathbf{V}_{\spt},\mathbf{f}_{\spt},\textcolor{blue}{\Hat{\mathbf{Y}}_{o,\formers}})
            $}
            \State
            {$
                        \Hat{\mathbf{J}}_{\followers\times\formers}
                \gets   \partial_{\formers}\Delta\Hat{\mathbf{I}}_{\followers}
                        (\Hat{\mathbf{I}}_{\formers},\Hat{\mathbf{V}}_{\followers},\mathbf{S}_{\spt},\textcolor{blue}{\Hat{\mathbf{Y}}_{o,\followers}})
            $}
            \State
            {$
                        \Hat{\mathbf{J}}_{\followers\times\followers}
                \gets   \partial_{\followers}\Delta\Hat{\mathbf{I}}_{\followers}
                        (\Hat{\mathbf{I}}_{\formers},\Hat{\mathbf{V}}_{\followers},\mathbf{S}_{\spt},\textcolor{blue}{\Hat{\mathbf{Y}}_{o,\followers}})
            $}
            \State{\texttt{\# Newton-Raphson iteration}}
            \State
            {$
                \begin{bmatrix}
                    \Delta\Hat{\mathbf{I}}_{\formers}\\
                    \Delta\Hat{\mathbf{V}}_{\followers}
                \end{bmatrix}
                \gets
                \begin{bmatrix}
                        \Hat{\mathbf{J}}_{\formers\times\formers}
                    &   \Hat{\mathbf{J}}_{\formers\times\followers}\\
                        \Hat{\mathbf{J}}_{\followers\times\formers}
                    &   \Hat{\mathbf{J}}_{\followers\times\followers}
                \end{bmatrix}^{-1}
                \begin{bmatrix}
                    \Delta\Hat{\mathbf{V}}_{\formers}\\
                    \Delta\Hat{\mathbf{I}}_{\followers}
                \end{bmatrix}
            $}
            \State
            {$
                \begin{bmatrix}
                    \Hat{\mathbf{I}}_{\formers}\\
                    \Hat{\mathbf{V}}_{\followers}
                \end{bmatrix}
                \gets
                    \begin{bmatrix}
                        \Hat{\mathbf{I}}_{\formers}\\
                        \Hat{\mathbf{V}}_{\followers}
                    \end{bmatrix}
                -   \begin{bmatrix}
                        \Delta\Hat{\mathbf{I}}_{\formers}\\
                        \Delta\Hat{\mathbf{V}}_{\followers}
                    \end{bmatrix}
            $}
            \State{\textcolor{blue}{\texttt{\# Update}}}
            \State
            {$
                \textcolor{blue}{
                \Hat{\mathbf{Y}}_{o,s}
                \gets
                    \hat{\mathbf{T}}_{o,s}\Hat{\mathbf{Y}}_{s} ~\forall s\in\formers
                }
            $}   \label{algo:opt:updt:form}
            \State
            {$
                \textcolor{blue}{
                \Hat{\mathbf{Y}}_{o,r}
                \gets
                    \hat{\mathbf{T}}_{o,r}\Hat{\mathbf{Y}}_{r} ~\forall r\in\followers
                }
            $}   \label{algo:opt:updt:follow}
        \EndWhile
    \EndProcedure
\end{algorithmic}
}
	\caption{Newton-Raphson solution of the \HPF problem.}
	\label{alg:newton-raphson}
\end{algorithm}
}

\section{Grid-Following \CIDER Including the DC-Side}
\label{sec:model-DC}

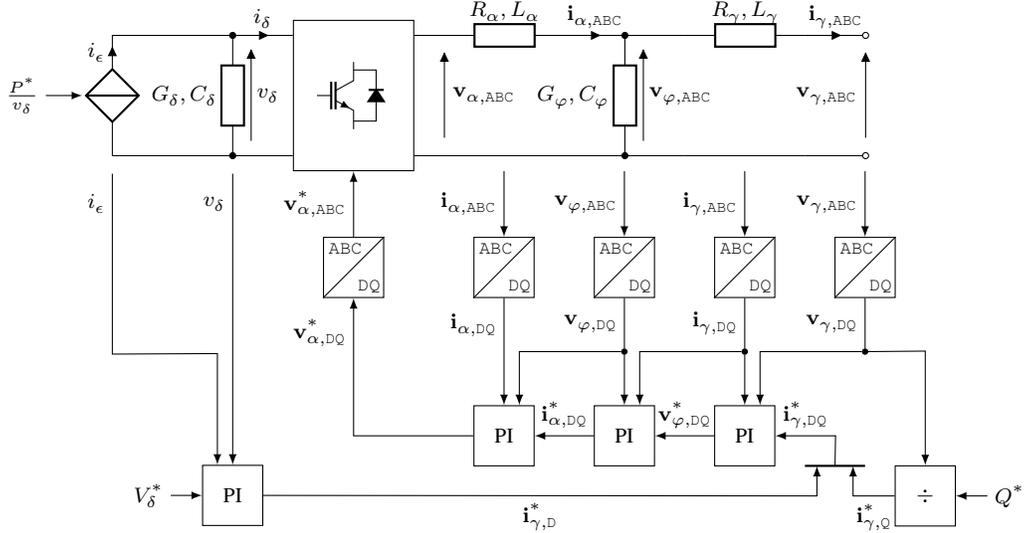
\begin{figure*}
	\centering
    {

\ctikzset{bipoles/length=1.0cm}
\ctikzset{bipoles/diode/height=.2}
\ctikzset{bipoles/diode/width=.15}
\tikzstyle{block}=[rectangle, draw=black,fill=white, minimum size=8mm, inner sep=0pt]
\tikzstyle{dot}=[circle, draw=black, fill=black, minimum size=2pt, inner sep=0pt]
\tikzstyle{measurement}=[rectangle,draw=black,minimum size=1mm,inner sep=0pt]
\tikzstyle{signal}=[-latex]

\def\transform#1#2
{%
\begin{scope}[shift={#2}]
    \node[block] (#1) at (0,0) {};
    \draw (#1.south west) to (#1.north east);
    \node at ($(#1.north west)+(0.3,-0.15)$) {\scriptsize$\phsABC$};
    \node at ($(#1.south east)+(-0.2,0.15)$) {\scriptsize$\cmpDQ$};
\end{scope}
}

\definecolor{blue}{rgb}{0.74, 0.83, 0.9}
\definecolor{green}{rgb}{0.78, 0.9, 0.82}
\definecolor{red}{rgb}{0.97, 0.71, 0.67}
\definecolor{yellow}{rgb}{1.0, 0.99, 0.82}

\footnotesize

\begin{circuitikz}
    
    \def\x{1.6}
    \def\y{1.6}
    
    
    
    \node[rectangle,draw=black,fill=white,minimum height=20mm,minimum width=16mm,inner sep=0pt] (C) at (0,0) {};
    \node[nigbt,scale=0.8] (IGBT) at (C) {};
    \draw (IGBT.E)++(0,0.1) -- ++(0.3,0) to[D*] ($(IGBT.C)+(0.3,-0.1)$)   -- ++(-0.3,0);

    \coordinate (AN) at ($(C.east)-(0*\x,0.5*\y)$);
    \coordinate (AP) at ($(C.east)+(0*\x,0.5*\y)$);
    
    \coordinate (FALN) at ($(AN)+(0.25*\x,0)$);
    \coordinate (FALP) at ($(FALN)+(0,\y)$);
    \coordinate (FARN) at ($(FALN)+(\x,0)$);
    \coordinate (FARP) at ($(FARN)+(0,\y)$);
    
    \coordinate (FGLN) at ($(FARN)+(\x,0)$);
    \coordinate (FGLP) at ($(FGLN)+(0,\y)$);
    \coordinate (FGRN) at ($(FGLN)+(\x,0)$);
    \coordinate (FGRP) at ($(FGRN)+(0,\y)$);
    
    \coordinate (GN) at ($(FGRN)+0.5*(\x,0)$);
    \coordinate (GP) at ($(GN)+(0,\y)$);

    \draw (FALN) to[open,v_=$\VT_{\act,\phsABC}$] (FALP);
    
    \draw (AN) to[short] (FALN);
    \draw (AP) to[short] (FALP);

    \draw (FALN) to[short] (FARN);
    \draw (FALP) to[generic=${R_\act,L_\act}$,fill=white] (FARP);
    
    \draw (FARN) to[short] (FGLN);
    \draw (FARP)
        to[short,i=$\IT_{\alpha,\phsABC}$] ($0.5*(FARP)+0.5*(FGLP)$)
        to[short] (FGLP); 
    
    \draw ($0.5*(FARN)+0.5*(FGLN)$) to[generic=${G_\fltr,C_\fltr}$,v=$\VT_{\fltr,\phsABC}$,*-*,fill=white] ($0.5*(FARP)+0.5*(FGLP)$);
    
    \draw (FGLN) to[short] (FGRN);
    \draw (FGLP) to[generic=${R_\grd,L_\grd}$,fill=white] (FGRP);
    
    \draw (FGRN) to[short,-o] (GN);
    \draw (FGRP) to[short,-o,i=$\IT_{\grd,\phsABC}$] (GP);
    
    \draw (GN) to[open,v^=$\VT_{\grd,\phsABC}$] (GP);
    
    
    
    \coordinate (IA) at ($(C.south)-0.0*(0,\y)$);
    \coordinate (OFA) at ($(IA)+(1.25*\x,0)$);
    \coordinate (OFI) at ($(OFA)+(\x,0)$);
    \coordinate (OFG) at ($(OFI)+(\x,0)$);
    \coordinate (OG) at ($(OFG)+(\x,0)$);
    
    \transform{TA}{($(IA)-0.8*(0,\y)$)}
    \transform{TFA}{($(OFA)-0.8*(0,\y)$)}
    \transform{TFI}{($(TFA)+(\x,0)$)}
    \transform{TFG}{($(TFI)+(\x,0)$)}
    \transform{TG}{($(TFG)+(\x,0)$)}
    
    \draw[signal] (TA.north) to node[midway,left]{$\VT_{\act,\phsABC}^{*}$} (IA.south);
    \draw[signal] (OFA) to node[midway,left]{$\IT_{\act,\phsABC}$} (TFA.north);
    \draw[signal] (OFI) to node[midway,left]{$\VT_{\fltr,\phsABC}$} (TFI.north);
    \draw[signal] (OFG) to node[midway,left]{$\IT_{\grd,\phsABC}$} (TFG.north);
    \draw[signal] (OG) to node[midway,left]{$\VT_{\grd,\phsABC}$} (TG.north);
    
    
    
    \node[block] (CFA) at ($(TFA)-1.4*(0,\y)$) {\ctrlPI};
    \node[block] (CFI) at ($(CFA)+(\x,0)$) {\ctrlPI};
    \node[block] (CFG) at ($(CFI)+(\x,0)$) {\ctrlPI};
    \coordinate (R) at ($(CFG)+(1*\x,0)$);
    \node[block] (Rb) at ($(CFG)+(1.5*\x,-0.5*\y)$) {$\div$};
    \node (S) at ($(Rb)+0.7*(\x,0)$) {$Q^{*}$};
    
    \node[dot] (XFI) at ($0.5*(TFI)+0.5*(CFI)$) {};
    \node[dot] (XFG) at ($0.5*(TFG)+0.5*(CFG)$) {};
    \node[dot] (XG) at ($0.5*(TG)+0.5*(R)$) {};
    
    \draw[signal] (TFA.south)
        to node[midway,left]{$\IT_{\act,\cmpDQ}$} ($0.5*(TFA.south)+0.5*(CFA.north)$)
        to (CFA.north);
    \draw[signal] (CFG.west) to node[midway,above]{$\VT^*_{\fltr,\cmpDQ}$} (CFI.east);
    \draw[signal] (CFA.west)
        to ($(CFA)-(1.25*\x,0)$)
        to node[near end,left]{$\VT_{\act,\cmpDQ}^{*}$} (TA.south);
    
    \draw[-] (TFI.south) to node[midway,left]{$\VT_{\fltr,\cmpDQ}$} (XFI.north);
    \draw[signal] (XFI.south) to (CFI.north);
    \draw[signal] (XFI.west)
        to ($0.5*(TFA.south)+0.5*(CFA.north east)$)
        to ($0.5*(CFA.north)+0.5*(CFA.north east)$);
    \draw[signal] (CFI.west) to node[midway,above]{$\IT^*_{\act,\cmpDQ}$} (CFA.east);

    \draw[-] (TFG.south) to node[midway,left]{$\IT_{\grd,\cmpDQ}$} (XFG.north);
    \draw[signal] (XFG.south) to (CFG.north);
    \draw[signal] (XFG.west) 
        to ($0.5*(TFI.south)+0.5*(CFI.north east)$)
        to ($0.5*(CFI.north)+0.5*(CFI.north east)$);
    
    \draw[-] (TG.south) to node[midway,left]{$\VT_{\grd,\cmpDQ}$} (XG.north);
    \draw[signal] (XG.east)
        to ($(Rb.north)+0.95*(0,\y)$)
        to (Rb.north);
    \draw[signal] (XG.west)
        to ($0.5*(TFG.south)+0.5*(CFG.north east)$)
        to ($0.5*(CFG.north)+0.5*(CFG.north east)$);
    
    \coordinate (Rm) at ($0.5*(Rb.north)+0.5*(CFG.south)$);
    \draw[-, thick] ($(Rm)-0.25*(\x,0)$) to ($(Rm)+0.25*(\x,0)$);
    
    \draw[signal] (Rb.west) 
        to node[midway,below]{$\IT^{*}_{\grd,\cmpQ}$} ($(Rm)+(0.15*\x,-0.25*\y)$)
        to ($(Rm)+0.15*(\x,0)$);
    
    \draw[signal] (S.west) to (Rb.east);
    
    \draw[signal] (Rm) 
        to ($(CFG.east)+0.5*(\x,0)$)
        to node[midway,above]{$\IT^{*}_{\grd,\cmpDQ}$} (CFG.east);
    
    
    
    \coordinate (DN) at ($(C.west)-(0*\x,0.5*\y)$);
    \coordinate (DP) at ($(C.west)+(0*\x,0.5*\y)$);
    
    \coordinate (FDRN) at ($(DN)-(0.5*\x,0)$);
    \coordinate (FDRP) at ($(FDRN)+(0,\y)$);
    \coordinate (FDLN) at ($(FDRN)-(\x,0)$);
    \coordinate (FDLP) at ($(FDLN)+(0,\y)$);
    
    \draw (DN) to[short] (FDRN);
    \draw (FDRP) to[short,i^=$i_{\delta}$] (DP);

    \draw ($(FDRN)$) to[generic=${G_\delta,C_\delta}$,v=$v_{\delta}$,*-*,fill=white] ($(FDRP)$);

    \draw (FDRN)
        to[short] (FDLN)
        to[cI,i^=$i_{\epsilon}$,fill=white] (FDLP)
        to[short] (FDRP);
        
    \coordinate (IECS) at($0.5*(FDLP)+0.5*(FDLN)-(0.25*\x,0)$);
    \node (IECL) at ($(IECS)-(0.5*\x,0)$) {$\frac{P^{*}}{v_{\delta}}$};
    \draw[signal] (IECL) to (IECS);
    
    
    
    \coordinate (ID) at ($(FDRN)-0.15*(0,\y)$);
    \coordinate (IE) at ($(FDLN)-0.15*(0,\y)$);
    
    \node[block] (CVD) at ($(ID)-2.675*(0,\y)$) {\ctrlPI};
    
    \draw[signal] (ID.south) 
        to node[midway,left]{$v_{\delta}$} ($(ID)-(0,0.5*\y)$)
        to (CVD.north);
    \draw[signal] (IE.south) 
        to node[midway,left]{$i_{\epsilon}$} ($(IE)-(0,0.5*\y)$)
        to ($(IE)-(0,1.5*\y)$)
        to ($0.5*(CVD.north)+0.5*(CVD.north west)+(0,0.925*\y)$)
        to ($0.5*(CVD.north)+0.5*(CVD.north west)$);
        
    \node (V) at ($(CVD)-0.7*(\x,0)$) {$V_{\delta}^{*}$};
    \draw[signal] (V.east) to (CVD.west);
 
    \draw[signal] (CVD.east) 
        to node[midway,below]{$\IT^{*}_{\grd,\cmpD}$} ($(Rm)-(0.15*\x,0.25*\y)$)
        to ($(Rm)-0.15*(\x,0)$);
        
\end{circuitikz}

}  
	\caption
	{%
		Overview of the grid-following \CIDER including \DC-side dynamics.
		The power hardware is connected through measurements and coordinate transformations (i.e., in case of \AC signals) to the control software.
	}
	\label{fig:CIDER:PQ}
\end{figure*}

In this section, the detailed model of a grid-following \CIDER including the \DC-side dynamics is developed.
Specifically, a \CIDER with \DC-voltage control is considered.
To this end, the \DC side is represented by the current-source model as introduced in \cref{sec:intro} and \cref{fig:DCSide:CS}.
The precise structure of the \CIDER is shown in \cref{fig:CIDER:PQ}.
The power hardware consists of an $LCL$ filter on the \AC side, plus a current source and a link capacitor on the \DC side.
The control software is composed of a cascade of controllers.
For the sake of illustration, \ctrlPI controllers are considered. 
Note that the measurements and control signals, that are exchanged between the power hardware and control software, pass through coordinate transformations.
In this particular case, the Park transform is employed (this is common for three-phase \CIDER[s]).
The reference calculation computes the current setpoint for the control software based on the power setpoint and the voltage at the point of connection.

\subsection{Power Hardware}

In line with the considerations in \cref{sec:intro}, and assuming that the \CIDER exhibits constant-power behaviour, the \DC equivalent can be described by a controlled current source \cite{Jrn:Yazdani:2005}.
\begin{Hypothesis}\label{hyp:ph:idc}
    The \DC equivalent current $i_{\epsilon}$ is computed in order to track the power setpoint~$P^{*}$ using the \DC-side voltage~$v_{\delta}$:
    \begin{equation}
        i_{\epsilon}(t) = \frac{P^{*}}{v_{\delta}(t)}
        \label{eq:ph:idc}
    \end{equation}
\end{Hypothesis}
\noindent
In general, it can be assumed that the \DC-side voltage control loop of the \CIDER is designed to track its reference with zero error (e.g., a well-tuned \ctrlPI controller \cite{Bk:2006:Franklin}).
Namely, in steady-state its \DC component $V_{\delta,0}$ follows the reference $V_{\delta}^{*}$.
\begin{Hypothesis} \label{hyp:ph:vdc:0}
    The \DC-voltage control tracks the \DC-voltage reference in the \DC component without steady-state error.
    That is,:
    \begin{equation}
            V_{\delta,0}
        =   V_{\delta}^{*}
    \end{equation}
\end{Hypothesis}
\noindent
Typically, the \DC-voltage harmonics are negligible when compared to the \DC component~\cite{Jrn:TEC:2011:Wu}.
The following assumption is made.
\begin{Hypothesis}\label{hyp:ph:vdc:harms}
    The time-variant signal content of $v_{\delta}(t)$, as given by $\xi(t)$ below, is low.
    \begin{alignat}{2}
                v_{\delta}(t)
        &=      V_{\delta,0}(1+\xi(t)),~
        &       \Abs{\xi(t)}
        &\ll    1
    \end{alignat}
\end{Hypothesis}
\noindent
As a consequence of \cref{hyp:ph:vdc:0} and \cref{hyp:ph:vdc:harms}, \eqref{eq:ph:idc} can be linearized around the \DC voltage reference $V_{\delta}^{*}$.
\begin{align}
        i_{\epsilon}(t) 
    &\approx     \frac{P^{*}}{V_{\delta}^{*}} - \frac{P^{*}}{{V_{\delta}^{*}}^2} (v_{\delta}(t) - V_{\delta}^{*})\\
    &=     2\frac{P^{*}}{V_{\delta}^{*}} - \frac{P^{*}}{{V_{\delta}^{*}}^2} v_{\delta}(t)
    \label{eq:ph:ieps}
\end{align}

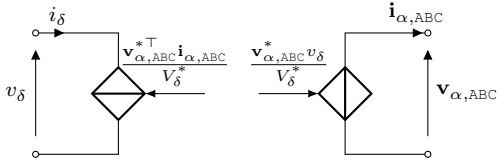
\begin{figure}
	\centering
	{

\footnotesize
\ctikzset{bipoles/length=1.0cm}
\tikzstyle{signal}=[-latex]

\begin{circuitikz}
    
	\def\x{1.0}
	\def\y{1.0}

    \node[rectangle, draw=white, minimum height=20mm, minimum width=40mm, inner sep=0pt] (C) at ($(0,0)$) {};
    
	\coordinate (IAN) at ($(C.south west)+(-0.6*\x,0.2*\y)$);
	\coordinate (IAP) at ($(C.north west)+(-0.6*\x,-0.2*\y)$);
	\coordinate (VAN) at ($(C.south east)+(0.6*\x,0.2*\y)$);
	\coordinate (VAP) at ($(C.north east)+(0.6*\x,-0.2*\y)$);
	
	\draw (IAP)
		to[short,o-,i^=$i_{\delta}$] ($(IAP)+(0.6*\x,0)$)
		to[short] ($(IAP)+(1.1*\x,0)$)
		to[cI] ($(IAN)+(1.1*\x,0)$)
		to[short,-o] (IAN);
	\draw (IAN) to[open,v^=$v_{\delta}$] (IAP);
	
	\draw (VAN)
		to[short,o-] ($(VAN)+(-1.1*\x,0)$)
		to[cV] ($(VAP)+(-1.1*\x,0)$)
		to[short] ($(VAP)-(0.3*\x,0)$)
		to[short,-o,i^=$\IT_{\act,\phsABC}$] (VAP);
	\draw (VAN) to[open,v_=$\VT_{\act,\phsABC}$] (VAP);
		
    \coordinate (I) at ($0.5*(IAP)+0.5*(IAN)+(1.45*\x,0)$);
    \node (IL) at ($(I)+(0.9*\x,0)$) {};
    \draw[signal] (IL) to node[midway,above]{$\frac{\VT_{\act,\phsABC}^{*\top}\IT_{\act,\phsABC}}{V_{\delta}^{*}}$} (I);
    
    \coordinate (V) at ($0.5*(VAP)+0.5*(VAN)-(1.45*\x,0)$);
    \node (VL) at ($(V)-(0.9*\x,0)$) {};
    \draw[signal] (VL) to node[midway,above]{$\frac{\VT_{\act,\phsABC}^{*}v_{\delta}}{V_{\delta}^{*}}$} (V);
    
\end{circuitikz}

}
	\caption
	{%
		Representation of the actuator by an average model, consisting of a controlled current source on the \DC side and a controlled voltage source on the \AC side.
	}
	\label{fig:actuator}
\end{figure}

As mentioned in \cref{sec:intro}, the actuator of a \CIDER is a \PWM switching converter.
Assuming that the \PWM generator has a high switching frequency (i.e., beyond the range of frequencies which are of interest for \HPF analysis), these high-frequency components do not need to be considered~\cite{Jrn:TPD:2010:Chiniforoosh}.
Thus, the actuator can be represented by an \emph{average model}~\cite{Dis:Peralta:2013}.
\begin{Hypothesis}\label{hyp:ph:act}
    In the frequency range of interest for the \HPF study, the switching losses and high-frequency components due to the converter switching are negligible.
    Therefore, the converter can be represented by an average model based on the instantaneous power balance equation between the \DC-side power $P_{\delta}$ and the \AC-side power $P_{\act}$.
    \begin{equation}
        P_{\delta} = P_{\act}
    \end{equation}
\end{Hypothesis}
The average model consists of a controlled current source on the DC side and a controlled voltage source on the AC side \cite{Dis:Peralta:2013} (see \cref{fig:actuator}).
The \AC-side voltage is derived from the \AC-side reference voltage $\mathbf{v}_{\act,\phsABC}^{*}$ and the \DC-side voltage $v_{\delta}$ through:
\begin{align}
        \mathbf{v}_{\act,\phsABC}(t) 
    =      \frac{1}{V_{\delta}^{*}}\mathbf{v}_{\act,\phsABC}^{*}(t) v_{\delta}(t)
    \label{eq:act:v:nonlin}
\end{align}
The \DC-side current $i_{\delta}$ is derived from $\mathbf{v}_{\act,\phsABC}^{*}$ and the \AC-side actuator current $\mathbf{i}_{\act,\phsABC}$:
\begin{align}
        \hspace{-1.6mm}i_{\delta}(t) 
    &=      \frac{1}{V_{\delta}^{*}} \sum_{j=\phsABC}v^{*}_{\act,j}(t)i_{\act,j}(t)
    =       \frac{1}{V_{\delta}^{*}} \mathbf{v}_{\act,\phsABC}^{*\top}(t)\mathbf{i}_{\act,\phsABC}(t)
    \label{eq:act:i:nonlin}
\end{align}
As previously mentioned in \cref{sec:model-rsc:nonlin}, these equations are nonlinear w.r.t. the point of view of the state-space variables.
Therefore, as proposed in \cref{sec:model-rsc:nonlin}, the actuator model is linearized around the operating point
\begin{align}
    \mathbf{y}_{o}(t) 
	&=	\left[
			\begin{array}{r}
				\bar{\mathbf{v}}_{\act,\phsABC}^{*}(t)	\\
				\bar{\mathbf{i}}_{\act,\phsABC}(t)	\\
				\bar{v}_{\delta}(t)	
			\end{array}
			\right]
    \label{eq:act:op}
\end{align}
which leads to the following linear representation (i.e., a Taylor expansion of \eqref{eq:act:v:nonlin} and \eqref{eq:act:i:nonlin} around \eqref{eq:act:op} truncated at the first term):
\begin{align}
		\mathbf{v}_{\act,\phsABC}(t)
    &\approx 	
		\begin{aligned}
			\frac{1}{V_{\delta}^{*}}
            \Big(   \bar{\mathbf{v}}_{\act,\phsABC}^{*}(t) v_{\delta}(t)
            +   \bar{v}_{\delta}(t)\mathbf{v}_{\act,\phsABC}^{*}(t)\\
            -   \bar{v}_{\delta}(t)\bar{\mathbf{v}}_{\act,\phsABC}^{*}(t)\Big)
		\end{aligned}
    \label{eq:act:voltage}
    \\
		i_{\delta}(t)
    &\approx 	
		\begin{aligned}
			\frac{1}{V_{\delta}^{*}}
            \Big(    \bar{\mathbf{v}}_{\act,\phsABC}^{*\top}(t) \mathbf{i}_{\act,\phsABC}(t)
            +   \bar{\mathbf{i}}_{\act,\phsABC}^{\top}(t)\mathbf{v}_{\act,\phsABC}^{*}(t) \\
            -   \bar{\mathbf{i}}_{\act,\phsABC}^{\top}(t)\bar{\mathbf{v}}_{\act,\phsABC}^{*}(t)\Big)
		\end{aligned}
    \label{eq:act:current}
\end{align}
Note that all quantities are linear time-periodic and thus the expression can be employed when deriving the \LTP model of the power hardware.

A general derivation of the filter equations was described by the authors of this paper in \cite{jrn:2020:kettner-becker:HPF-2}.
An inductive and capacitive filter stage are described by the following differential equations, respectively:
\begin{align}
		\hspace{-3mm}\VT_{\lambda-1,\phsABC} - \VT_{\lambda+1,\phsABC}
	&=	\RP_{\lambda,\phsABC}\IT_{\lambda,\phsABC} + \LP_{\lambda,\phsABC}\frac{d}{dt}\IT_{\lambda,\phsABC}
	\label{eq:fltr:ind:diffeq}\\
		\hspace{-3mm}\IT_{\lambda-1,\phsABC} - \IT_{\lambda+1,\phsABC}
	&=	\GP_{\lambda,\phsABC}\VT_{\lambda,\phsABC} + \CP_{\lambda,\phsABC}\frac{d}{dt}\VT_{\lambda,\phsABC}
	\label{eq:fltr:cap:diffeq}
\end{align}
$\RP_{\lambda,\phsABC},\LP_{\lambda,\phsABC}$ and $\GP_{\lambda,\phsABC},\CP_{\lambda,\phsABC}$ are the compound electrical parameters of the inductive and capacitive filter stage, respectively. $\IT_{\lambda,\phsABC}$ is the current flowing through the inductive filter stage, and $\VT_{\lambda-1,\phsABC},\VT_{\lambda+1,\phsABC}$ are the voltages at the start and end node of that stage, respectively.
Similarly, $\VT_{\lambda,\pwr}$ is the voltage across the capacitive filter stage, and $\IT_{\lambda-1,\pwr},\IT_{\lambda+1,\pwr}$ are the currents flowing into and out of it, respectively.

To obtain the state-space model of the power hardware, the equations of the \AC and the \DC filter stages are combined with the linearized actuator model and the current of the \DC equivalent.
More precisely, one needs to insert \eqref{eq:act:voltage} and \eqref{eq:act:current} as well as \eqref{eq:ph:ieps} into the equations of the filter stages and, then, derive the state-space format.

Thus, the state of the combined power hardware is given by the inductor currents $\IT_{\act,\phsABC}$ and $\IT_{\grd,\phsABC}$ and the capacitor voltage $\VT_{\fltr,\phsABC}$ on the \AC side and the capacitor voltage $v_{\delta}$ on the \DC side.
The input is the actuator voltage reference $\VT_{\act,\phsABC}^{*}$.
The disturbances are the grid voltage $\VT_{\grd,\phsABC}$, the average \DC-side equivalent current described by $\frac{P^{*}}{V_{\delta}^{*}}$ and the operating point of the actuator voltage reference $\bar{\VT}_{\act,\phsABC}^{*}$.
The output includes the state and the grid voltage as well as the \DC-side equivalent current $i_\epsilon$.
Namely,
\begingroup
\allowdisplaybreaks
\begin{align}
			\XT_{\pwr}(t)
	&=	\left[
			\begin{array}{r}
				\IT_{\act,\phsABC}(t)	\\
				\VT_{\fltr,\phsABC}(t)	\\
				\IT_{\grd,\phsABC}(t)    \\
				v_{\delta}(t)	
			\end{array}
			\right]
	\\
			\UT_{\pwr}(t)
	&=	\VT^{*}_{\act,\phsABC}(t)     	\\
			\WT_{\pwr}(t)
	&=	\left[
			\begin{array}{r}
			    \VT_{\grd,\phsABC}(t)       \\
				\frac{P^{*}}{V_{\delta}^{*}}\\
				\bar{\VT}^{*}_{\act,\phsABC}(t)       
			\end{array}
			\right]
    \\
		\YT_\pwr(t)
	&=	\begin{bmatrix}
	        \XT_\pwr(t)\\
			\VT_{\grd,\phsABC}(t)     \\
			i_{\epsilon}(t)
	   \end{bmatrix}
\end{align}
\endgroup
As one can see from \eqref{eq:act:voltage} and \eqref{eq:act:current} the third term is bilinear w.r.t. the components of the operating point (i.e., $\bar{v}_{\delta}\bar{\mathbf{v}}_{\act,\phsABC}^{*}$ and $\bar{\mathbf{i}}_{\act,\phsABC}^{\top}\bar{\mathbf{v}}_{\act,\phsABC}^{*}$).
In order to put this in the form of an \LTP system, one of the quantities (i.e., the voltage reference $\bar{\VT}^{*}_{\act,\phsABC}$) is added to the disturbance vector, while $\bar{v}_{\delta}$ and $\bar{\IT}_{\act,\phsABC}$ enter the matrices.
Similarly, the first term of \eqref{eq:ph:ieps} (i.e., $\frac{P^{*}}{V_{\delta}^{*}}$) is defined as a disturbance to the state-space model.

Recall that the matrices can be separated in two terms: one being invariant w.r.t. the operating point and one being a function of it.
\begin{align}
		\AP_{\pwr}(t)
	=       \tilde{\AP}_{\pwr}(t)+\bar{\AP}_{\pwr}(t,\mathbf{y}_{o,n}(t))
\end{align}
Each of these matrices itself can be described by its Fourier coefficients as in \eqref{eq:TP:matrix}.
The matrices of the state-space model that are invariant w.r.t. the operating point are given by
\begingroup
\allowdisplaybreaks
\begin{align}
        \tilde{\AP}_{\pwr,0}
	&=		\left[
            \setlength{\arraycolsep}{2.0pt}
			\begin{array}{llll}
					-\LP^{-1}_{\act}\RP_{\act}
				&	-\LP^{-1}_{\act}
				&	\phantom{-}\mathbf{0}_{3\times3}
				&	\phantom{-}\mathbf{0}_{3\times1}
				\\
					\phantom{-}\CP^{-1}_{\fltr}
				&	-\CP^{-1}_{\fltr}\GP_{\fltr}
				&	-\CP^{-1}_{\fltr}
				&	\phantom{-}\mathbf{0}_{3\times1}
				\\
					\phantom{-}\mathbf{0}_{3\times3}
				&	\phantom{-}\LP^{-1}_{\grd}
				&	-\LP^{-1}_{\grd}\RP_{\grd}
				&	\phantom{-}\mathbf{0}_{3\times1}
				\\
					\phantom{-}\mathbf{0}_{1\times3}
				&	\phantom{-}\mathbf{0}_{1\times3}
				&	\phantom{-}\mathbf{0}_{1\times3}
				&	(\tilde{\AP}_{\pwr,0})_{44}
			\end{array}
			\right]\\
	    &(\tilde{\AP}_{\pwr,0})_{44}
	=      -C^{-1}_{\delta}G_{\delta} - C^{-1}_{\delta} \frac{P^{*}}{{V_{\delta}^{*}}^2}
	\\
	\tilde{\BP}_{\pwr,0}
	&=		\mathbf{0}_{10\times3}
	\\
	\tilde{\EP}_{\pwr,0}
	&=		\left[
			\begin{array}{lll}
				    \phantom{-}\mathbf{0}_{6\times3}
				&   \phantom{-}\mathbf{0}_{6\times1}
				&   \phantom{-}\mathbf{0}_{6\times3}\\
				    -\LP^{-1}_{\grd}
				&   \phantom{-}\mathbf{0}_{3\times1}
				&   \phantom{-}\mathbf{0}_{3\times3}\\
				    \phantom{-}\mathbf{0}_{1\times3}
				&   -2C^{-1}_{\delta}
				&   \phantom{-}\mathbf{0}_{1\times3}
			\end{array}
			\right]
	\\
	\tilde{\CP}_{\pwr,0}
	&=		\left[
			\begin{array}{ll}
				    \diag(\mathbf{1}_{9})
				&   \phantom{-}\mathbf{0}_{9\times1}	\\
				    \mathbf{0}_{1\times9}
				&   \phantom{-}1	\\
				    \mathbf{0}_{3\times9}
				&   \phantom{-}\mathbf{0}_{3\times1}	\\
				    \mathbf{0}_{1\times9}
				&   -\frac{P^{*}}{{V_{\delta}^{*}}^2}	
			\end{array}
			\right]
	\\
	\tilde{\DP}_{\pwr,0}
	&=		\mathbf{0}_{14\times3}		
	\\
	\tilde{\FP}_{\pwr,0}
	&=		\left[
			\begin{array}{lll}
				    \mathbf{0}_{10\times3}	
				&   \mathbf{0}_{10\times1}	
				&   \mathbf{0}_{10\times3}\\
				    \diag(\mathbf{1}_{3})	
				&   \mathbf{0}_{3\times1}	
				&   \mathbf{0}_{3\times3}\\
				    \mathbf{0}_{1\times3}	
				&   2	
				&   \mathbf{0}_{1\times3}
			\end{array}
			\right]
\end{align}
\endgroup
and all other Fourier coefficients equal to zero.
The matrices that are a function of the operating point are given by
\begingroup
\allowdisplaybreaks
\begin{alignat}{1}
    \bar{\AP}_{\pwr,h}
	&=		\left[
			\begin{array}{llll}
					\phantom{-}\mathbf{0}_{3\times3}
				&	\phantom{-}\mathbf{0}_{3\times6}
				&	(\bar{\AP}_{\pwr,h})_{13}
				\\
					\phantom{-}\mathbf{0}_{6\times3}
				&	\phantom{-}\mathbf{0}_{6\times6}
				&	\phantom{-}\mathbf{0}_{6\times1}
				\\
					(\bar{\AP}_{\pwr,h})_{31}
				&	\phantom{-}\mathbf{0}_{1\times6}
				&	\phantom{-}0
			\end{array}
			\right]\\
	    &(\bar{\AP}_{\pwr,h})_{13}
	=      \frac{1}{V_{\delta}^{*}} \LP^{-1}_{\act} \mathbf{\bar{V}}_{\act,\phsABC,h}^{*}\\
	    &(\bar{\AP}_{\pwr,h})_{31}
	=      -\frac{1}{V_{\delta}^{*}} C^{-1}_{\delta} \mathbf{\bar{V}}_{\act,\phsABC,h}^{*\top}
	\\
	\bar{\BP}_{\pwr,h}
	&=		\left[
			\begin{array}{l}
				    \phantom{-}\frac{1}{V_{\delta}^{*}}\LP^{-1}_{\act}\bar{V}_{\delta,h}\\
				    \phantom{-}\mathbf{0}_{6\times3}\\
				    -\frac{1}{V_{\delta}^{*}} C^{-1}_{\delta} \mathbf{\bar{I}}_{\act,\phsABC,h}^{\top}
			\end{array}
			\right]
	\\
	\bar{\EP}_{\pwr,h}
	&=		\left[
			\begin{array}{ll}
				    \phantom{-}\mathbf{0}_{3\times4}
				&   -\frac{1}{V_{\delta}^{*}}\LP^{-1}_{\act}\bar{V}_{\delta,h}\\
				    \phantom{-}\mathbf{0}_{3\times4}
				&   \phantom{-}\mathbf{0}_{3\times3}\\
				    \phantom{-}\mathbf{0}_{3\times4}
				&   \phantom{-}\mathbf{0}_{3\times3}\\
				    \phantom{-}\mathbf{0}_{1\times4}
				&   \frac{1}{V_{\delta}^{*}} C^{-1}_{\delta} \mathbf{\bar{I}}_{\act,\phsABC,h}^{\top}
			\end{array}
			\right]
	\\
	\bar{\CP}_{\pwr,h}
	&=		\mathbf{0}_{14\times10}
	\\
	\bar{\DP}_{\pwr,h}
	&=		\mathbf{0}_{14\times3}		
	\\
	\bar{\FP}_{\pwr,h}
	&=		\mathbf{0}_{14\times7}
\end{alignat}
\endgroup

\subsection{Control Software}

The state-space model of the control software is obtained employing the theory described in \cite{jrn:2020:kettner-becker:HPF-2}.
The \DC-side controller provides the reference for the \DC current $i_{\delta}^{*}$, which is used for the grid current reference in the $\cmpD$-component $i_{\grd,\cmpD}^{*}$.
Namely,
\begin{align}
    i_{\grd,\cmpD}^{*}(t) = i_{\delta}^{*}(t)
    \label{eq:ctrl:ref:DCAC}
\end{align}

The generic description of a controller stage that corresponds to a filter stage of the power hardware was given in \cite{jrn:2020:kettner-becker:HPF-2}.
Let $\KP_{\ctrlFB,\lambda}$, $\KP_{\ctrlFF,\lambda},\KP_{\ctrlFT,\lambda}$ be the proportional gains and $T_{\ctrlFB,\lambda}$ the integration time of \ctrlPI controller, respectively.
The control law of an inductive filter stage is given by
\begin{align}
		\VT^{*}_{\lambda-1,\cmpDQ}
	&=	\left[~
			\begin{aligned}
				    &\KP_{\ctrlFB,\lambda}\left(\Delta\IT_{\lambda,\cmpDQ}+\frac{1}{T_{\ctrlFB,\lambda}}\int\Delta\IT_{\lambda,\cmpDQ}\,dt\right)\\
			    +   &\KP_{\ctrlFT,\lambda}\VT_{\lambda+1,\cmpDQ} + \KP_{\ctrlFF,\lambda}\IT^{*}_{\lambda,\cmpDQ}
			\end{aligned}
			\right.
	\label{eq:ctrl:ind:law}\\
				\Delta\IT_{\lambda,\cmpDQ}
	&\coloneqq	\IT^{*}_{\lambda,\cmpDQ}-\IT_{\lambda,\cmpDQ}
	\label{eq:ctrl:ind:error}
\end{align}
$\VT^{*}_{\lambda-1,\cmpDQ},\IT^{*}_{\lambda,\cmpDQ}$ are the reference voltage at the input of the controller stage and the reference current at its output, respectively.
$\VT_{\lambda+1,\cmpDQ},\IT_{\lambda,\cmpDQ}$ are the voltage at the output of the filter stage and the current through it, respectively.
The control law for a capacitive filter stage is obtained analogously by replacing voltages with currents and vice versa.

Combining \eqref{eq:ctrl:ref:DCAC} with the equations of the controller stages from the \AC and \DC side, leads to the state-space model of the control software.
As can be seen from \cref{fig:CIDER:PQ} the control software is composed of \ctrlPI controllers.
Its state is then given by the temporal integrals of the errors w.r.t. the inductor currents $\Delta\IT_{\act,\cmpDQ}$ and $\Delta\IT_{\grd,\cmpDQ}$, and the capacitor voltage $\Delta\VT_{\fltr,\cmpDQ}$.
The input and output vectors are defined by the interconnection with the power hardware as shown in \cref{fig:CIDER:PQ}.
The disturbance is the \cmpQ-component of the grid reference current $i^*_{\grd,\cmpQ}$ and the \DC-voltage reference $V^{*}_{\delta}$.
Accordingly
\begingroup
\allowdisplaybreaks
\begin{align}
						\XT_{\ctrl}(t)
	&\coloneqq	\int
						\left[
						\begin{array}{r}
							\Delta\IT_{\act,\cmpDQ}(t)	\\
							\Delta\VT_{\fltr,\cmpDQ}(t)	\\
							\Delta\IT_{\grd,\cmpDQ}(t)   \\
							\Delta v_{\delta}(t)
						\end{array}
						\right]
						dt
	=					\left[
						\begin{array}{l}
							\XT_{\ctrl,1}(t)	\\
							\XT_{\ctrl,2}(t)	\\
							\XT_{\ctrl,3}(t)    \\
							x_{\ctrl,4}(t)
						\end{array}
						\right]
	\\
						\UT_{\ctrl}(t)
	&\coloneqq	\left[
						\begin{array}{r}
							\IT_{\act,\cmpDQ}(t)	    \\
							\VT_{\fltr,\cmpDQ}(t)	\\
							\IT_{\grd,\cmpDQ}(t)	    \\
							v_{\delta}(t)	        \\
							\VT_{\grd,\cmpDQ}(t)     \\
							i_{\epsilon}(t)
						\end{array}
						\right]
	=					\left[
						\begin{array}{l}
							\UT_{\ctrl,1}(t)	\\
							\UT_{\ctrl,2}(t)	\\
							\UT_{\ctrl,3}(t)	\\
							u_{\ctrl,4}(t)      \\
							\UT_{\ctrl,5}(t)    \\
							u_{\ctrl,6}(t)
						\end{array}
						\right]
	\\
						\WT_{\ctrl}(t)
	&\coloneqq	\left[
        		\begin{array}{c}
        			i^{*}_{\grd,\cmpQ}(t)	\\
        			V^{*}_{\delta}  
        		\end{array}
        		\right]	
    =			\left[
				\begin{array}{l}
					w_{\ctrl,1}(t)	\\
					w_{\ctrl,2}(t)	
				\end{array}
				\right]
	\\
					    \YT_{\ctrl}(t)
	&\coloneqq	\VT^{*}_{\act,\cmpDQ}(t)
\end{align}
\endgroup
As opposed to the power hardware, the state-space model of the control software does purely depend on constant parameters, since no linearization needs to be performed.
Therefore, all matrices can be directly described by \eqref{eq:CIDER:ctrl:Atilde}--\eqref{eq:CIDER:ctrl:Ftilde}.
For the sake of clarity the following substitutions have been employed, $(K_{\FF} + K_{\FB}) = K_{\FFB}$ and similarly $(\KP_{\FF}+\KP_{\FB}) = \KP_{\FFB}$.
\begin{figure*}
\begin{align}
	\hspace{-3mm}	\tilde{\AP}_{\ctrl,0}
	&=		\left[
			\begin{array}{cccc}
					\mathbf{0}_{2\times2}
				&	\frac{\KP_{\FB,\fltr}}{T_{\FB,\fltr}}
				&	\KP_{\FFB,\fltr} \frac{\KP_{\FB,\grd}}{T_{\FB,\grd}}
				&	\KP_{\FFB,\fltr} \KP_{\FFB,\grd}\mathbf{e}_1 \frac{K_{\FB,\delta}}{T_{\FB,\delta}}
				\\
					\mathbf{0}_{2\times2}
				&	\mathbf{0}_{2\times2}
				&	\frac{\KP_{\FB,\grd}}{T_{\FB,\grd}}
				&	\KP_{\FFB,\grd}\mathbf{e}_1 \frac{K_{\FB,\delta}}{T_{\FB,\delta}}
				\\
					\mathbf{0}_{1\times2}
				&	\mathbf{0}_{1\times2}
				&	\mathbf{0}_{1\times2}
				&	\frac{K_{\FB,\delta}}{T_{\FB,\delta}}
				\\
					\mathbf{0}_{1\times2}
				&	\mathbf{0}_{1\times2}
				&	\mathbf{0}_{1\times2}
				&	0
				\\
					\mathbf{0}_{1\times2}
				&	\mathbf{0}_{1\times2}
				&	\mathbf{0}_{1\times2}
				&	0
			\end{array}
			\right]
	\label{eq:CIDER:ctrl:Atilde}
	\\
	\hspace{-3mm}	\tilde{\BP}_{\ctrl,0}
	&=		\left[
            \setlength{\arraycolsep}{1.8pt}
			\begin{array}{cccccc}
					-\diag(\mathbf{1}_{2})
				&	- \KP_{\FB,\fltr}
				&	\KP_{\FT,\fltr} {-} \KP_{\FFB,\fltr} \KP_{\FB,\grd}
				&	- \KP_{\FFB,\fltr} \KP_{\FFB,\grd}\mathbf{e}_1 K_{\FB,\delta}
				&	\KP_{\FFB,\fltr} \KP_{\FT,\grd}
				&	\KP_{\FFB,\fltr} \KP_{\FFB,\grd}\mathbf{e}_1 K_{\FT,\delta}
				\\
					\mathbf{0}_{2\times2}
				&	-\diag(\mathbf{1}_{2})
				&	-	\KP_{\FB,\grd}
				&	-	\KP_{\FFB,\grd}\mathbf{e}_1 K_{\FB,\delta}
				&	\KP_{\FT,\grd}
				&	\KP_{\FFB,\grd}\mathbf{e}_1 K_{\FT,\delta}
				\\
					\mathbf{0}_{1\times2}
				&	\mathbf{0}_{1\times2}
				&	-\mathbf{e}_1^{\top}
				&	-	K_{\FB,\delta}
				&	\mathbf{0}_{1\times2}
				&	K_{\FT,\delta}
				\\
					\mathbf{0}_{1\times2}
				&	\mathbf{0}_{1\times2}
				&	-\mathbf{e}_2^{\top}
				&	0
				&	\mathbf{0}_{1\times2}
				&	0
				\\
					\mathbf{0}_{1\times2}
				&	\mathbf{0}_{1\times2}
				&	\mathbf{0}_{1\times2}
				&	-1
				&	\mathbf{0}_{1\times2}
				&	0
			\end{array}
			\right]
	\\
	\hspace{-3mm}	\tilde{\EP}_{\ctrl,0}
	&=		\left[
			\begin{array}{cc}
					\KP_{\FFB,\fltr} \KP_{\FFB,\grd}\mathbf{e}_2
				&	\KP_{\FFB,\fltr} \KP_{\FFB,\grd}\mathbf{e}_1 K_{\FFB,\delta}
				\\
					\KP_{\FFB,\grd}\mathbf{e}_2
				&	\KP_{\FFB,\grd}\mathbf{e}_1 K_{\FFB,\delta}
				\\
					0
				&	K_{\FFB,\delta}
				\\
					1
				&	0
				\\
					0
				&	1
			\end{array}
			\right]
	\\
	\hspace{-3mm}    \tilde{\CP}_{\ctrl,0} 
	&= 	    \left[
			\begin{array}{cccc}
					\frac{\KP_{\FB,\act}}{T_{\FB,\act}}
				&	\KP_{\FFB,\act}\frac{\KP_{\FB,\fltr}}{T_{\FB,\fltr}}	
				&	\KP_{\FFB,\act}\KP_{\FFB,\fltr} \frac{\KP_{\FB,\grd}}{T_{\FB,\grd}}	
				&	\KP_{\FFB,\act}\KP_{\FFB,\fltr} \KP_{\FFB,\grd}\mathbf{e}_1 \frac{K_{\FB,\delta}}{T_{\FB,\delta}}	
			\end{array}
			\right]
    \\
    \hspace{-3mm}    \tilde{\DP}_{\ctrl,0} 
    &=      \left[
            \setlength{\arraycolsep}{4pt}
			\begin{array}{cccccc}
					-\KP_{\FB,\act}
				&	\KP_{\FT,\act}{-}\KP_{\FFB,\act}\KP_{\FB,\fltr}
				&	(\tilde{\DP}_{\ctrl,0})_{3}
				&	(\tilde{\DP}_{\ctrl,0})_{4}
				&	\KP_{\FFB,\act} \KP_{\FFB,\fltr} \KP_{\FT,\grd}
				&	\KP_{\FFB,\act} \KP_{\FFB,\fltr} \KP_{\FFB,\grd}\mathbf{e}_1 K_{\FT,\delta}
			\end{array}
			\right]\\
	\hspace{-3mm}    &(\tilde{\DP}_{\ctrl,0})_{3}
    =	    \KP_{\FFB,\act}\left(\KP_{\FT,\fltr} {-} \KP_{\FFB,\fltr} \KP_{\FB,\grd}\right)\\
	\hspace{-3mm}    &(\tilde{\DP}_{\ctrl,0})_{4}	
    =	    - \KP_{\FFB,\act} \KP_{\FFB,\fltr} \KP_{\FFB,\grd}\mathbf{e}_1 K_{\FB,\delta}
    \\
	\hspace{-3mm}    \tilde{\FP}_{\ctrl,0} 
    &=      \left[
			\begin{array}{cc}
					\KP_{\FFB,\act} \KP_{\FFB,\fltr} \KP_{\FFB,\grd}\mathbf{e}_2
				&	\KP_{\FFB,\act} \KP_{\FFB,\fltr} \KP_{\FFB,\grd}\mathbf{e}_1 K_{\FFB,\delta}
			\end{array}
			\right]
	\label{eq:CIDER:ctrl:Ftilde}
\end{align}
\end{figure*}


In \cref{fig:CIDER:PQ} the calculation of the \cmpQ-component of the grid reference current $i_{\grd,\cmpQ}^{*}(t)$ is performed using the reactive power setpoint $Q^{*}$ and the \cmpD-component of the grid voltage.
\begin{equation}
        i^*_{\grd,\cmpQ}(t)
    =   \frac{Q^*}{v_{\grd,\cmpD}(t)}
\end{equation}
In \cite{jrn:2020:kettner-becker:HPF-2} it is described how this reference calculation is incorporated in the \HPF method.

\section{Validation of the Resource Model}
\label{sec:val-rsc}

\subsection{Methodology and Key Performance Indicators}
\label{sec:val-rsc:method}



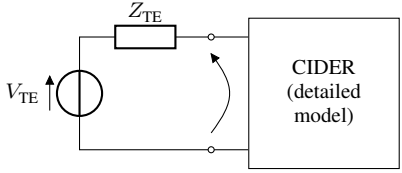
\begin{figure}[t]
	\centering
	{

\footnotesize
\ctikzset{bipoles/length=1.0cm}

\begin{circuitikz}[european]
    
	\def\x{1.0}
	\def\y{1.0}
	
	\coordinate (SN) at (0,0);
	\coordinate (SP) at ($(SN)+(0,1.5*\y)$);
	\coordinate (ON) at ($(SN)+(1.75*\x,0)$);
	\coordinate (OP) at ($(SP)+(1.75*\x,0)$);
	\coordinate (I) at ($0.5*(SN)+0.5*(SP)-(\x,0)$);
	\coordinate (R) at ($(I)+(4.25*\x,0)$);
	
	
	
	\node[rectangle,draw=black,minimum size=2.0cm] (Res) at (R)
	{%
	    \begin{tabular}{c}
	        \CIDER\\
	        (detailed\\ model)
	    \end{tabular}
	};
	\draw ($(Res.west)+(0,0.75*\y)$) to[short] (OP);
	\draw ($(Res.west)+(0,-0.75*\y)$) to[short] (ON);
	
	
	\draw (ON)
	    to[short] (SN)
	    to[voltage source = $V_{\TE}$] (SP)
	    to[R = $Z_{\TE}$] (OP);
	\draw (ON) to[open,o-o,v=$ $] (OP);
	
\end{circuitikz}

}
	\caption
	{%
	    Test setup for the validation of the \HPF method on individual \CIDER[s].
	    The resource is represented by a detailed state-space model (see \cref{sec:model-DC}), and the power system by a Th{\'e}venin equivalent (see \cref{tab:TE:parameters,tab:TE:harmonics}).
	}
	\label{fig:val-rsc:setup}
\end{figure}

\begin{table}[t]
    \centering
    \caption{Short-Circuit Parameters of the Th{\'e}venin Equivalent.}
    \label{tab:TE:parameters}
    {

\renewcommand{\arraystretch}{1.1}

\begin{tabular}{cccl}
    \hline
        Parameter
    &   Resource
    &   System
    &   Description
    \\
    &   Validation
    &   Validation
    \\
    \hline
        $V_{n}$
    &   230\,V-\RMS
    &   230\,V-\RMS
    &   Nominal voltage 
    \\
        $S_{\mathit{sc}}$
    &   267\,kW
    &   3.85\,MW
    &   Short-circuit power
    \\
        $\Abs{Z_{\mathit{sc}}}$
    &   195\,m$\Omega$
    &   13.7\,m$\Omega$
    &   Short-circuit impedance
    \\
        $R_{\mathit{sc}}/X_{\mathit{sc}}$
    &   6.207
    &   0.271
    &   Resistance-to-reactance ratio
    \\
    \hline
\end{tabular}
}
\end{table}

\begin{table}[t]
	\centering
	\caption{Harmonic Voltages of the Th{\'e}venin Equivalent (see \cite{Std:BSI-EN-50160:2000}).}
	\label{tab:TE:harmonics}
	{
\renewcommand{\arraystretch}{1.1}
\begin{tabular}{ccr}
    \hline
        $h$
    &   $|V_{\TE,h}|$
    &   \multicolumn{1}{c}{$\angle V_{\TE,h}$}
    \\
    \hline
        1
    &   1.0\,p.u.
    &   0\,rad
    \\
        5
    &   6.0\,\%
    &   $\pi$/8\,rad
    \\
        7
    &   5.0\,\%
    &   $\pi$/12\,rad
    \\ 
        11
    &   3.5\,\%
    &   $\pi$/16\,rad
    \\
        13
    &   3.0\,\%
    &   $\pi$/8\,rad
    \\
        17
    &   2.0\%
    &   $\pi$/12\,rad
    \\
        19
    &   1.5\,\%
    &   $\pi$/16\,rad
    \\
        23
    &   1.5\,\%
    &   $\pi$/16\,rad
    \\
    \hline
\end{tabular}
}

\end{table}

The validation of the individual resource transfer function is performed using the test setup shown in \cref{fig:val-rsc:setup}.
The \CIDER is directly connected to a \emph{Th{\'e}venin Equivalent} (\TE), which is described by the short-circuit parameters given in \cref{tab:TE:parameters}.
Furthermore, the \TE injects harmonics with levels shown in \cref{tab:TE:harmonics} based on \cite{Std:BSI-EN-50160:2000}.

The parameters of the grid-following \CIDER that includes the \DC-side characteristics are given in \cref{tab:CIDER-following:parameters}.
The parameters of the \AC-filter and the \DC-link capacitor are derived based on the design guidelines proposed in \cite{Jrn:Liserre:2004}.
The active and reactive power setpoints are $P^{*}=50\,\text{kW}$ and $Q^{*}=16.4\,\text{kVAr}$, respectively, and $V_{\delta}^{*}=900\,\text{V}$.

In the context of this paper the Matlab project of the \HPF method proposed in \cite{jrn:2020:kettner-becker:HPF-1, jrn:2020:kettner-becker:HPF-2} is extended to include the \DC-side modelling.
As comparison \emph{Time-Domain Simulations} (\TDS) in Simulink are conducted using averaged models of the \CIDER[s].
Additionally, the Simulink models are updated to include the \DC-side characteristics.
A \emph{Discrete Fourier Transform} (\DFT) of the last 5 periods of the fundamental frequency in steady state is performed.
All signals are normalized w.r.t. the base power $P_{b}=50\,\text{kW}$ and the base voltage $V_{b}=230\,\text{V-\RMS}$.

\begin{table}[t]
	\centering
	\caption
	{%
	    Parameters of the Grid-Following Resource\linebreak
	    (Rated Power $60\,\text{kVA}$).
	}
	\label{tab:CIDER-following:parameters}
	{

\renewcommand{\arraystretch}{1.1}
\setlength{\tabcolsep}{0.15cm}

\begin{tabular}{lccccc}
	\hline
		Filter stage
	&	$L$/$C$
	&	$R$/$G$
	&	$K_{\FB}$
	&	$T_{\FB}$
	&	$K_{\FT}$
	\\
	\hline
	    DC-Link Capacitor ($\delta$)
	&	310~$\mu$F
	&	0~S
	&	10
	&	3e-3
	&	1
	\\
		Actuator-side inductor ($\alpha$)
	&	325~$\mu$H
	&	1.02~m$\Omega$
	&	9.56
	&	1.47e-4
	&	1
	\\
	    Capacitor ($\varphi$)
	&	90.3~$\mu$F
	&	0~S
	&	0.569
	&	8.97e-4
	&	0
	\\
		Grid-side inductor ($\gamma$)
	&	325~$\mu$H
	&	1.02~m$\Omega$
	&	0.23
	&	1e-3
	&	1
	\\
	\hline
\end{tabular}

}
\end{table}


In order to assess the accuracy of the \HPF method \emph{Key Performance Indicators} (\KPI[s]) are defined.
That is, the accuracy is defined as the errors of the harmonic phasors between the \DFT of the \TDS and the results of the \HPF.
If the Fourier coefficient of a three-phase electrical quantity (i.e., voltage or current) is denoted as $\mathbf{X}_{h}$, the \KPI[s] are defined as follows:
\begin{align}
                e_{\textup{abs}}(\mathbf{X}_{h})
    &\coloneqq  \max_{p} \Abs{\Abs{X_{h,p,\HPF}}-\Abs{X_{h,p,\TDS}} }\\
                e_{\textup{arg}}(\mathbf{X}_{h})
    &\coloneqq  \max_{p} \Abs{ \angle X_{h,p,\HPF}- \angle X_{h,p,\TDS} }
\end{align}
In other words, $e_{\textup{abs}}(\mathbf{X}_{h})$ and $e_{\textup{arg}}(\mathbf{X}_{h})$ describe the maximum absolute errors over all phases $p\in\phases$ in magnitude and phase, respectively.


\subsection{Results and Discussion}
\label{sec:val-rsc:results}

\begin{figure}[t]
    \centering
    \subfloat[]
    {%
        \centering
        \includegraphics[width=1\linewidth]{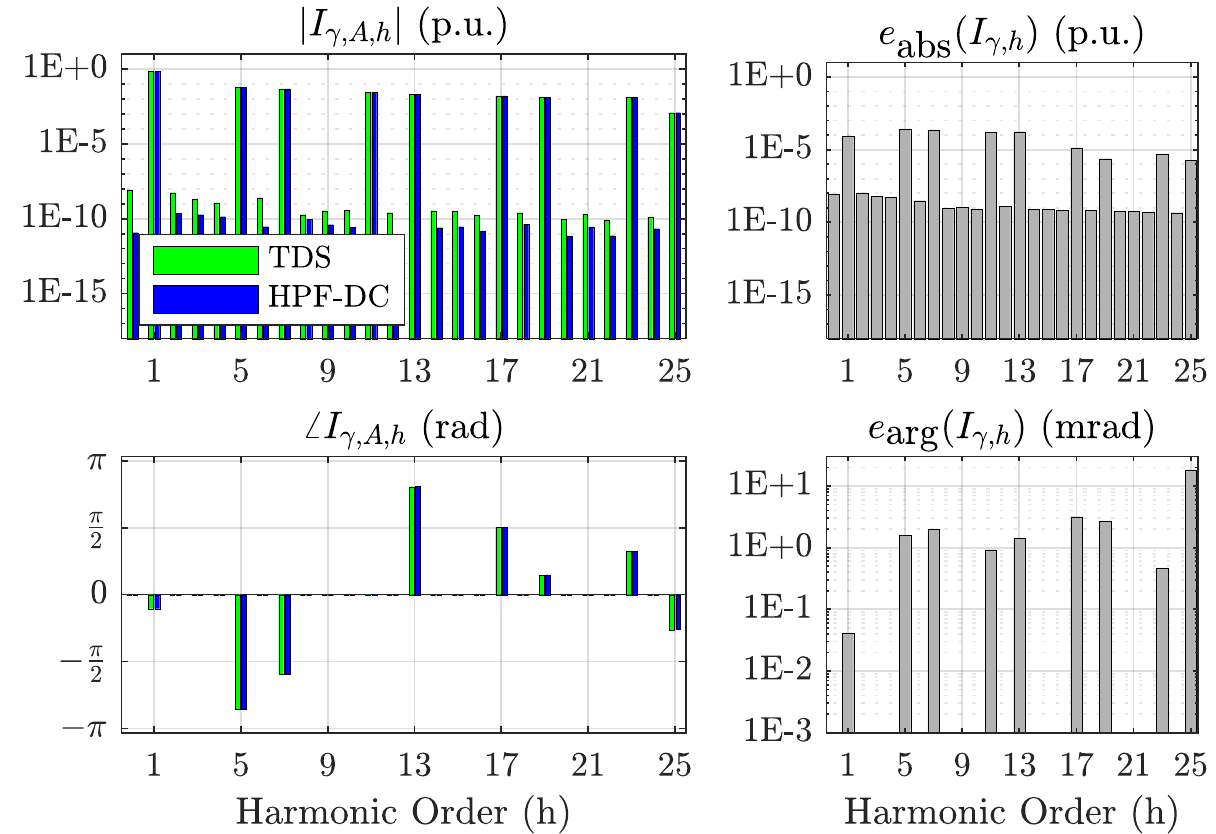}
        \label{fig:resource:results:IG}
    }
    
    \subfloat[]
    {%
        \centering
        \includegraphics[width=1\linewidth]{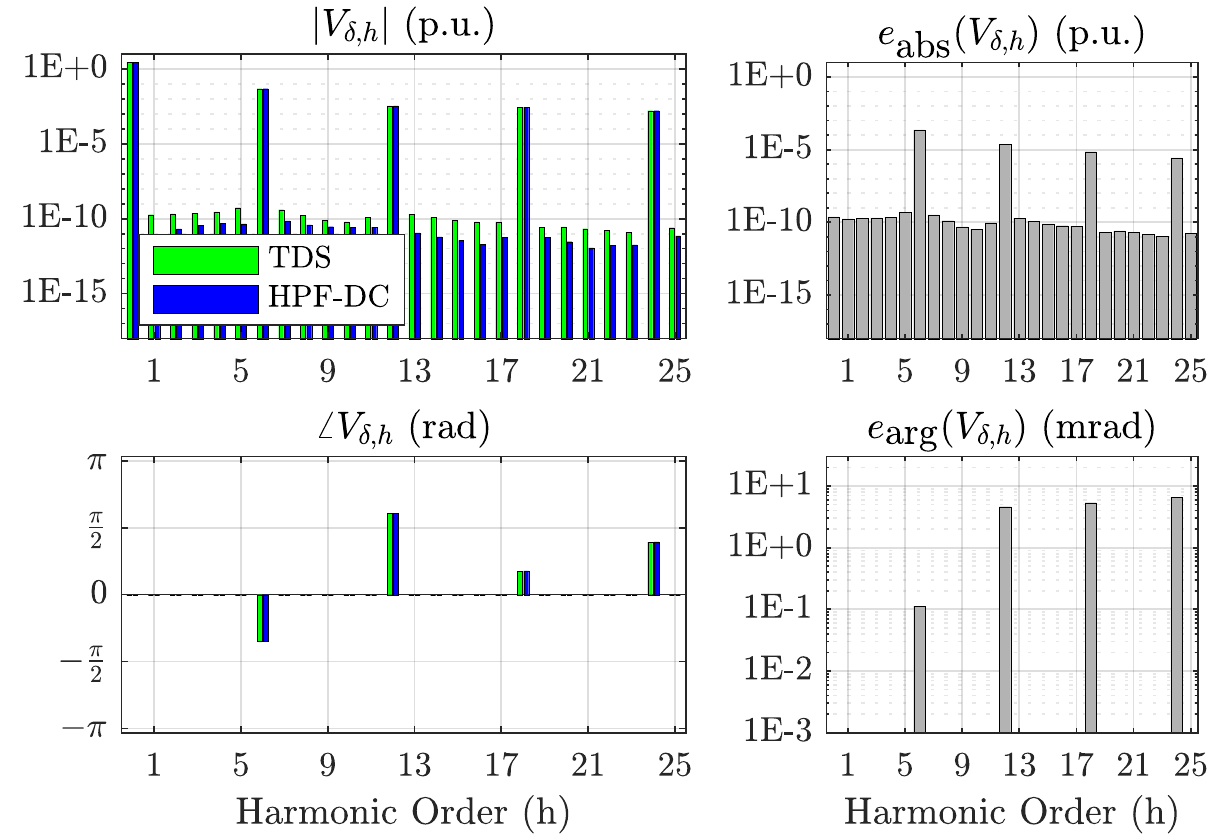}
        \label{fig:resource:results:VD}
    }
    \caption
    {%
        Results of the validation on individual grid-following \CIDER.
        The grid current (\ref{fig:resource:results:IG}) and \DC-side voltage (\ref{fig:resource:results:VD}) are shown.
        The plots on the left-hand side show the spectra (i.e., for phase A in case of the grid current), and the ones on the right-hand side the error defined through the \KPI[s].
    }
    \label{fig:resource:results}
\end{figure}

In \cref{fig:resource:results} the results for the controlled quantities (i.e., the grid-current and the \DC-link voltage) of the individual \CIDER is shown.
The left-hand side of the plots depict the magnitude and angle of the Fourier coefficients obtained with the \HPF and \TDS, respectively.
The quantities are almost identical, yielding small \KPI[s] as shown on the right-hand side of the plots.
The highest errors are $e_{\textup{abs}}(\mathbf{I}_{\grd,5})=2.62\text{ E-4}$~p.u. and $e_{\textup{arg}}(\mathbf{I}_{\grd,25})=18.3$~mrad for the grid current and $e_{\textup{abs}}(\mathbf{V}_{\delta,6})=2.27$~E-4~p.u. and $e_{\textup{arg}}(\mathbf{V}_{\delta,24})=6.6$~mrad for the \DC-link voltage.
Notably, since the applied \TE voltage does not include unbalances, the \DC-link voltage exhibits only even harmonics.

\section{Validation of the Extended Algorithm}
\label{sec:val-sys}


\subsection{Methodology and Key Performance Indicators}
\label{sec:val-sys:method}


The extension of the \HPF algorithm that includes the \DC side of the \CIDER[s] is validated on a test system adapted from \CIGRE low-voltage benchmark microgrid \cite{Rep:2014:CIGRE}.
It is composed of a substation at node N1, five grid-following \CIDER[s] at nodes N11, and N15-N18 and four unbalanced loads at nodes N19-N22.
The feeding substation is modelled as a \TE described by parameters depicted in \cref{tab:TE:parameters}.
The \TE injects harmonics with the same parameters as for the validation in \cref{sec:val-rsc:method} (i.e., see \cref{tab:TE:harmonics}).
Notably, these parameters account for both the substation transformer and the upstream grid.
The lines are described by the sequence parameters given in \cref{tab:grid:parameters}.
The parameters of the grid-following \CIDER[s] are identical to the ones used in the individual resource validation (i.e., they are given in \cref{tab:CIDER-following:parameters}.
Unbalances are introduced into the grid through the unbalanced wye-connected constant-impedance loads at nodes N19-N22.
The amount of unbalance is expressed by phase weights, that describe the distribution of the load over the phases.
All setpoints and phase weights of the grid-following resources are given in \cref{tab:resources:references}.

The validation of the \HPF algorithm is three-fold.
Firstly, similar to \cref{sec:val-rsc}, the results of the \HPF method for the test system including the \DC sides of the \CIDER[s] are compared against the results from Simulink (\TDS).
Secondly, the \HPF method that includes the \DC side of the \CIDER[s], and the one that excludes it (i.e., using the models of the \CIDER[s] introduced in \cite{jrn:2020:kettner-becker:HPF-2}), are compared.
Thirdly, a comparison of the \HPF method with a classical decoupled \HPF is performed.
For the decoupled \HPF, the resources are represented by independent and superposed harmonic current sources and the system equations are solved independently at each harmonic frequency (see Appendix~\ref{app:D-HPF} for more details).

In both cases the Newton-Raphson algorithm is initialized as described in \cref{sec:algo}.
That is, initial values of \AC and \DC quantities are based on balanced, sinusoidal conditions and nominal values or setpoints.
In order to assess the robustness of the convergence, this initial point is distorted with random positive, negative and homopolar sequence components.
The magnitudes and phases are chosen from uniform distributions within $\pm~20\%$ and $[0,2\pi)$, respectively.

The accuracy of the result is assessed using the same \KPI[s] as for the validation of the individual resource model.
Additionally, the computation time of the \HPF method is compared against the execution time of the \TDS.
Notably, the initial transients of the \TDS are not taken into account in this context.
Hence, it is computed as the sum of the simulation time for $5$ periods of fundamental frequency and the time needed for the Fourier analysis.
In order to compare the two versions of the \HPF method (i.e., the one including and excluding the \DC side of the \CIDER[s]) the \emph{Total Harmonic Distortion} (\THD) of the nodal voltages and injected currents are analysed.
To reduce the amount of data to be displayed, the maximum \THD over all phases is used.

\begin{figure}[t]
	\centering
    {
\tikzstyle{bus}=[circle,fill=black,minimum size=1.5mm,inner sep=0mm]
\tikzstyle{UG1}=[]
\tikzstyle{UG3}=[dashed]
\tikzstyle{load}=[-latex]

\begin{circuitikz}
	\scriptsize
	
	\def\BlockSize{1.0}	
	\def\X{3.8}	
	\def\Y{3.6}	
	\def\dlX{0.77}
	\def\dlY{1.1}
	\def\dload{0.4}
	
	
	\node[bus,label={left:N1}] (N1) at (0,0) {};
	\node[bus,label={below right:N2}] (N2) at (\dlX,0) {};
	\node[bus,gray,label={below:N21}] (N21) at (\dlX,-\dlY) {};
	\node[bus,label={below:N3}] (N3) at (2*\dlX,0) {};
	\node[bus,label={above:N11}] (N11) at (2*\dlX,\dlY) {};
	\node[bus,label={below:N4}] (N4) at (3*\dlX,0) {};
	\node[bus,label={below right:N5}] (N5) at (4*\dlX,0) {};
	\node[bus,label={left:N12}] (N12) at (4*\dlX,-\dlY) {};
	\node[bus,gray,label={below:N19}] (N19) at (5*\dlX,-\dlY) {};
	\node[bus,label={below:N13}] (N13) at (4*\dlX,-2*\dlY) {};
	\node[bus,gray,label={below:N22}] (N22) at (3*\dlX,-2*\dlY) {};
	\node[bus,label={below:N14}] (N14) at (5*\dlX,-2*\dlY) {};
	\node[bus,label={below:N15}] (N15) at (6*\dlX,-2*\dlY) {};
	\node[bus,label={below:N6}] (N6) at (5*\dlX,0) {};
	\node[bus,label={above:N16}] (N16) at (5*\dlX,\dlY) {};
	\node[bus,label={below:N7}] (N7) at (6*\dlX,0) {};
	\node[bus,label={below:N8}] (N8) at (7*\dlX,0) {};
	\node[bus,gray,label={above:N20}] (N20) at (7*\dlX,\dlY) {};
	\node[bus,label={below right:N9}] (N9) at (8*\dlX,0) {};
	\node[bus,label={below:N10}] (N10) at (8*\dlX,-\dlY) {};
	\node[bus,label={below:N17}] (N17) at (9*\dlX,0) {};
	\node[bus,label={below:N18}] (N18) at (9*\dlX,-\dlY) {};

	
	\draw[UG1] (N1) to node[midway,above]{35m} (N2) {};
	\draw[UG1] (N2) to node[midway,above]{35m} (N3) {};
	\draw[UG1] (N3) to node[midway,above]{35m} (N4) {};
	\draw[UG1] (N4) to node[midway,above]{35m} (N5) {};
	\draw[UG1] (N5) to node[midway,above]{35m} (N6) {};
	\draw[UG1] (N6) to node[midway,above]{35m} (N7) {};
	\draw[UG1] (N7) to node[midway,above]{35m} (N8) {};
	\draw[UG1] (N8) to node[midway,above]{35m} (N9) {};
	\draw[UG1] (N10) to node[sloped,anchor=center,above]{35m} (N9) {};
	
	\draw[UG3] (N3) to node[sloped,anchor=center,above]{30m} (N11) {};
	
	\draw[UG3] (N12) to node[sloped,anchor=center,above]{35m} (N5) {};
	\draw[UG3] (N13) to node[sloped,anchor=center,above]{35m} (N12) {};
	\draw[UG3] (N13) to node[midway,above]{35m} (N14) {};
	\draw[UG3] (N14) to node[midway,above]{30m} (N15) {};
	
	\draw[UG3] (N6) to node[sloped,anchor=center,above]{30m} (N16) {};
	\draw[UG3] (N9) to node[sloped,anchor=center,above]{30m} (N17) {};
	\draw[UG3] (N18) to node[sloped,anchor=center,above]{30m} (N10) {};

	\filldraw[gray,UG3] (N12) to node[sloped,anchor=center,above]{30m} (N19) {};
	\draw[gray,UG3] (N8) to node[sloped,anchor=center,above]{30m} (N20) {};
	\draw[gray,UG3] (N21) to node[sloped,anchor=center,above]{30m} (N2) {};
	\draw[gray,UG3] (N22) to node[midway,above]{30m} (N13) {};
	
	\draw[load] (N11) to node[right,align=left]{~P/Q}
	($(N11)+\dload*(1,0)$);
	\draw[load] (N15) to node[right,align=left]{~P/Q}
	($(N15)+\dload*(1,0)$);
	\draw[load] (N16) to node[right,align=left]{~P/Q}
	($(N16)+\dload*(1,0)$);
	\draw[load] (N17) to node[right,align=left]{~P/Q}
	($(N17)+\dload*(1,0)$);
	\draw[load] (N18) to node[right,align=left]{~P/Q}
	($(N18)+\dload*(1,0)$);
	
	\draw[load] (N19) to node[right,align=left]{~Z}
	($(N19)+\dload*(1,0)$);
	\draw[load] (N20) to node[right,align=left]{~Z}
	($(N20)+\dload*(1,0)$);
	\draw[load] (N21) to node[right,align=left]{~Z}
	($(N21)+\dload*(1,0)$);
	\draw[load] (N22) to node[left,align=right]{Z~}
    ($(N22)-\dload*(1,0)$);

	\draw[-] ($(N1)+\dload*(0,-1)$) to ($(N1)+\dload*(0,1)$);
	\node[label={above:Substation}] (Substation) at ($(N1)+\dload*(0,1)$) {};

	
    \coordinate (Leg) at (-0.3*\dlX,-1.5*\dlY);
    \matrix [draw,below right] at (Leg) {
        \node [UG1,label=right:~~~~UG1] {}; \\
        \node [UG3,label=right:~~~~UG3] {}; \\
    };
    \draw[UG1] ($(Leg)+(0.15*\dlX,-0.25*\dlY)$) to node[]{} ($(Leg)+(0.75*\dlX,-0.25*\dlY)$);
    \draw[UG3] ($(Leg)+(0.15*\dlX,-0.55*\dlY)$) to node[]{} ($(Leg)+(0.75*\dlX,-0.55*\dlY)$);
    
\end{circuitikz}
}
	\caption
	{%
	    Schematic diagram of the test system, which is based on the \CIGRE low-voltage benchmark microgrid \cite{Rep:2014:CIGRE} (in black) and extended by unbalanced impedance loads (in grey).
	    The resources are composed of constant impedance loads (Z) and constant power loads (P/Q), parameters given in \cref{tab:resources:references}.
	}
	\label{fig:grid:schematic}
\end{figure}
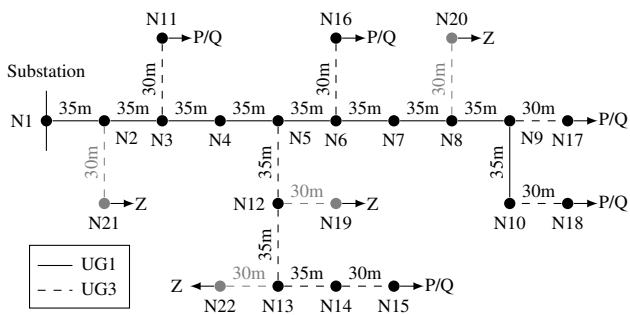

\begin{table}[t]
    \centering
    \caption{Sequence Parameters of the Lines in the Test System.}
    \label{tab:grid:parameters}
	{
\renewcommand{\arraystretch}{1.1}
\setlength{\tabcolsep}{0.15cm}

\begin{tabular}{ccccccc}
    \hline
        ID
    &   $R_{+}/R_{-}$ 
    &   $R_{0}$ 
    &   $L_{+}/L_{-}$ 
    &   $L_{0}$ 
    &   $C_{+}/C_{-}$ 
    &   $C_{0}$ 
    \\
    \hline
        UG1
    &   0.162~$\Omega$
    &   0.529~$\Omega$
    &   0.262~mH
    &   1.185~mH
    &   637~nF
    &   388~nF
    \\
        UG3
    &   0.822~$\Omega$
    &   1.794~$\Omega$
    &   0.270~mH
    &   3.895~mH
    &   637~nF
    &   388~nF
    \\
    \hline
\end{tabular}
}
\end{table}

\begin{table}[t]
    \centering
    \caption{Parameters of the Grid-Following Resources and Loads\newline in the Test System.}
    \label{tab:resources:references}
	{

\renewcommand{\arraystretch}{1.2}
\begin{tabular}{ccccc}
	\hline
		Node
	&	S
	&	pf
	&	Type
	&   Phase weights
	\\
	\hline
		N11
	&	\phantom{-}15.0~kW
	&	0.95
	&   P/Q
	&   [0.33 0.33 0.33]
	\\
		N15
	&	\phantom{-}52.0~kW
	&	0.95
	&   P/Q
	&   [0.33 0.33 0.33]
	\\
		N16
	&	\phantom{-}55.0~kW
	&	0.95
	&   P/Q
	&   [0.33 0.33 0.33]
	\\
		N17
	&	\phantom{-}35.0~kW   
	&	0.95
	&   P/Q
	&   [0.33 0.33 0.33]
	\\
		N18
	&	\phantom{-}47.0~kW   
	&	0.95
	&   P/Q
	&   [0.33 0.33 0.33]
	\\
		N19
	&	-51.2~kW   
	&	0.95
	&   Z
	&   [0.31 0.50 0.19]
	\\
		N20
	&	-51.7~kW   
	&	0.95
	&   Z
	&   [0.45 0.23 0.32]
	\\
		N21
	&	-61.5~kW   
	&	0.95
	&   Z
	&   [0.24 0.39 0.37]
	\\
		N22
	&	-61.9~kW   
	&	0.95
	&   Z
	&   [0.31 0.56 0.13]
	\\
	\hline
\end{tabular}
}
\end{table}


\subsection{Results and Discussion}
\label{sec:val-sys:results}

\begin{figure}[t]
    \centering
    \includegraphics[width=1\linewidth]{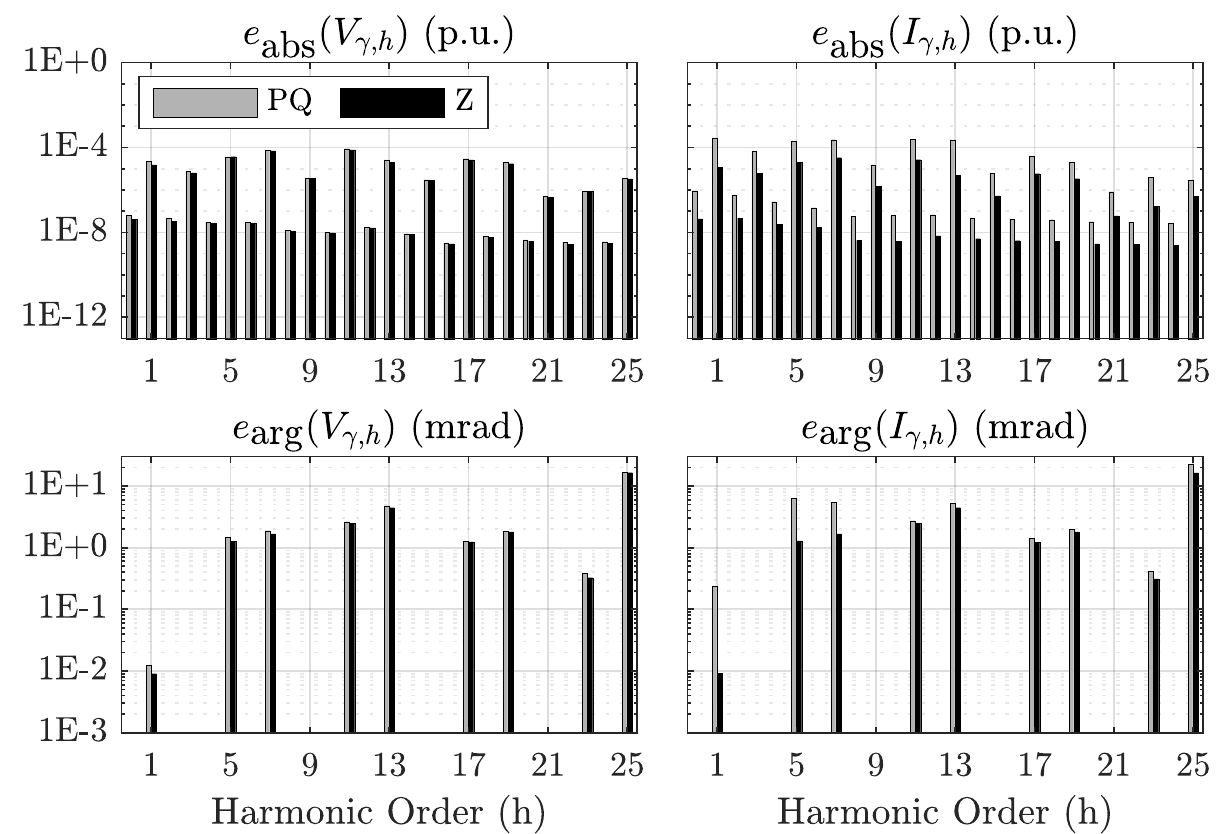}
    \caption
    {%
        Results of the validation on the benchmark system for the grid-following \CIDER[s] (PQ) and the constant impedance loads (Z).
        The plots show the maximum absolute errors over all nodes and phases, for voltages (left column) and currents (right column), in magnitude (top row) and phase (bottom row).
    }
    \label{fig:system:error:VI}
\end{figure}

The accuracy of the \HPF method including the \DC-side characteristics compared to the \TDS is shown in \cref{fig:system:error:VI}.
The highest errors at every harmonic frequency and over all nodes and phases are shown for the set of all grid-following \CIDER[s] and the set of passive impedance loads.
Notably, the errors for the third set (i.e., the zero-injection nodes) can be inferred directly from the hybrid parameters of the grid and the other nodal quantities (i.e., as a linear superposition).
Accordingly, if high errors between HPF and TDS were to be observed at a zero-injection node, the origin of this issue would be at nodes with \CIDER[s], where these problems would be expected to be even more prominent.
The maximum errors in \cref{fig:system:error:VI} occur in the set of grid-following \CIDER[s].
More precisely, the maximum errors regarding the voltage in magnitude and angle are $e_{\textup{abs}}(\mathbf{V}_{11})=8.71$E-5~p.u. and $e_{\textup{arg}}(\mathbf{V}_{25})=16.7$~mrad, respectively.
In terms of errors w.r.t. current magnitude and angle the highest values are $e_{\textup{abs}}(\mathbf{I}_{1})=2.84$E-4~p.u. and $e_{\textup{arg}}(\mathbf{I}_{25})=22.5$~mrad, respectively.
Notably, the observed errors are very low, i.e., lower than the accuracy of standard measurement devices (i.e., compared to a 0.5~class instrument transformer, as defined in the standards \cite{Std:IEC_61869_3:2011,Std:IEC_61869_2:2012,Std:IEC_61869_6:2016}, and shortly discussed in Appendix~\ref{app:IT_accuracy}).
Therefore, the extended \HPF method is regarded as more precise and will be used as a benchmark in the subsequent analyses of this paper.

All simulations are run on the same laptop computer, namely a MacBook Pro 2019 with a 2.4 GHz Intel Core i9 CPU and 32 GB 2400 MHz DDR4 RAM. 
The \HPF method takes 9~iterations and 16.7~sec, while the \TDS takes 52.3~sec, out of which 0.5~sec are used for the \DFT.
The \TDS takes roughly three times as long as the \HPF.
Note that the implementation of the \HPF was not done with a strong focus on numerical optimization.

\begin{figure}[ht]
    \centering
    
    \subfloat[]
    {%
        \centering
        \includegraphics[width=\linewidth]{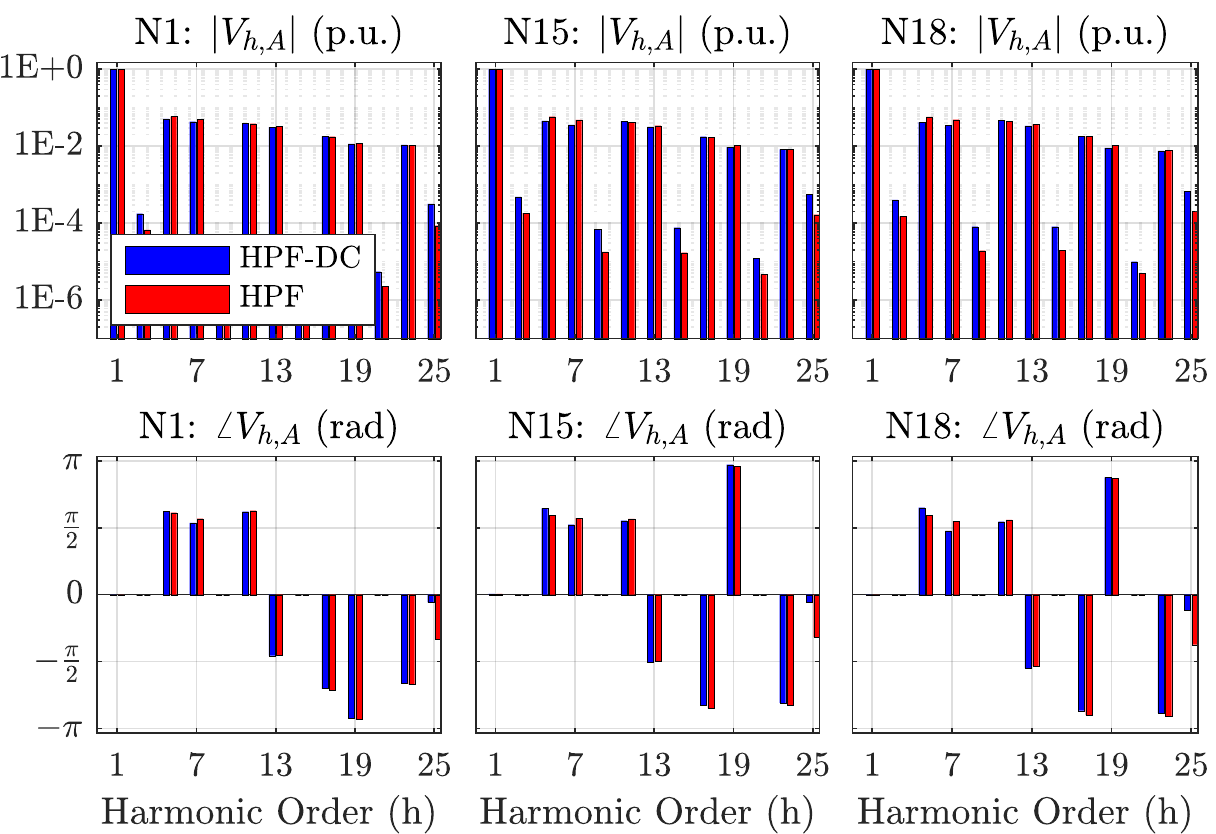}
        \label{fig:cmpDCside:voltage}
    }
    
    \subfloat[]
    {%
        \centering
        \includegraphics[width=\linewidth]{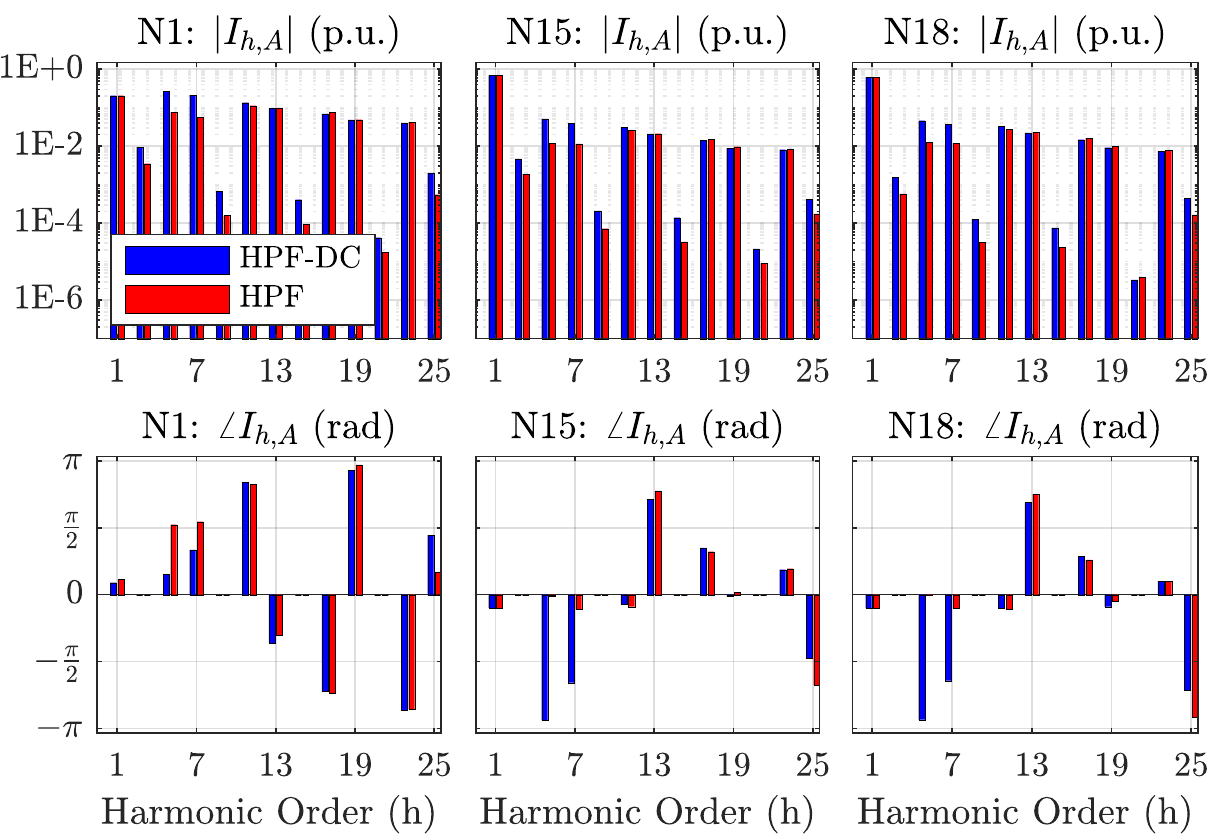}
        \label{fig:cmpDCside:current}
    }
    
    \caption{%
        Comparison of the \HPF study including (i.e., HPF-DC) and excluding the \DC side of the \CIDER[s] at three nodes throughout the benchmark system.
        The voltages for Phase A are given in (\ref{fig:cmpDCside:voltage}) and the currents in (\ref{fig:cmpDCside:current}).
    }
    \label{fig:cmpDCside}
\end{figure}

In order to assess the impact of the \DC-side modelling on the propagation of harmonics, analyses on the benchmark system were conducted using models of the \CIDER[s] that either in- or exclude the \DC side.
The obtained results are shown in \cref{fig:cmpDCside}, where the comparision is done at three nodes throughout the system.
The spectra differ significantly between the two versions, particular high differences are visible in the angles of the currents.
Similar observations can be drawn from \cref{tab:system:THD}, where the maximum \THD in voltages and currents show significant differences between the two methods.
One can conclude that the inclusion of the \DC side into the \CIDER modelling does have a strong impact on the harmonic propagation through the system.

\begin{table}[t]
    \centering
    \caption{
        Maximum \THD at nodes with resources for the \HPF method including and excluding the \DC side of \CIDER[s].
    }
    \label{tab:system:THD}
	{
\renewcommand{\arraystretch}{1.1}

\begin{tabular}{ccccc}
    \hline
    &   \multicolumn{2}{c}{$\mathrm{THD}_{\textup{max}}(\mathbf{V}_{\grd})$}
    &   \multicolumn{2}{c}{$\mathrm{THD}_{\textup{max}}(\mathbf{I}_{\grd})$}
    \\
        Node
    &   HPF
    &   HPF-DC
    &   HPF
    &   HPF-DC
    \\
    \hline
    N01 & 9.64 & 8.86 & 99.06 & 189.52\\
    N11 & 9.71 & 8.59 & 21.28 & \phantom{1}25.38\\
    N15 & 9.63 & 8.30 & \phantom{0}6.24 & \phantom{1}11.27\\
    N16 & 9.75 & 8.36 & \phantom{0}6.09 & \phantom{1}10.99\\
    N17 & 9.88 & 8.35 & 10.06 & \phantom{1}14.55\\
    N18 & 9.88 & 8.33 & \phantom{0}7.47 & \phantom{1}12.04\\
    N19 & 9.82 & 8.57 & \phantom{0}4.16 & \phantom{10}3.41\\
    N20 & 9.93 & 8.46 & \phantom{0}5.28 & \phantom{10}4.17\\
    N21 & 9.69 & 8.75 & \phantom{0}4.14 & \phantom{10}3.59\\
    N22 & 9.82 & 8.53 & \phantom{0}4.17 & \phantom{10}3.41\\
    \hline
\end{tabular}

} 

\end{table}

\begin{figure}[ht]
    \centering
    
    \subfloat[]
    {%
        \centering
        \includegraphics[width=\linewidth]{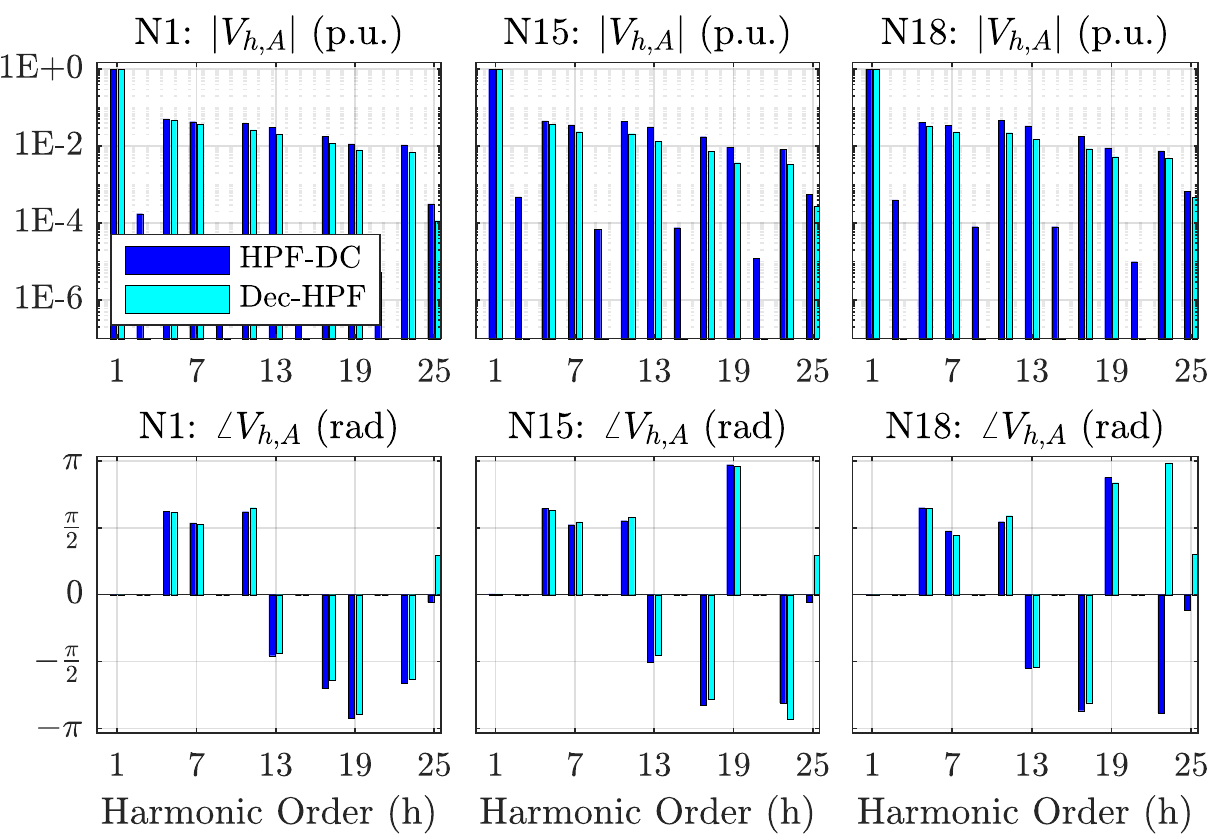}
        \label{fig:D-HPF:voltage}
    }
    
    \subfloat[]
    {%
        \centering
        \includegraphics[width=\linewidth]{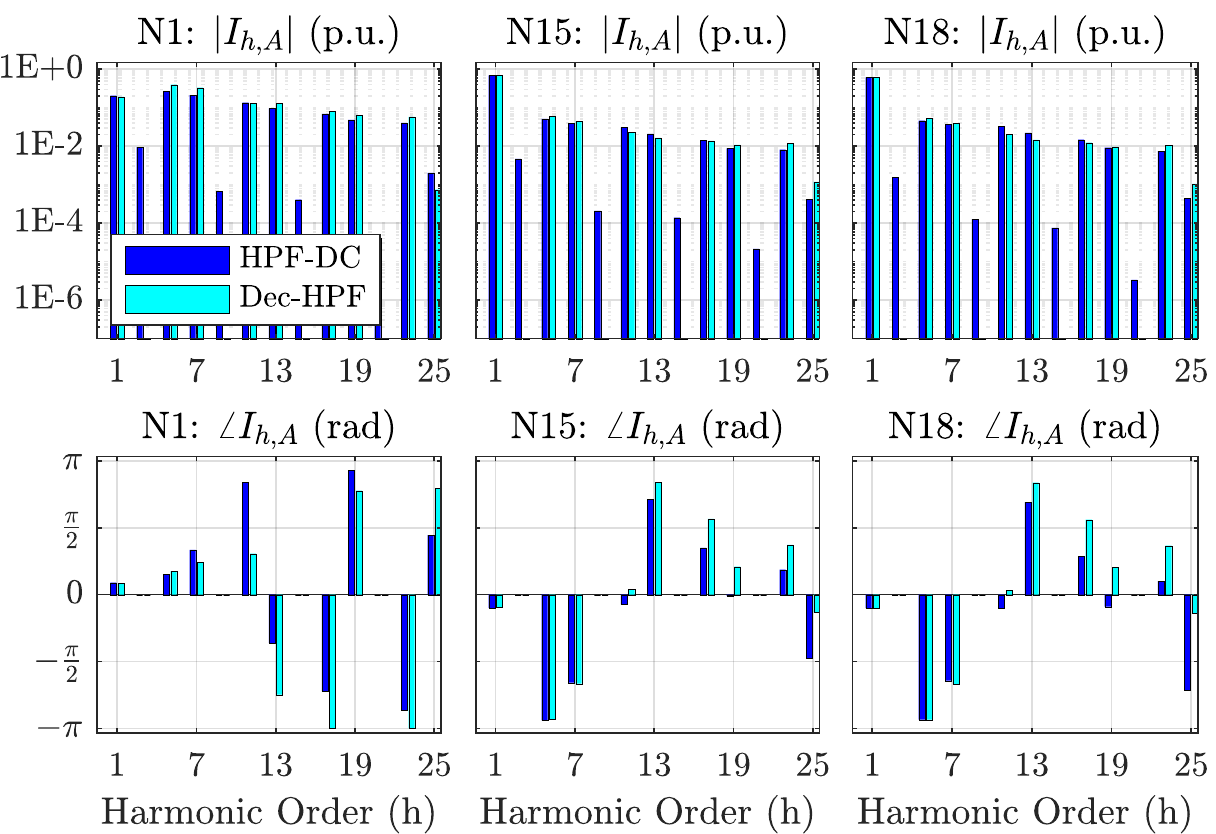}
        \label{fig:D-HPF:current}
    }
    
    \caption{%
        Comparison of the \HPF study with the decoupled \HPF study.
        The voltages for Phase A are given in (\ref{fig:D-HPF:voltage}) and the currents in (\ref{fig:D-HPF:current}).
    }
    \label{fig:D-HPF}
\end{figure}

Lastly, the proposed extended \HPF framework is benchmarked w.r.t. a classical decoupled \HPF.
In the classical decoupled \HPF the \CIDER[s] are represented by independent and superposed harmonic current sources.
The complex ratios of the harmonic currents (see \eqref{eq:dhpf:ratios} in the Appendix~\ref{app:D-HPF}) are derived beforehand for a \CIDER operating at rated power.
For this purpose, the \CIDER is connected directly to the \TE of the system validation (similarly to the setup shown in \cref{fig:val-rsc:setup}).
\cref{fig:D-HPF} shows the comparison between the proposed \HPF method and the classical decoupled \HPF method.
More precisely, the spectra of voltages and currents for three nodes throughout the system are presented.
The spectra show non-negligible differences in both magnitude and phase.
One can conclude that the proposed \HPF is more accurate than the classical decoupled version.

\newpage
\section{Conclusions}
\label{sec:conclusion}

This paper proposes an extension for the \HPF study of three-phase power grids with \CIDER[s] introduced in \cite{jrn:2020:kettner-becker:HPF-1,jrn:2020:kettner-becker:HPF-2}.
More specifically, it is shown in this paper how the \DC-side of a \CIDER is included into the \CIDER model and the \HPF study.
The \DC side is represented by the current-source model together with the \DC-link capacitor and is connected to the \AC side through an \AC/\DC converter.
Including the latter into the \CIDER model introduces a nonlinearity that needs to be approximated for the numerical solution of \HPF study.
To this end, it is shown how nonlinearities within the internal structure of the \CIDER can be linearized and incorporated into the \HPF algorithm.
In particular, the linearization takes into account the entire harmonic content of a signal, as opposed to the fundamental component only.
The extended \HPF method allows to study \AC/\DC and converter interactions in a power system.
The validation of the individual \CIDER model confirms that the linearization is indeed accurate.
The validation of the extended algorithm is performed on an entire system (i.e., the \CIGRE low-voltage benchmark microgrid), and provides high precisions too.
The largest observed errors are 2.8E-4~p.u. w.r.t. current magnitude, 8.7E-5~p.u. w.r.t. voltage magnitude, and 22.5~mrad w.r.t. phase.
The execution of the \HPF method for the benchmark system is up to three times faster as compared to the \TDS (incl. the Fourier analysis).
Additionally, the extended \HPF method is compared against its predecessor which neglects the \DC side of the \CIDER[s] and against an existing method (i.e., a decoupled \HPF).
The findings presented in this paper show that it is possible to accurately analyze \AC/\DC and converter interactions using the extended \HPF method.

\appendices

\section{Underlying Hypotheses of the \HPF Framework}
\label{app:HPF_hyp}

Brief summaries of the hypotheses from \cite{jrn:2020:kettner-becker:HPF-1,jrn:2020:kettner-becker:HPF-2} are shown in \cref{tab:HPF_hyp:Part1,tab:HPF_hyp:Part2}.

\begin{table}[t]
    \centering
    \caption{
        Summary of Hypotheses from \cite{jrn:2020:kettner-becker:HPF-1}.
    }
    \label{tab:HPF_hyp:Part1}
	{
\renewcommand{\arraystretch}{1.1}

\begin{tabular}{c p{0.8\linewidth}}
    \hline
        Hyp.
    &   Summary
    \\
    \hline
        1
    &   The lumped elements of the grid model are linear and passive.
        Therefore its circuit equations can be formulated independently at each frequency $f$ using either impedance or admittance parameters.
    \\
        2
    &   The compound branch impedance and shunt admittance matrices are symmetric, invertible, and lossy at all frequencies.
    \\
        3
    &   There exists a steady state in which all time-variant quantities are periodic.
        Therefore, all spectra can be characterized by the fundamental frequency $f_1$ and the harmonic orders $h \in \harmonics$.
    \\
        4
    &   The control software is a digital discrete-time system.
        The conversion from analogue to digital signals is done such that the frequency band of interest for \HPF studies can be reconstructed exactly.
    \\
        5 \& 6
    &   The closed-loop model of the \CIDER[s] and the associated transfer functions exist.
        That is, the matrix inversion in the closed-loop gain can be evaluated.
    \\
        7
    &   The function which approximates the reference calculation in the harmonic domain is differentiable.
    \\
    \hline
\end{tabular}
}
\end{table}

\begin{table}[t]
    \centering
    \caption{
        Summary of Hypotheses from \cite{jrn:2020:kettner-becker:HPF-2}.
    }
    \label{tab:HPF_hyp:Part2}
	{
\renewcommand{\arraystretch}{1.1}

\begin{tabular}{c p{0.8\linewidth}}
    \hline
        Hyp.
    &   Summary
    \\
    \hline
        1
    &   In the frequency range of interest for the \HPF study, switching effects are negligible.
        Therefore, the actuator can be regarded as an ideal voltage source.
    \\
        2
    &   The compound electrical parameters of the filter stages are diagonal matrices with equal nonzero entries.
    \\
        3
    &   Each controller stage consists of a PI controller for feed-back control, and two proportional controllers for feed-forward and feed-through controls.
    \\
        4
    &   The feed-back and feed-through gains are diagonal matrices with equal nonzero entries.
    \\
        5
    &   The feed-forward gains can be set in order to achieve zero error in steady-state.
    \\
        6
    &   The reference angle $\theta$, w.r.t. which the \cmpDQ~frame is defined, is synchronized with the fundamental tone.
    \\
        7
    &   In steady state, the frequency setpoints of all grid-forming \CIDER[s] are equal to the fundamental frequency.
    \\
        8
    &   The reference voltage for the grid-forming \CIDER is calculated 
    as $\VT^*_{\varphi,\cmpDQ}(t) = \sqrt{\frac{3}{2}} \left[V_\sigma~0\right]^\top$.
    \\
        9
    &   The synchronization units of the grid-following \CIDER[s] lock to the fundamental positive-sequence component of the grid voltage.
    \\
        10
    &   The time-variant signal content of the grid voltage in the \cmpDQ~reference frame is small compared to the time-invariant signal content.
    \\
        11
    &   For the calculation of the reference current in the grid-following \CIDER[s], the reciprocal of the $\cmpD$-component of the grid voltage is approximated by a second-order Taylor series.
    \\
    \hline
\end{tabular}
}
\end{table}

\section{Measurement Accuracy of Instrument Transformers}
\label{app:IT_accuracy}

The accuracy of instrument transformers, is defined by several standards \cite{Std:IEC_61869_6:2016,Std:IEC_61869_3:2011,Std:IEC_61869_2:2012}. 
For the measurements of harmonics, it is refered to commonly used 0.5~class instrument transformers, whose accuracies are given in \cref{tab:IEC_accuracies}.
The values are defined in percentage of the rated voltage magnitude.
\begin{table}[t]
	\centering
	\caption{Accuracy requirements for instrument transformers for voltage measurements including harmonics \cite{Std:IEC_61869_3:2011,Std:IEC_61869_6:2016}.}
	\label{tab:IEC_accuracies}
	
\renewcommand{\arraystretch}{1.1}
\begin{tabular}{crr}
    \hline
        $h$
    &   \multicolumn{1}{c}{$e(\Abs{V})$}
    &   \multicolumn{1}{c}{$e(\Arg{V})$}
    \\
    \hline
        1
    &   0.5\,\%
    &   6\,mrad
    \\
        2-4
    &   5\,\%
    &   87.3\,mrad
    \\
        5-6
    &   10\,\%
    &   174.5\,mrad
    \\
        7-9
    &   20\,\%
    &   349.1\,mrad
    \\
        10-13
    &   20\,\%
    &   349.1\,mrad
    \\
        above 13
    &   20-100\,\%
    &   \multicolumn{1}{c}{-}
    \\
    \hline
\end{tabular}
\end{table}

\section{Decoupled Harmonic Power Flow}
\label{app:D-HPF}

In the decoupled \HPF, the \CIDER[s] are represented by harmonic current sources.
The harmonic content of these current sources is determined a priori through appropriate simulations or measurements.
More precisely, the harmonic current phasors are characterized by a complex ratio w.r.t. the fundamental current phasor:
\begin{align}
    \alpha_{h} = \frac{I_{h}}{I_{1}} \in \mathbb{C}
    \label{eq:dhpf:ratios}
\end{align}
For the \HPF analysis, a power flow study is first performed at the fundamental frequency.
Then, the harmonic currents are inferred from the calculated fundamental currents and the available harmonic ratios.
The system equations are solved independently at each harmonic frequency using the hybrid parameters of the grid.
For further details please refer to~\cite{Ulinuha:AUPEC:2007}.



\bibliographystyle{IEEEtran}
\bibliography{Bibliography}

\begin{IEEEbiography}[{\includegraphics[width=1in,height=1.25in,clip,keepaspectratio]{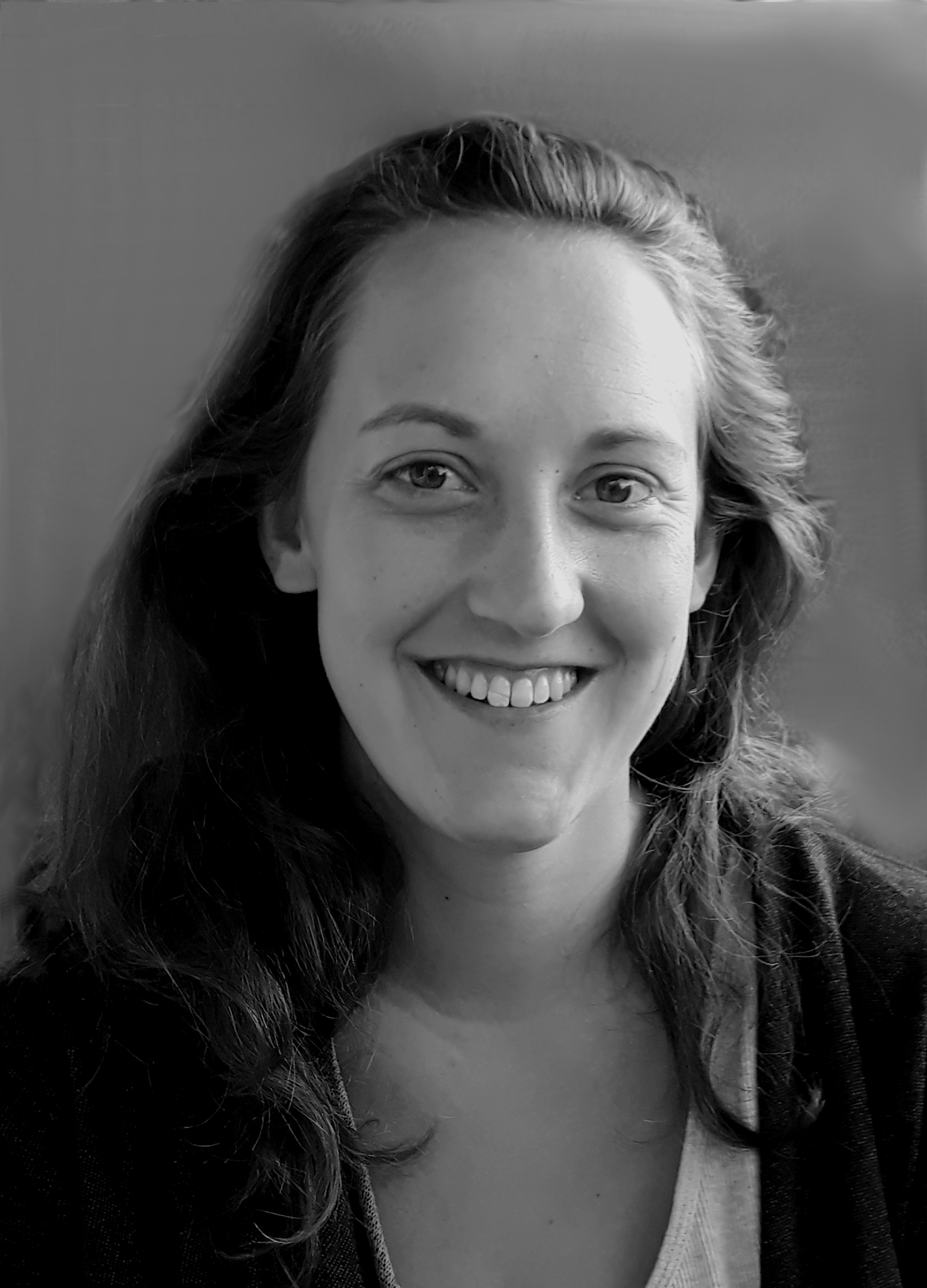}}]{Johanna Kristin Maria Becker}
	(S’19) received the B.Sc. degree in Microsystems Engineering from Freiburg University, Germany in 2015 and the M.Sc. degree in Electrical Engineering from the Swiss Federal Institute of Technology of Lausanne (EPFL), Lausanne, Switzerland in 2019.
    She is currently pursuing a Ph.D. degree at the Distributed Electrical System Laboratory, EPFL, with a focus on robust control and stability assessment of active distribution systems in presence of harmonics.
\end{IEEEbiography}

\begin{IEEEbiography}[{\includegraphics[width=1in,height=1.25in,clip,keepaspectratio]{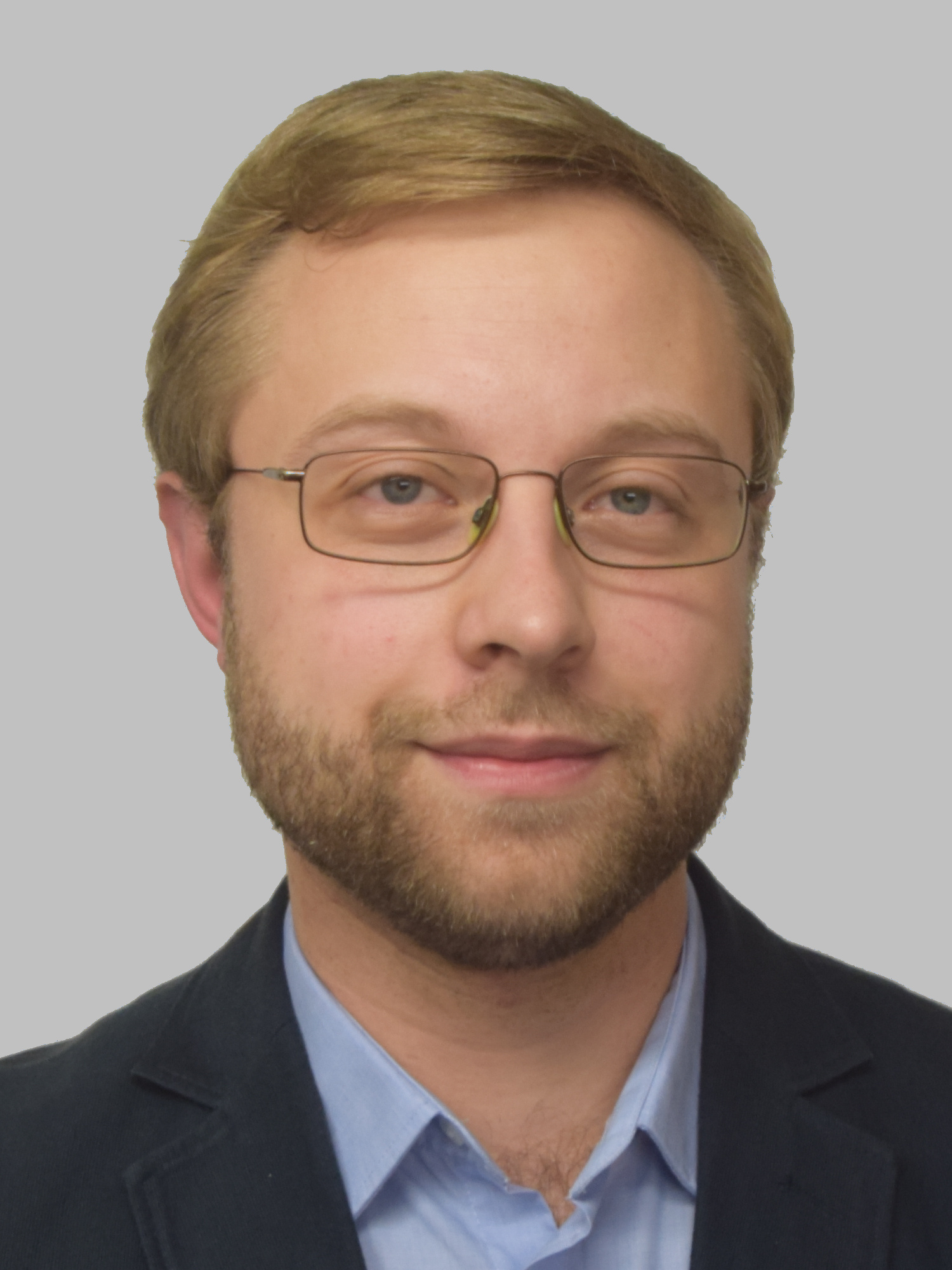}}]{Andreas Martin Kettner}
	(S’15-M’19) received the M.Sc. degree in electrical engineering and information technology from the Swiss Federal Institute of Technology of Zürich (ETHZ), Zürich, Switzerland in 2014, and the Ph.D. degree in power systems engineering from the Swiss Federal Institute of Technology of Lausanne (EPFL), Lausanne, Switzerland in 2019.
	He was a postdoctoral researcher at the Distributed Electrical Systems Laboratory (DESL) of EPFL from 2019 to 2020.
	Since then, he has been working as a project engineer for PSI NEPLAN AG in K{\"u}snacht, Switzerland and continues to collaborate with DESL.
\end{IEEEbiography}

\begin{IEEEbiography}[{\includegraphics[width=1in,height=1.25in,clip,keepaspectratio]{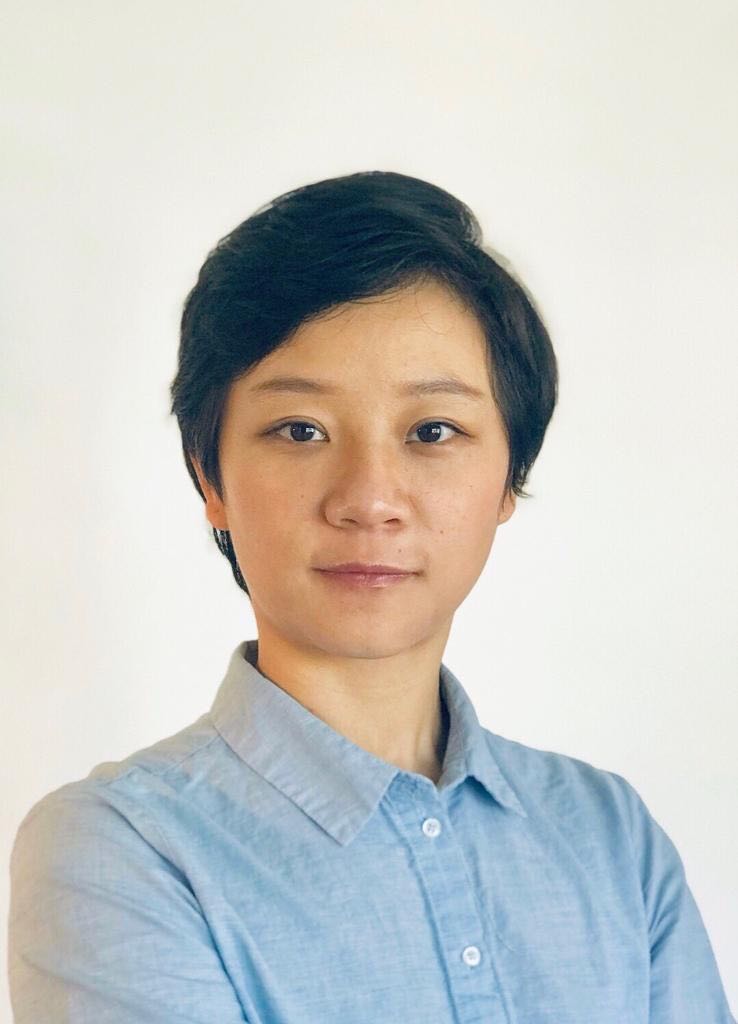}}]{Yihui Zuo}
    (M’18) received the B.Sc. and M.Sc. degrees in electrical engineering from North China Electric Power University, Beijing, China, in 2013 and 2016, respectively, and the Ph.D. degree in electrical engineering from the Swiss Federal Institute of Technology (EPFL), Lausanne, Switzerland, in 2021. 
    She is currently a postdoctoral researcher with the Distributed Electrical System Laboratory, EPFL. 
    Her research interests focus on power grids with increasing converter-interfaced energy resources, including the development of system-level control for electrical grids and the device control of distributed resources. 
\end{IEEEbiography}

\begin{IEEEbiography}[{\includegraphics[width=1in,height=1.25in,clip,keepaspectratio]{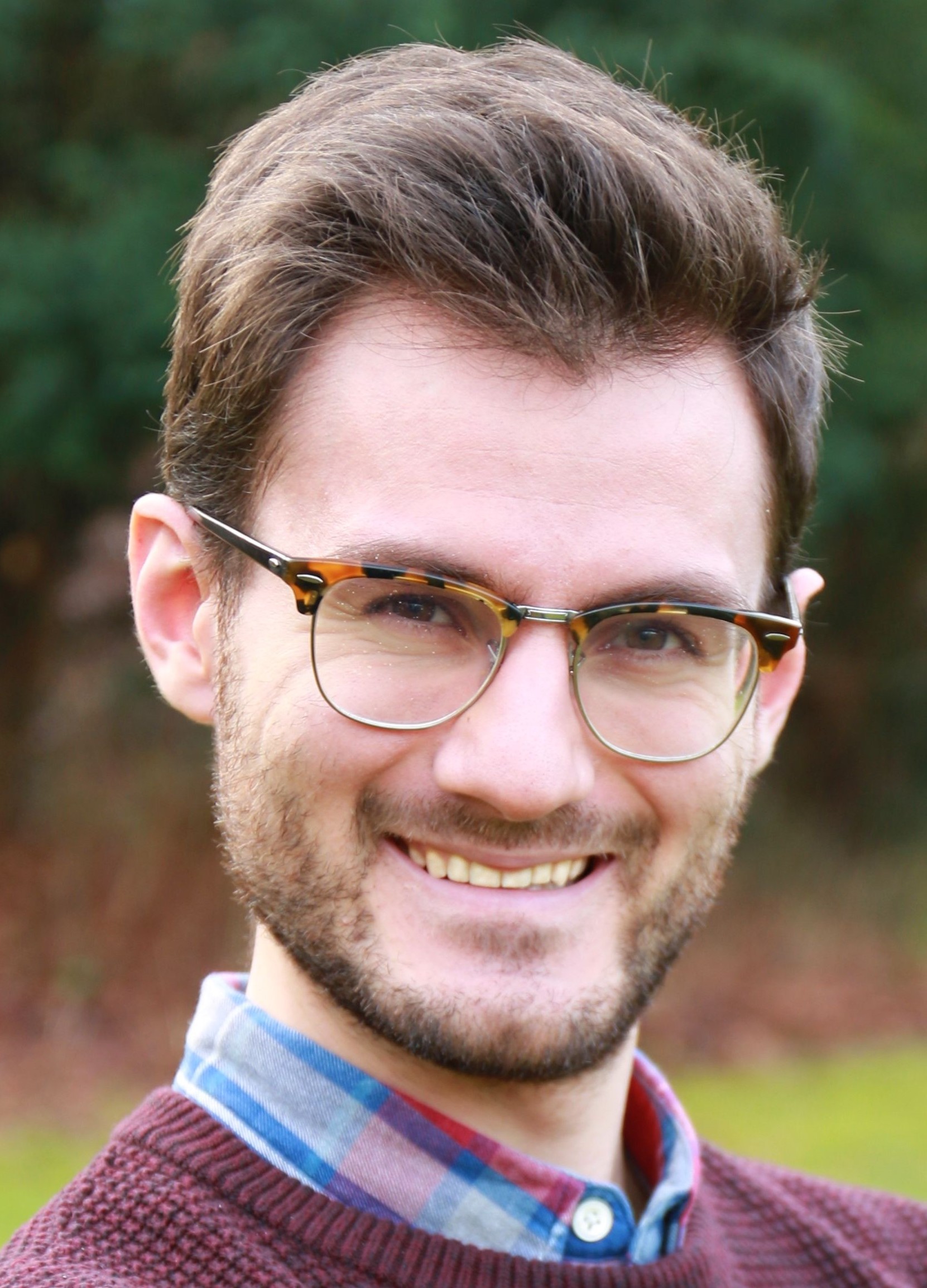}}]{Federico Cecati}
    (S'18) received the B.Sc. and M.Sc. degree in Automatic Control Engineering from the University of L'Aquila, L'Aquila, Italy in 2015 and 2017, respectively. 
    Since 2018 he is a Ph.D. researcher at the Chair of Power Electronics, Kiel University, Kiel, Germany. 
    From September 2020 to February 2021 he was a guest researcher at the Institute of Energy Technology, Aalborg University, Aalborg, Denmark. 
    His research interests include modelling, stability analysis, harmonic propagation and control in power electronics-based power systems.
\end{IEEEbiography}

\begin{IEEEbiography}[{\includegraphics[width=1in,height=1.25in,clip,keepaspectratio]{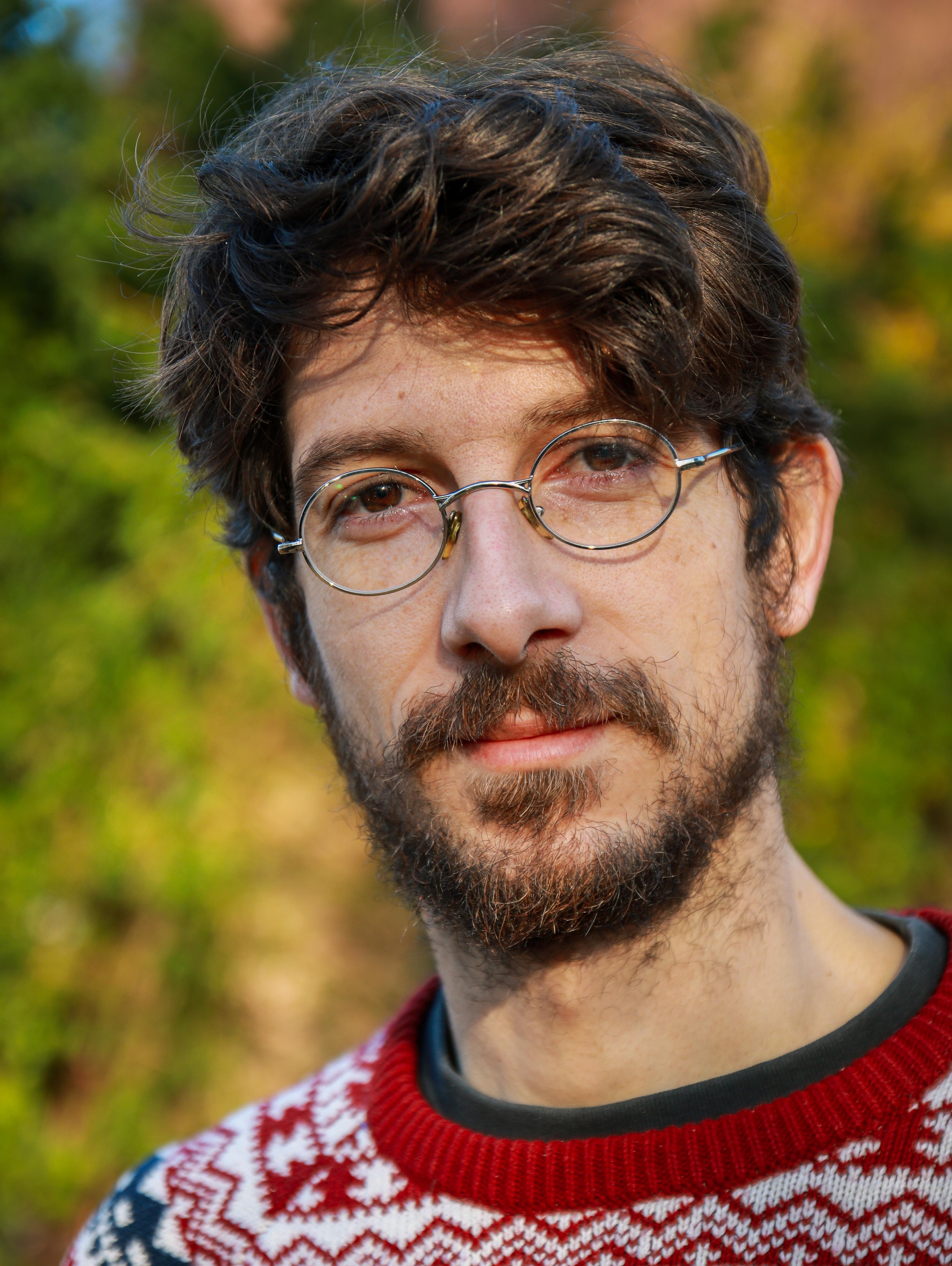}}]{Sante Pugliese}
    (M’18) received the M.Sc. degree in automation engineering and the Ph.D. degree in electrical and information engineering from the Politecnico di Bari, Bari, Italy, in 2013 and 2018, respectively. 
    In 2017, he was a Visiting Scholar with the Chair of Power Electronics, Kiel, Germany, where he is currently a Post-Doctoral Researcher. 
    In 2018-2019, he was Post-doc responsible in the EEMSWEA (0325797A) research project, Medium Voltage Grid Analyzer - Mittel Spannungs Netz Analyse. 
    In 2020 he was Post-doc responsible in the Add-On (0350022B) research project funded by the Bundesministerium für Wirtschaft und Energie. 
    His research interests include power converters and control techniques for distributed power generation systems based on renewable energies.
\end{IEEEbiography}

\begin{IEEEbiography}[{\includegraphics[width=1in,height=1.25in,clip,keepaspectratio]{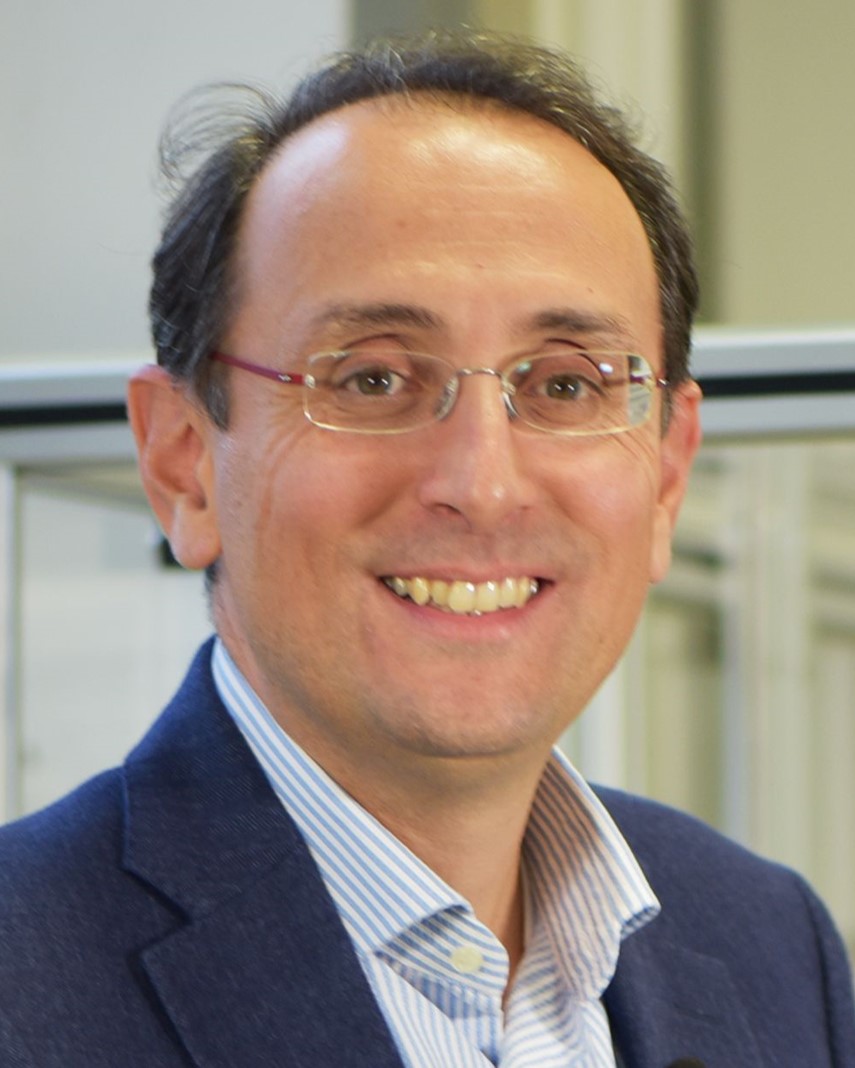}}]{Marco Liserre}
    (S'00-M'02-SM'07-F'13) received the MSc and PhD degree in Electrical Engineering from the Bari Polytechnic, respectively in 1998 and 2002. 
    He has been Associate Professor at Bari Polytechnic and from 2012 Professor in reliable power electronics at Aalborg University (Denmark). 
    From 2013 he is Full Professor and he holds the Chair of Power Electronics at Kiel University (Germany). He has published 500 technical papers (1/3 of them in international peer-reviewed journals) and a book. These works have received more than 35000 citations. 
    Marco Liserre is listed in ISI Thomson report “The world’s most influential scientific minds” from 2014. 
    He has been awarded with an ERC Consolidator Grant for the project “The Highly Efficient And Reliable smart Transformer (HEART), a new Heart for the Electric Distribution System”.
    He is member of IAS, PELS, PES and IES. 
    He has been serving all these societies in different capacities. 
    He has received the IES 2009 Early Career Award, the IES 2011 Anthony J. Hornfeck Service Award, the 2014 Dr. Bimal Bose Energy Systems Award, the 2011 Industrial Electronics Magazine best paper award in 2011 and 2020 and the Third Prize paper award by the Industrial Power Converter Committee at ECCE 2012, 2012, 2017 IEEE PELS Sustainable Energy Systems Technical Achievement Award and the 2018 IEEE-IES Mittelmann Achievement Award.
\end{IEEEbiography}

\begin{IEEEbiography}[{\includegraphics[width=1in,height=1.25in,clip,keepaspectratio]{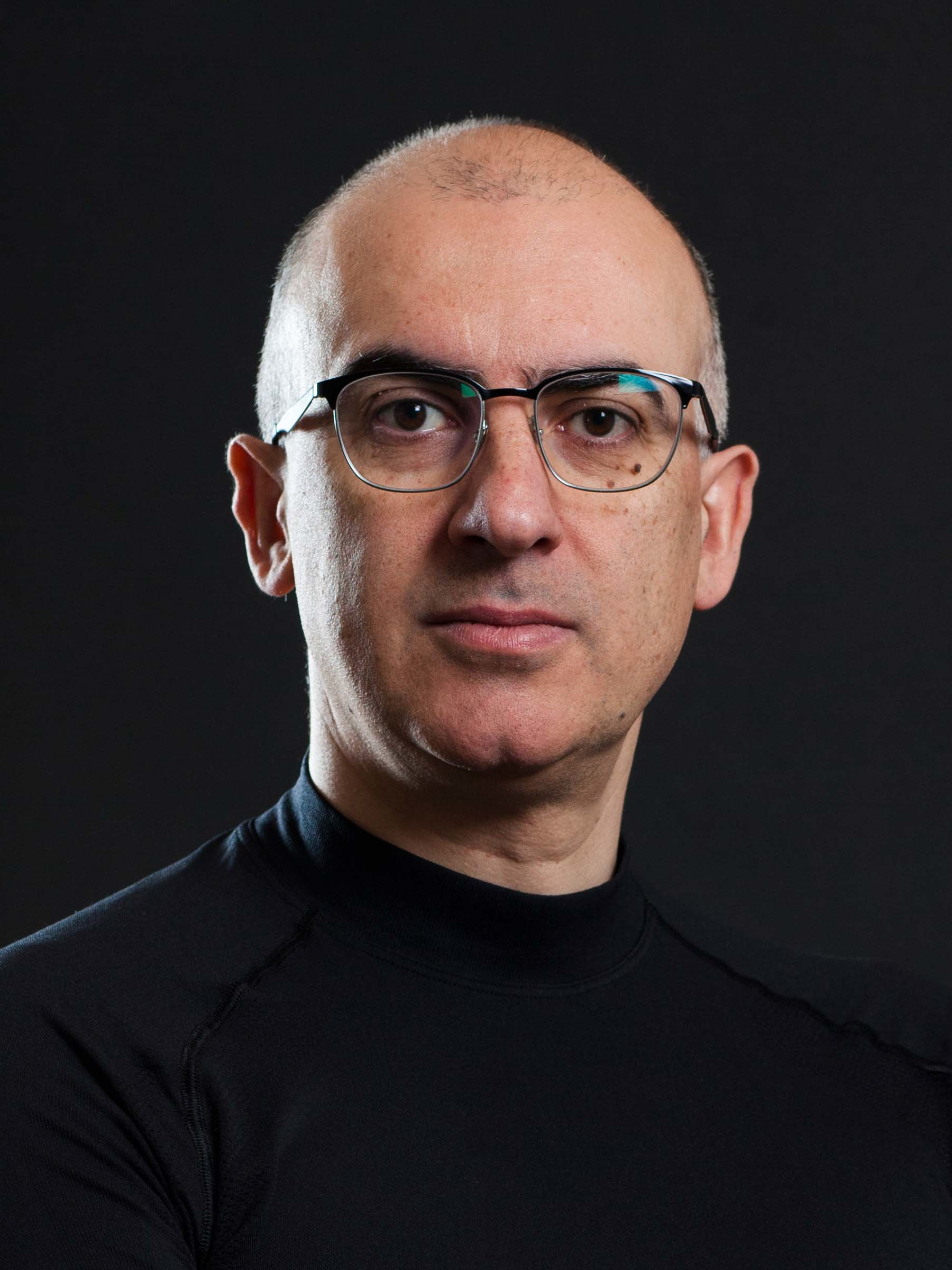}}]{Mario Paolone}
	(M’07–SM’10-F'22) received the M.Sc. (Hons.) and Ph.D. degrees in electrical engineering from the University of Bologna, Italy, in 1998 and 2002. 
    From 2005 to 2011, he was an Assistant Professor in power systems with the University of Bologna. 
    Since 2011, he has been with the Swiss Federal Institute of Technology, Lausanne, Switzerland, where he is Full Professor and the Chair of the Distributed Electrical Systems Laboratory. 
    His research interests focus on power systems with particular reference to real-time monitoring and operational aspects, power system protections, dynamics and transients. 
    Dr. Paolone’s most significant contributions are in the field of PMU-based situational awareness of active distribution networks (ADNs) and in the field of exact, convex and computationally-efficient methods for the optimal planning and operation of ADNs. 
    Dr. Paolone is Fellow of the IEEE and was the founder Editor-in-Chief of the Elsevier journal Sustainable Energy, Grids and Networks.
\end{IEEEbiography}

\end{document}